\documentclass{article}
\usepackage{amsmath}
\usepackage{graphicx}
\usepackage{amssymb}

\numberwithin{equation}{section}

\begin{document}

\title{Nonperturbative calculational method \\
in quantum field theory}

\author{Vladimir Dzhunushaliev
\thanks{e-mail: dzhun@hotmail.kg}\\
\textit{Dept. Phys. and Microel. Engineer.,} \\
\textit{Kyrgyz-Russian Slavic University}\\
\textit{Bishkek, Kievskaya Str. 44, 720021, Kyrgyz Republic} \\
~ \\
Douglas Singleton\thanks{e-mail: dougs@csufresno.edu}\\
\textit{Physics Dept., CSU Fresno,} \\
\textit{M/S MH37 Fresno, CA 93740-8031, USA}\\
\textit{and}\\
\textit{ Dept. of Theo. Phys., PFUR,} \\
\textit{6 Miklukho-Maklaya St., Moscow 117198, Russia} \\
~ \\
Tatyana Nikulicheva \\
\textit{Dept. Phys. and Microel. Engineer.,} \\
\textit{Kyrgyz-Russian Slavic University}\\
\textit{Bishkek, Kievskaya Str. 44, 720021, Kyrgyz Republic}}

\date{}

\maketitle

\pagestyle{myheadings} 
\thispagestyle{plain}         
\markboth{V. Dzhunushaliev, D. Singleton and T. Nikulicheva}
{Nonperturbative calculational method in quantum field theory} 
\setcounter{page}{1}         

\begin{abstract}
An approximate procedure for performing nonperturbative
calculations in quantum field theories is presented. The focus
will be quantum non-Abelian gauge theories with the goal of
understanding some of the open questions of these theories such as
the confinement phenomenon and glueballs. One aspect of this
nonperturbative method is the breaking down of the non-Abelian
gauge group into smaller pieces. For example $SU(2) \rightarrow
U(1) + coset$ or $SU(3) \rightarrow SU(2) + coset$. The procedure
also uses some aspects of an old method by Heisenberg to calculate
the n-point Green's function of a strongly interacting, non-linear
theory. Using these ideas we will give approximate calculations of
the 2 and 4-points Green's function of the theories considered.
\end{abstract}

\section{Introduction}

Modern perturbative quantum field theories (QFT) are based on the concept of a
particle or field quanta. This perturbative paradigm has given rise to the
most accurate description of some aspects of the physical world. The
standard example being the theoretical calculation of the anomalous magnetic moment
of the electron where theory and experiment are in agreement to the
order of $10^{-9}$. Mathematically the elementary particles that
arise in perturbative quantum field theory are harmonic excitations
of fundamental  fields. The quanta are defined
through the creation and annihilation operators, $a^{\dagger}$ and
$a$, of the ``second-quantized" theory. For instance, in quantum
electrodynamics (QED), the entire electromagnetic field can be
seen as a collection of superposed quanta, each with an energy
$\omega_k$. The Hamiltonian operator of the electromagnetic field (omitting
the zero-point energy) can be written
\begin{equation}
\hat H  =  \sum_k \hat N_k \omega_k,
\label{intr-10}
\end{equation}
where
\begin{equation}
 \hat N_k = \hat a_k^{\dagger} \hat a_k ,
\label{intr-20}
\end{equation}
is the ``number operator'', {\it i.e.} giving the number of quanta
with a specific four-momentum $k$  when operating on a free state
\begin{equation}
 \hat N_k |... n_k ... \rangle = n_k |... n_k... \rangle.
\label{intr-30}
\end{equation}
where $n_k$ is a positive integer, the number of quanta with that
particular momentum. The energy in the electromagnetic field is
thus the eigenvalue of the Hamiltonian (1). The reasoning for
fermion fields is the same, but then the number of quanta in any
given state can be only 0 or 1 due to Fermi statistics.
\par
Usually, in quantum field theories (for example, in QED) ``fields''
and ``particles'' are used interchangeably. This is
because the almost universal usage of Feynman diagrams gives the
false impression that a field is always equivalent to a ``cloud" of quanta.
The Feynman diagram technique (based on propagators and vertices)
can be justified as an approximation for mildly nonlinear theories
({\it i.e.} in the weak coupling limit) but breaks down for strongly coupled
Abelian and non-Abelian theories.
\par
The physical reason of the appearance of quanta is that in linear
theories (for example, in electrodynamics) a general solution can be
obtained by making the Fourier expansion:
\begin{equation}
A_{\mu}^{linear} = \int d^3 k \sum_{\lambda = 0}^3 [a_k (\lambda)
\epsilon_{\mu}(k, \lambda) e^{-i k \cdot x} + a_k^{\dagger}
(\lambda) \epsilon_{\mu}^{\ast} (k, \lambda) e^{i k \cdot x}],
\label{intr-50}
\end{equation}
where $\epsilon_{\mu}$ is the polarization vector.
\par
However, for a theory based on a non-Abelian group, like quantum
chromodynamics (QCD), this can no longer be done in the strong
coupling limit, due to the nonlinear nature of Yang-Mills
equations
\begin{eqnarray}
A_{\mu}^{b \, QCD} \neq   \int d^3 k \sum_{\lambda = 0}^3 [a_k^b
(\lambda) \epsilon_{\mu}(k, \lambda) e^{-i k \cdot x}  + a_k^{b \,
\dagger} (\lambda) \epsilon_{\mu}^{\ast} (k, \lambda) e^{i k \cdot
x}].
\label{intr-60}
\end{eqnarray}
Formally one can change this inequality to an equality
but this is only a good approximation in the limit when the
coupling is weak -- $g_s \ll 1$. Due to the nonlinear
character of non-Abelian gauge theories the superposition
principle does not hold in general. There are plane wave solutions
in non-Abelian gauge theories [1] however one can not superpose different
plane waves as is done in \eqref{intr-50}. Thus, color fields can be represented
by harmonic oscillators (gluons) only in the limit when the
strong interaction coupling constant tends to zero, $g_s
\rightarrow 0$ (or equivalently when
$Q^2 \rightarrow \infty$ because of asymptotic freedom).
There are certain nonperturbative field configurations (for example, the
hypothesized color electric flux tube that is thought to stretch between
quarks and lead to the dual superconducting picture of confinement) in which the
field distribution  can not be explained as a cloud of quanta. The
{\it fields} are there, but their {\it quanta} -- gluons and
quarks -- are relevant only when probed at sufficiently
short distances (or large momentum). One way of looking at this
situation is that the fields of these nonperturbative configurations
are split into ordered fields (for example, the fields inside flux tube stretched
between quark and antiquark) and disordered fields (the fields outside the
flux tube). These field can not be interpreted as a (perturbative) cloud
of quanta. In such situations the fields play a primary role over the
particles \cite{hansson}.
\par
The need for nonperturbative techniques in strongly
interacting, nonlinear quantum field theories is an old
problem that has been around since the beginning of
the study of quantum fields. Much effort has gone into trying to
resolve  this puzzle. The different approaches that have been tried include:
(i) lattice QCD  \cite{karsch}-\cite{Jansen}, (ii) the dual
Meissner effect in the QCD-vacuum \cite{Pandey}-\cite{Agasian},
(iii) instantons \cite{Suganuma} \cite{Shifman},  (iv) path integration
\cite{Kondo1}-\cite{Kondo6}, (v) analytic calculations \cite{Simonov},
(vi) Dyson-Schwinger equations \cite{Gentles} (vii) and various other
methods \cite{Ferreira}-\cite{Simonov3}. Despite this the problem is not yet fully resolved.
In the 1950's Heisenberg \cite{Heisenberg1} \cite{Heisenberg2}
studied a nonlinear spinor field
and worked out nonperturbative techniques for quantizing
the nonlinear spinor field. In this method one writes down an infinite
system of equations which connects together all the
n-point Green's functions of the theory (this can be compared to the
infinite number of Feynman diagrams which must, in
principle, be calculated for a given process in pertubative
quantum field theory). In order to solve this system of equations
one must find some physically reasonable approximations
for cutting off this system of equations at
some point so that one reduces the infinite system of equations
into a finite system. Some preliminary work \cite{dzhsin1}-\cite{dzhsin4}
has been done in
applying these nonperturbative quantization techniques,
in conjunction with ideas from Abelian projection, to
study quantum SU(2) and SU(3) gauge theories, and also to examine
the possible application of these techniques to the high-$T_c$ superconductors
of condensed matter physics.  The advantage of this approach is that it
is nonperturbative. The problem is to find physically reasonable motivations
for cutting off  the infinite equations set for Green's functions.
\par
The preliminary work in this nonperturbative approach
was devoted to a group of classical singular solutions of the
Yang-Mills field equations, and their quantization
\cite{dzhsin1}-\cite{dzhsin4}. In these papers some singular classical solutions
were found, and they were then quantized using a nonperturbative
method with the approximation that
some of the degrees of freedom of the non-Abelian gauge theory
could be treated as almost classical degrees of freedom
(this was the approximation which cut off the infinite
system of equations). After the quantization procedure was
applied  the bad asymptotic behavior (the divergence of some
components of the Yang-Mills fields) of the original classical
solution was modified into acceptable asymptotic behavior.
Physically the nonperturbative quantization smoothed out
the bad classical behavior of the solutions.
\par
In later work \cite{dzhsin5} the role of the other,
non-classical degrees of freedom was investigated.
In Ref. \cite{dzhsin5} the above approach was applied to the
SU(2) non-Abelian gauge theory. The assumption was made
that the SU(2) gauge field $A^a_\mu$ could be split into
components which had different qualitative behavior: one
component, $A^3_\mu$ , was taken to be in an ordered phase, while the
remaining two components, $A^{1,2}_\mu$, were in a disordered
phase. Physically  the ordered phase described the U(1)
electric field in a hypothesized flux tube stretched
between sources ({\it i.e} quarks) of this field, and the disordered
phase described a condensate which ``pushes out" ({\it i.e.} exhibits the
Meissner effect)  the ordered phase. The calculations of this \cite{dzhsin5}
indicated that such a scenario did occur.
In this paper the dynamics of the disordered phase were ``frozen" or
taken to be nondynamical.
\par
In the present paper we would like to bring together and extend
several results of these analytical calculations in
nonperturbative QFT. Two general types of calculations will be
presented: (i) the decomposition of the SU(3) gauge field
components into SU(2) + coset components. This will lead to flux
tube solutions which are important in the dual superconducting
model of confinement. (ii) The simplification of the SU(3)
quantized fields using a scalar field. This will lead to a simple
model for a glueball.

\section{Heisenberg quantization method}
\label{heis1}

\subsection{Initial Heisenberg idea for the quantization \\
of a nonlinear spinor field}
\label{heis2}

Heisenberg's basic idea is that the n-point Green's
functions can be found from some infinite set of coupled
differential equations, which are
derived from the field equations for the field operators. As an example
we show how this method of quantization works for a spinor field with
a nonlinear self interaction \cite{Heisenberg1} \cite{Heisenberg2}.
The basic equation for the spinor field is :
\begin{equation}
\label{sec1-10}
\gamma ^{\mu} \partial _{\mu} {\hat \psi} (x) -
l^2 \Im [{\hat \psi}(x) ({\hat {\bar \psi}}(x)
{\hat \psi}(x)) ] = 0
\end{equation}
where $\gamma ^{\mu}$ are Dirac matrices;
${\hat \psi} (x), {\hat {\bar \psi}}(x)$ are
the operators of the spinor field and its adjoint respectively;
$\Im [{\hat \psi} ({\hat {\bar \psi}}
{\hat \psi)} ] = {\hat \psi} ({\hat {\bar \psi}}
{\hat \psi})$ or ${\hat \psi} \gamma ^5
({\hat {\bar \psi}} \gamma ^5{\hat \psi})$ or
${\hat \psi} \gamma ^{\mu}
({\hat {\bar \psi}} \gamma _{\mu} {\hat \psi})$ or \\
${\hat \psi} \gamma ^{\mu} \gamma
^5 ({\hat {\bar \psi}} \gamma _{\mu} \gamma ^5{\hat \psi} )$.
The constant
$l$ has units of length, and sets the scale for the strength of the
interaction. Heisenberg emphasized that the 2-point Green's function,
$G_2 (x_2, x_1)$, in this theory differs strongly from the propagator in
a linear theory in its behavior on the light
cone : in the nonlinear theory $G_2 (x_2 , x_1)$ oscillates strongly on
the light cone in contrast to the propagator of the linear theory
which has a $\delta$-like singularity. Heisenberg defined
$\tau$ functions
\begin{equation}
\label{sec1-20}
\tau (x_1 x_2 ... | y_1 y_2 ...) = \langle 0 |
T[{\hat \psi} (x_1) {\hat \psi} (x_2) ...
{\hat {\bar \psi }} (y_1) {\hat {\bar \psi }} (y_2) ...] |
\Phi \rangle
\end{equation}
where $T$ is the time ordering operator; $| \Phi \rangle$ is a state for
the system described by Eq. (\ref{sec1-10}). Eq. (\ref{sec1-20})
establishes a one-to-one correspondence between the system state,
$| \Phi \rangle$, and the set of functions $\tau$. This state can be defined
using the infinite function set of Eq. (\ref{sec1-20}). Using
equation (\ref{sec1-10}) and (\ref{sec1-20}) we obtain the following infinite
system of equations for the $\tau $'s
\begin{eqnarray}
\label{sec1-30}
&&l^{-2} \gamma ^{\mu} _{(r)} \frac{\partial}{\partial x^{\mu} _{(r)}}
\tau (x_1 ...x_n |y_1 ... y_n ) = \Im [ \tau (x_1 ... x_n x_r |
y_1 ... y_n y_r)] + \nonumber \\
&&\delta (x_r -y_1) \tau ( x_1 ... x_{r-1} x_{r+1} ... x_n |
y_2 ... y_{r-1} y_{r+1} ... y_n ) + \nonumber \\
&&\delta (x_r - y_2) \tau (x_1 ... x_{r-1} x_{r+1} ... x_n |
y_1 y_2 ... y_{r-1} y_{r+1} ... y_n ) + ...
\end{eqnarray}
Eq. (\ref{sec1-30}) represents one of an infinite set of coupled equations
which relate various orders (given by the index $n$) of the $\tau$
functions to one another. To make some head way toward solving
this infinite set of equations Heisenberg employed the Tamm-Dankoff
method whereby he only considered $\tau$ functions up to a certain
order. This effectively turned the infinite set of coupled equations
into a finite set of coupled equations.
\par
Heisenberg used the procedure sketched above to study the
Dirac equation with a nonlinear coupling. Here we apply this
procedure to nonlinear, bosonic field theories such as QCD in
the low energy limit. In particular we will apply this method
to several solutions of the Yang-Mills field equations which
have divergent fields and infinite energy. By making certain
assumptions analogous to the Tamm-Dankoff cut-off,
the unphysical asymptotic behavior of these classical Yang-Mills
solutions will be ``smoothed'' out.

\subsection{Heisenberg quantization for QCD}

In this subsection we will apply a version of Heisenberg's quantization
method sketched above to QCD.  The classical SU(3) Yang-Mills equations are
\begin{equation}
    \partial_\nu \mathcal F^{B\mu\nu} = 0
\label{sec2-1-10}
\end{equation}
where $\mathcal F^B_{\mu \nu} = \partial_\mu \mathcal A^B_\nu -
\partial_\nu \mathcal A^B_\mu + g f^{BCD} \mathcal A^C_\mu \mathcal A^D_\nu$
is the field strength; $B,C,D = 1, \ldots ,8$ are the SU(3) color indices;
$g$ is the coupling constant; $f^{BCD}$ are the structure constants for
the SU(3) gauge group.
In quantizing the system given in Eqs. (\ref{sec2-1-10}) - via Heisenberg's method
one first replaces the
classical fields by field operators
$\mathcal A^B_{\mu} \rightarrow \widehat{\mathcal A}^B_\mu$. This yields the
following differential equations for the operators
\begin{equation}
    \partial_\nu \widehat {\mathcal F}^{B\mu\nu} = 0.
\label{sec2-1-20}
\end{equation}
These nonlinear equations for the field operators of
the nonlinear quantum fields can be used to determine
expectation values for the field operators
$\widehat {\mathcal A}^B_\mu$, where
$\langle \cdots \rangle = \langle Q | \cdots | Q \rangle$ and
$| Q \rangle$ is some quantum state). One can also use these
equations to determine the expectation values of operators
that are built up from the fundamental operators
$\widehat {\mathcal A}^B_\mu$. For example, the ``electric'' field
operator, $\widehat {\mathcal E}^B_z = \partial _0 \widehat {\mathcal A}^B_z -
\partial _z \widehat {\mathcal A}^B_0 + g f^{BCD} \mathcal A^C_0 \mathcal A^D_z$
giving the expectation $\langle \widehat {\mathcal E}^B_z \rangle$.
The simple gauge field expectation values,
$\langle \mathcal{A}_\mu (x) \rangle$, are obtained by
average Eq. \eqref{sec2-1-20} over some quantum state $| Q \rangle$
\begin{equation}
  \left\langle Q \left|
  \partial_\nu \widehat {\mathcal F}^{B\mu\nu}
  \right| Q \right\rangle = 0.
\label{sec2-1-30}
\end{equation}
One problem in using these equations to obtain expectation values
like $\langle \mathcal A^B_\mu \rangle$, is that these equations
involve not only powers or derivatives of $\langle \mathcal
A^B_\mu \rangle$ ({\it i.e.} terms like $\partial_\alpha \langle
\mathcal A^B_\mu \rangle$ or $\partial_\alpha
\partial_\beta \langle \mathcal A^B_\mu \rangle$) but also contain
terms like $\mathcal{G}^{BC}_{\mu\nu} = \langle \mathcal A^B_\mu
\mathcal A^C_\nu \rangle$. Starting with Eq. \eqref{sec2-1-30} one
can generate an operator differential equation for the product
$\widehat {\mathcal A}^B_\mu \widehat {\mathcal A}^C_\nu$ thus
allowing the determination of the Green's function
$\mathcal{G}^{BC}_{\mu\nu}$
\begin{equation}
  \left\langle Q \left|
  \widehat {\mathcal A}^B(x) \partial_{y\nu} \widehat {\mathcal F}^{B\mu\nu}(x)
  \right| Q \right\rangle = 0.
\label{sec2-1-40}
\end{equation}
However this equation will in turn contain other, higher order
Green's functions \cite{dzhsin1} -\cite{dzhsin3}.
Repeating these steps leads to an infinite set
of equations connecting Green's functions of ever increasing
order. This construction, leading to an infinite set of coupled,
differential equations, does not have an exact, analytical solution
and so must be handled using some approximation.
\par
Operators which are involved in equation \eqref{sec2-1-20}
are only well determined if there is a Hilbert space of quantum
states. Thus we need to look into the question of the definition of the quantum states
$| Q \rangle$ in the above construction. The resolution to this
problem is as follows: There is an one-to-one correspondence
between a given quantum state $| Q \rangle$ and the infinite set
of quantum expectation values over any product of field operators,
$\mathcal{G}^{mn \cdots}_{\mu\nu \cdots}(x_1, x_2 \ldots) =
\langle Q | \mathcal A^m_\mu (x_1) \mathcal A^n_\nu (x_2) \ldots
| Q \rangle$. So if all the Green's functions
-- $\mathcal{G}^{mn \cdots}_{\mu\nu \cdots}(x_1, x_2 \ldots)$ --
are known then the quantum states, $| Q \rangle$ are known,
\textit{i.e.} the action of $| Q \rangle$ on any product
of field operators
$\widehat {\mathcal A}^m_\mu (x_1) \widehat {\mathcal A}^n_\nu (x_2) \ldots$
is known. The Green's functions are determined from the above,
infinite set of equations (following Heisenberg's idea).
\par
Another problem associated with products of field operators like
$\widehat {\mathcal A}^m_\mu (x) \widehat {\mathcal A}^n_\nu (x)$
which occur in Eq. \eqref{sec2-1-20} is that the two operators
occur at the same point. For \textit{non-interacting} field it is
well known that such products have a singularity. In this paper we
are considering \textit{interacting} fields so it is not known if
a singularity would arise for such products of operators evaluated
at the same point. Heisenberg in his investigations of a quantized
nonlinear spinor field repeatedly underscored that in a quantum
field theory with strong interaction the singularities of
propagators can be essentially smoothed out. In this situation the
problem of the renormalization of nonrenormalizable QFT will be
crucially changed. This problem appears when we have the product
of propagators of some quanta, but the nonrenormalizable theories
have to be considered on the nonperturbative level which probably
gives rise to smoothed out, nonsingular Green's functions. Thus
one might suppose that the nonperturbative quantization methods
may resolve the problem of the quantization of nonrenomalizable
QFT. Physically it is hypothesized that there are situations in
interacting field theories where these singularities do not occur
({\it e.g.} for flux tubes in Abelian or non-Abelian theory
quantities like the ``electric'' field inside the tube, $\langle
\mathcal E^a_z \rangle < \infty$, and energy density $\varepsilon
(x) = \langle (\mathcal E^a_z)^2 \rangle < \infty$ are
nonsingular). Here we take as an assumption that such
singularities do not occur.

\section{The calculation of 2 and 4-points Green's functions}

The quantization of equations \eqref{sec2-1-30}-\eqref{sec2-1-40}
in the spirit of section \ref{heis1} evidently is very hard and
deriving exact results is probably impossible. In order to do some
calculations we give an approximate method which leads to the 2
and 4-points Green's functions only. In order to derive the
equations describing the quantized field we average the Lagrangian
over a quantum state $\left.\left. \right| Q \right\rangle$
\begin{equation}
\begin{split}
    &\left\langle Q \left| \widehat {\mathcal L}_{SU(3)}(x) \right| Q \right\rangle =
    \left\langle \widehat {\mathcal L}_{SU(3)} \right\rangle = \\
    &\frac{1}{2}
    \left\langle
      \left( \partial_\mu \widehat A^B_\nu (x) \right)
      \left( \partial^\mu \widehat A^{B\nu} (x) \right) -
      \left( \partial_\mu \widehat A^B_\nu (x) \right)
      \left( \partial^\nu \widehat A^{B\mu} (x) \right)
    \right\rangle + \\
    &\frac{1}{2}    g f^{BCD}
    \left\langle
      \left( \partial_\mu \widehat A^B_\nu (x)-
      \partial_\nu \widehat A^B_\mu (x)\right)
      \widehat A^{C \mu} (x)\widehat A^{D \nu}(x)
    \right\rangle + \\
    &\frac{1}{4}g^2 f^{BC_1D_1} f^{BC_2D_2}
    \left\langle
      \widehat A^{C_1}_\mu (x)\widehat A^{D_1}_\nu (x)
      \widehat A^{C_2 \mu} (x)\widehat A^{D_2\nu} (x)
    \right\rangle .
\end{split}
\label{sec3-5}
\end{equation}
Now we will detail the kind of physical situations we wish to
describe. The model given here is similar to stationary turbulence
when there are time dependent fluctuations in any point of the
liquid but all averaged quantities are time independent. For a QFT
this means that all Green's functions are time independent and
there is a correlation between quantum fields in different points
at one moment
\begin{eqnarray}
    \left\langle
    \mathcal A^{B_1}_{\mu_1}(x^\alpha_1) \cdots \mathcal
    A^{B_n}_{\mu_n}(x^\alpha_2)
    \right\rangle &\neq &0 ,
\label{sec3-10}\\
  t_1 &=&  \cdots = t_n ,
\label{sec3-20}\\
  \vec{r}_1 &\neq& \cdots \neq \vec{r}_n .
\label{sec3-30}
\end{eqnarray}
In linear and perturbative QFT this is not the case because the
interaction is carried by quanta which move with a speed less than
or equal to the speed of light. In these theories the correlation
between quantum fields in different points at the same time is
zero.
\begin{eqnarray}
    \left\langle
    \mathcal A^{B_1}_{\mu_1}(x^\alpha_1) \cdots \mathcal
    A^{B_n}_{\mu_n}(x^\alpha_2)
    \right\rangle &= &0 ,
\label{sec3-40}\\
  t_1 &=&  \cdots = t_n ,
\label{sec3-50}\\
  \vec{r}_1 &\neq& \cdots \neq \vec{r}_n .
\label{sec3-60}
\end{eqnarray}
In this sense one can say that nonperturbative QFT in some
physical situations is very close to turbulence, {\it i.e.} in
nonperturbative QFT there may exist extended objects where
quantized fields at all points are correlated between themselves
(example from QCD are flux tubes and glueballs). Such objects fall
into two categories
\begin{enumerate}
    \item
    The averaged value of all quantized fields are zero
    \begin{equation}
        \left\langle
        \mathcal A^B_\mu(x)
        \right\rangle = 0 .
    \label{sec3-70}
    \end{equation}
    But the square of these fields are nonzero
    \begin{equation}
        \left\langle
        \left( \mathcal A^B_\mu(x) \right)^2
        \right\rangle \neq 0
    \label{sec3-80}
    \end{equation}
    \item
    Some components of the quantized fields are nonzero and some zero,
    and the square of some components is nonzero
    \begin{eqnarray}
        \left\langle
        \mathcal A^{B_1}_\mu(x)
        \right\rangle &\neq & 0
        \quad \text{for some} \quad {B_1} \in 1,2, \cdots , 8
    \label{sec3-90}\\
    \left\langle
        \mathcal A^{B_2}_\mu(x)
        \right\rangle &=& 0
        \quad \text{for remaining} \quad {B_2} \in 1,2, \cdots , 8
    \label{sec3-100}\\
        \left\langle
        \left( \mathcal A^{B_3}_\mu(x) \right)^2
        \right\rangle &\neq & 0
        \quad \text{for some} \quad {B_3} \in 1,2, \cdots , 8
    \label{sec3-105}
    \end{eqnarray}
    The most natural case is when $\mathcal A^{B_1}_\mu$ belongs to a small subgroup
    of SU(3) gauge group (for example, to SU(2)) and $\mathcal A^{B_2}_\mu$ are
    the coset components and $B_3 = B_2$. For example,
    $B_1 = 1,2,3$ and $B_2 = 4,5,6,7,8$.
\end{enumerate}
In the first case the quantized fields are in a completely
disordered phase. In the second case one has both ordered and disordered
 phases. In this paper we present some calculations
which give illustrate each case. The first case of the completely disordered phase will
be illustrated by a glueball configuration; the second case of both ordered and disordered
phases will be illustrated by a flux tube solution.

\section{The first case: completely disordered phase -- glueball}
\label{glueball}

\subsection{The vector to scalar field replacement approximation -- $A^B_\mu \rightarrow \phi^B$}

In any quantum field theory the Green's functions contain the full
information about the quantized fields. In this section we will
present equations which describe only the 2 and 4-points Green's functions
for SU(3) gauge theory ({\it i.e.} QCD) using the
approximation where all the components of the gauge potential
are in a disordered phase. To begin we average the SU(3)
Lagrangian using some approximate expressions for the 2
and 4-points Green's functions. The SU(3) Lagrangian is
\begin{equation}
  \widehat {\mathcal L}_{SU(3)} = \frac{1}{4}
  \widehat {\mathcal F}^A_{\mu \nu}\widehat {\mathcal F}^{A \mu \nu}
\label{sec4-10}
\end{equation}
where $\widehat {\mathcal F}^B_{\mu \nu} = \partial_\mu \widehat {\mathcal A}^B_\nu -
\partial_\nu \widehat {\mathcal A}^B_\mu +
g f^{BCD} \widehat {\mathcal A}^C_\mu \widehat {\mathcal A}^D_\nu$
is the field strength operator; $B,C,D = 1, \ldots ,8$ are the SU(3) color indices;
$g$ is the coupling constant; $f^{BCD}$ are the structure constants for
the SU(3) gauge group; $\widehat {\mathcal A}^B_\mu$ is the gauge potential operator.
In order to derive equations describing the quantized field we average
the Lagrangian over a quantum state $\left.\left. \right| Q \right\rangle$
\begin{equation}
\begin{split}
    \left\langle Q \left| \widehat {\mathcal L} \right| Q \right\rangle =&
    \left\langle \widehat {\mathcal L} \right\rangle =
    \frac{1}{2}
    \left\langle
      \left( \partial_\mu \widehat {\mathcal A}^B_\nu  \right)
      \left( \partial^\mu \widehat {\mathcal A}^{B\nu}  \right) -
      \left( \partial_\mu \widehat {\mathcal A}^B_\nu  \right)
      \left( \partial^\nu \widehat {\mathcal A}^{B\mu}  \right)
    \right\rangle + \\
    &\frac{1}{2}    g f^{BCD}
    \left\langle
      \left( \partial_\mu \widehat {\mathcal A}^B_\nu -
      \partial_\nu \widehat {\mathcal A}^B_\mu \right)
      \widehat {\mathcal A}^{C \mu} \widehat {\mathcal A}^{D \nu}
    \right\rangle + \\
    &\frac{1}{4}g^2 f^{BC_1D_1} f^{BC_2D_2}
    \left\langle
      \widehat {\mathcal A}^{C_1}_\mu \widehat {\mathcal A}^{D_1}_\nu
      \widehat {\mathcal A}^{C_2 \mu} \widehat {\mathcal A}^{D_2} \nu
    \right\rangle
\end{split}
\label{sec4-30}
\end{equation}
This expression contains the following 2, 3 and 4-points
Green's functions:
$\left\langle \left( \partial \mathcal A \right)^2 \right\rangle$,
$\left\langle \left( \partial \mathcal A \right) \mathcal A^2 \right\rangle$ and
$\left\langle \left( \mathcal A \right)^4\right\rangle$. Defining
the 2-point Green's function
\begin{equation}
    \left\langle
      \widehat {\mathcal A}^B_\alpha (x) \widehat {\mathcal A}^C_\beta (y)
    \right\rangle =
    \mathcal{G}^{BC}_{\alpha \beta} (x,y) ~,
\label{sec4-40}
\end{equation}
the first term on the r.h.s of equation \eqref{sec4-30} becomes
\begin{equation}
\begin{split}
    &\left( \partial_\mu \widehat {\mathcal A}^B_\nu (x) \right)
    \left( \partial^\mu \widehat {\mathcal A}^{B\nu}(x) \right) =
    \partial_{x\mu} \partial_y^\mu
    \left( \widehat {\mathcal A}^B_\nu (x) \right)
    \left( \widehat {\mathcal A}^{B\nu}(y) \right) \Bigr |_{y \rightarrow x} =
   \\
    &\eta^{\alpha \beta} \partial_{x\mu} \partial_y^\mu
    \mathcal{G}^{BB}_{\alpha \beta} (x,y) \Bigr |_{y \rightarrow x} 
\end{split}
\label{sec4-45}
\end{equation}
where we denote $x^\mu$ as $x$ and $y^\mu$ as $y$. 
For simplicity we consider the case with $x^0=y^0$. For this
Green's function we will make the ``one-function approximation"
\cite{dzhsin6}
\begin{equation}
    \mathcal{G}^{AB}_{\alpha \beta} (x,y) \approx
    -\eta_{\alpha \beta} f^{ACD} f^{BCE} \phi^D (x) \phi^E(y)
\label{sec4-50}
\end{equation}
where $\phi^A(x)$ is the scalar field which describes the 2-point
Green's function. Other authors have used a similar technique of
replacing a vector, gauge field, by some combination of scalar
fields. For example Corrigan and Fairlie \cite{corr} and Wilczek
\cite{wilc} were able to transform the Yang-Mills equations into
those of a massless scalar $\lambda \phi ^4$ theory by writing the
SU(2) gauge fields in terms of of some combination of effective
scalar fields. Physically this approximation means that the
quantum properties of the field ${\mathcal A}^B_\mu$ can be
approximately described by a scalar field $\phi^B(x)$, {\it i.e.}
in this approximation the Lorentz character and Lorentz index
$\mu$ of the vector gauge fields are treated in a simple way
through \eqref{sec4-50}. Taking into account this approximation we
have
\begin{equation}
  \left\langle
    \left( \partial_\mu \widehat {\mathcal A}^B_\nu  \right)
    \left( \partial^\mu \widehat {\mathcal A}^{B\nu} \right)
    \right\rangle =
    - \eta^\nu_\nu f^{BAC} f^{BAD}
    \left( \partial_\mu \phi^C \right)
    \left( \partial^\mu \phi^D \right) =
    - 12 \left( \partial_\mu \phi^A \right)
    \left(\partial^\mu \phi^A\right)
\label{sec4-60}
\end{equation}
and
\begin{equation}
  \left\langle
      \left( \partial_\mu \widehat {\mathcal A}^B_\nu  \right)
      \left( \partial^\nu \widehat {\mathcal A}^{B\mu}  \right)
    \right\rangle =
    - 3 \left( \partial_\mu \phi^A \right)
    \left(\partial^\mu \phi^A \right) ,
\label{sec4-70}
\end{equation}
Next we make the assumption that the odd Green's functions average to zero
\begin{equation}
    \left\langle
      \widehat {\mathcal A}^B_\alpha(x) \widehat {\mathcal A}^C_\beta(y)
      \widehat {\mathcal A}^D_\gamma(z)
    \right\rangle = 0 .
\label{sec4-80}
\end{equation}
This gives
\begin{equation}
    \left\langle
      \left( \partial_\mu \widehat {\mathcal A}^B_\alpha (x) \right)
      \widehat {\mathcal A}^C_\beta (x) \widehat {\mathcal A}^D_\gamma (x)
    \right\rangle =
    \partial_{x\mu}
    \left\langle
      \widehat {\mathcal A}^B_\alpha (x) \widehat {\mathcal A}^C_\beta (y)
      \widehat {\mathcal A}^D_\gamma (z)
    \right\rangle \Bigr |_{y,z \rightarrow x} = 0
\label{sec4-90}
\end{equation}
which finally leads to
\begin{equation}
    \left\langle
      \left( \partial_\mu \widehat {\mathcal A}^B_\nu -
      \partial_\nu \widehat {\mathcal A}^B_\mu \right)
      \widehat {\mathcal A}^{C \mu} \widehat {\mathcal A}^{D \nu}
    \right\rangle = 0 .
\label{sec4-100}
\end{equation}
Last for the quartic term on the r.h.s of equation \eqref{sec4-10}
we make the approximation that
\begin{equation}
\begin{split}
    &\left\langle
      \widehat {\mathcal A}^B_\alpha (x) \widehat {\mathcal A}^C_\beta (y)
      \widehat {\mathcal A}^D_\gamma (z) \widehat {\mathcal A}^R_\delta (u)
    \right\rangle \approx
    \left\langle
      \widehat {\mathcal A}^B_\alpha (x) \widehat {\mathcal A}^C_\beta (y)
    \right\rangle
    \left\langle
      \widehat {\mathcal A}^D_\gamma (z) \widehat {\mathcal A}^R_\delta (u)
    \right\rangle + \\
    &\left\langle
      \widehat {\mathcal A}^B_\alpha (x) \widehat {\mathcal A}^D_\gamma (z)
    \right\rangle
    \left\langle
      \widehat {\mathcal A}^C_\beta (y) \widehat {\mathcal A}^R_\delta (u)
    \right\rangle +
    \left\langle
      \widehat {\mathcal A}^B_\alpha (x) \widehat {\mathcal A}^R_\gamma (u)
    \right\rangle
    \left\langle
      \widehat {\mathcal A}^C_\beta (y) \widehat {\mathcal A}^D_\gamma (z)
    \right\rangle .
\end{split}
\label{sec4-110}
\end{equation}
This assumption is that the 4-point Greens function is simply the sum of products
of the 2-point Green's function. In this approximation the l.h.s of
\eqref{sec4-110} is
\begin{equation}
\begin{split}
    &\left\langle
      \widehat {\mathcal A}^B_\mu (x) \widehat {\mathcal A}^C_\nu (x)
      \widehat {\mathcal A}^{D\mu} (x) \widehat {\mathcal A}^{R\nu} (x)
    \right\rangle = \\
  &\lambda_{1,2;(P_{1,2},Q_{1,2})}
    \left(
      f^{BE_1P_1} f^{CE_1 Q_1} \phi^{P_1}(x) \phi^{Q_1}(x)
    \right)
    \left(
      f^{DE_2P_2} f^{RE_2 Q_2} \phi^{P_2}(x) \phi^{Q_2}(x)
    \right) \eta_{\mu\nu} \eta^{\mu\nu} + \\
    &\lambda_{1,2;(P_{1,2},Q_{1,2})}
    \left(
      f^{BE_1P_1} f^{DE_1 Q_1} \phi^{P_1}(x) \phi^{Q_1}(x)
    \right)
    \left(
      f^{CE_2P_2} f^{RE_2 Q_2} \phi^{P_2}(x) \phi^{Q_2}(x)
    \right) \eta^\mu_\mu \eta^\nu_\nu + \\
    &\lambda_{1,2;(P_{1,2},Q_{1,2})}
    \left(
      f^{BE_1P_1} f^{RE_1 Q_1} \phi^{P_1}(x) \phi^{Q_1}(x)
    \right)
    \left(
      f^{CE_2P_2} f^{DE_2 Q_2} \phi^{P_2}(x) \phi^{Q_2}(x)
    \right) \eta^\nu_\mu \eta^\mu_\nu
\end{split}
\label{sec4-120}
\end{equation}
We have taken the spacetime variables of the scalar field, $\phi$, such that $x=y=z=u$.
In the above expression $\lambda_{1,2;(P_{1,2},Q_{1,2})}$ are some parameters depending
on the values of the indices $P_{1,2},Q_{1,2}$
\begin{equation}
    \lambda_{1,2; \left (P_{1,2},Q_{1,2} \right )}=
    \begin{cases}
      \lambda_1, & \text{if all indices } P_{1,2},Q_{1,2} = 1,2,3, \\
      \lambda_2, & \text{if all indices } P_{1,2},Q_{1,2} = 4,5,6,7,8, \\
      1,         & \text{otherwise}
    \end{cases}
\label{sec4-130}
\end{equation}
where $\lambda_{1,2}$ are constants. Equation \eqref{sec4-130}
essentially introduces three different coupling strengths
($\lambda_1 , \lambda_2$, and  $1$) in place of the original
single coupling. This step is necessary in order to insure that
the solutions we find are at least quasi-stable. This will be
explained in more detail when we discuss the numerical solutions
of this system in the next section. From the definition of the
index $\lambda_{1,2; \left (P_{1,2},Q_{1,2} \right )}$ one can see
that this approximate quantization procedure acts differently for
the scalar field components which belong to the small subgroup
$SU(2) \in SU(3)$ versus the coset $SU(3)/SU(2)$. The first term
in eq. \eqref{sec4-120} can be written as
\begin{equation}
\begin{split}
    &\left(
      f^{BE_1P_1} f^{CE_1 Q_1} \phi^{P_1} \phi^{Q_1}
    \right)
    \left(
      f^{DE_2P_2} f^{RE_2 Q_2} \phi^{P_2} \phi^{Q_2}
    \right) = \\
    &\lambda_1
    \left(
      f^{BE_1a} f^{CE_1 b} \phi^{a} \phi^{b}
    \right)
    \left(
      f^{DE_2c} f^{RE_2 d} \phi^{c} \phi^{d}
    \right) + \\
    &\lambda_2
    \left(
      f^{BE_1m} f^{CE_1 n} \phi^{m} \phi^{n}
    \right)
    \left(
      f^{DE_2p} f^{RE_2 q} \phi^{p} \phi^{q}
    \right) + 
     (\text{other terms} )
\end{split}
\label{sec4-140}
\end{equation}
A straightforward calculation using the antisymmetry property of the structure
constants shows that
\begin{eqnarray}
    f^{ABC} f^{ADR}
    \left(
      f^{BE_1P_1} f^{CE_1 Q_1} \phi^{P_1} \phi^{Q_1}
    \right)
    \left(
      f^{DE_2P_2} f^{RE_2 Q_2} \phi^{P_2} \phi^{Q_2}
    \right) & = & 0,
\label{sec4-150}\\
  f^{ABC} f^{ADR}
    \left(
      f^{BE_1a} f^{CE_1 b} \phi^{a} \phi^{b}
    \right)
    \left(
      f^{DE_2c} f^{RE_2 d} \phi^{c} \phi^{d}
    \right) & = & 0,
\label{sec4-160}\\
 f^{ABC} f^{ADR}
    \left(
      f^{BE_1m} f^{CE_1 n} \phi^{m} \phi^{n}
    \right)
    \left(
      f^{DE_2p} f^{RE_2 q} \phi^{p} \phi^{q}
    \right) & = & 0.
\label{sec4-170}
\end{eqnarray}
which consequently yields
\begin{equation}
\begin{split}
    \lambda_{1,2; \left( P_{1,2},Q_{1,2} \right)}
    f^{ABC} f^{ADR}
     \left(
      f^{BE_1P_1} f^{CE_1 Q_1} \phi^{P_1} \phi^{Q_1}
    \right) &\\
    \left(
      f^{DE_2P_2} f^{RE_2 Q_2} \phi^{P_2} \phi^{Q_2}
    \right) \eta_{\mu\nu} \eta^{\mu\nu} = 0 &.
\end{split}
\label{sec4-180}
\end{equation}
A similar calculation for the second term in eq. \eqref{sec4-120} shows that
\begin{eqnarray}
    f^{ABC} f^{ADR}
    \left(
      f^{BE_1P_1} f^{DE_1 Q_1} \phi^{P_1}(x) \phi^{Q_1}(x)
    \right) && \\
\nonumber 
    \left(
      f^{CE_2P_2} f^{RE_2 Q_2} \phi^{P_2}(x) \phi^{Q_2}(x)
    \right)  & = &
    \frac{27}{8}\left(
      \phi^a \phi^a + \phi^m \phi^m
    \right)^2 , 
\label{sec4-190}\\
  f^{ABC} f^{ADR}
    \left(
      f^{BE_1 a} f^{DE_1 b} \phi^a \phi^b
    \right)
    \left(
      f^{CE_2 c} f^{RE_2 d} \phi^c \phi^d
    \right)  & = &
    \frac{27}{8}\left(
      \phi^a \phi^a
    \right)^2,
\label{sec4-200}\\
 f^{ABC} f^{ADR}
    \left(
      f^{BE_1 m} f^{DE_1 n} \phi^m \phi^n
    \right)
    \left(
      f^{CE_2 p} f^{RE_2 q} \phi^p \phi^q
    \right)  & = &
    \frac{27}{8}\left(
      \phi^m \phi^m
    \right)^2.
\label{sec4-210}
\end{eqnarray}
Consequently
\begin{equation}
\begin{split}
    \lambda_{1,2; \left( P_{1,2},Q_{1,2} \right)}
f^{ABC} f^{ADR}
    \left(
      f^{BE_1P_1} f^{DE_1 Q_1} \phi^{P_1} \phi^{Q_1}
    \right)
    \left(
      f^{CE_2P_2} f^{RE_2 Q_2} \phi^{P_2} \phi^{Q_2}
    \right) &= \\
  \frac{27}{8}\lambda_1 \left(
      \phi^a \phi^a
    \right)^2 +
    \frac{27}{8}\lambda_2
    \left(
      \phi^m \phi^m
    \right)^2 +
    \frac{27}{4} \left(
      \phi^a \phi^a
    \right)
    \left(
      \phi^m \phi^m
    \right) . &
\end{split}
\label{sec4-220}
\end{equation}
Analogously the third term in eq. \eqref{sec4-120} can be shown to give
\begin{equation}
\begin{split}
    \lambda_{1,2; \left( P_{1,2},Q_{1,2} \right)}
f^{ABC} f^{ADR}
    \left(
      f^{BE_1P_1} f^{RE_1 Q_1} \phi^{P_1} \phi^{Q_1}
    \right)
    \left(
      f^{CE_2P_2} f^{DE_2 Q_2} \phi^{P_2} \phi^{Q_2}
    \right) &= \\
  \frac{27}{8}\lambda_1 \left(
      \phi^a \phi^a
    \right)^2 +
    \frac{27}{8}\lambda_2
    \left(
      \phi^m \phi^m
    \right)^2 +
    \frac{27}{4} \left(
      \phi^a \phi^a
    \right)
    \left(
      \phi^m \phi^m
    \right) . &
\end{split}
\label{sec4-230}
\end{equation}
Combining these results gives the following expression for the quartic term
\begin{equation}
  f^{ARB} f^{ACD}
    \left\langle
      \widehat {\mathcal A}^{R}_\mu \widehat {\mathcal A}^{B}_\nu
      \widehat {\mathcal A}^{C \mu} \widehat {\mathcal A}^{D} \nu
    \right\rangle =
    \frac{81}{2} \lambda_1 \left( \phi^a \phi^a \right)^2 +
    \frac{81}{2} \lambda_2 \left( \phi^m \phi^m \right)^2 +
    81 \left( \phi^a \phi^a \right) \left( \phi^m \phi^m \right) .
\label{sec4-240}
\end{equation}
Bringing together all the results gives the following effective
Lagrangian which describes the 2 and 4-points Green's functions
\begin{equation}
  \mathcal{L}_{eff} = - \frac{9}{2}
  \left( \partial_\mu \phi^A \right) \left( \partial^\mu \phi^A \right) +
  \frac{g^2}{4} \left[ 
  \frac{81}{2} \lambda_1 \left( \phi^a \phi^a \right)^2 +
    \frac{81}{2} \lambda_2 \left( \phi^m \phi^m \right)^2 +
    81 \left( \phi^a \phi^a \right) \left( \phi^m \phi^m \right) 
  \right].
\label{sec4-250}
\end{equation}
This Lagrangian can be put in the standard form by making the
following redefinitions: $\phi^a \rightarrow 2\phi^a /(3g)$ and
$\lambda_{1,2} \rightarrow \lambda_{1,2} /2$ which yields
\begin{equation}
    \frac{g^2}{4} \mathcal {L}_{eff} = - \frac{1}{2}
    \left( \partial_\mu \phi^A \right) \left( \partial^\mu \phi^A \right)+
    \frac{\lambda_1}{4} \left( \phi^a \phi^a \right)^2 +
    \frac{\lambda_2}{4} \left( \phi^m \phi^m \right)^2 +
    \left( \phi^a \phi^a \right) \left( \phi^m \phi^m \right) .
\label{sec4-260}
\end{equation}
By splitting up the kinetic energy term into parts for $\phi ^a$
and $\phi ^m$ one sees that this Lagrangian is of the form of two,
massless scalar fields with $\lambda \phi ^4$ self interaction and
an interaction term between the two scalar field ({\it i.e.} the
last term in \eqref{sec4-260}). Now in \cite{coleman2} it was
shown that radiative corrections could introduce a symmetry
breaking ({\it i.e.} negative) mass term into a scalar Lagrangian
similar to that in \eqref{sec4-260}. This effect is called
dimensional transmutation. For the pure scalar case one can not
rigourously justify a radiatively generated symmetry breaking term
since the scale at which the symmetry breaking occurs lies outside
the region where pertubation theory is valid. Nevertheless it was
postulated that a nonpertbative calculation would yield a similar
negative mass, symmetry breaking term. In the present paper we
will also make this assumption and add negative mass terms to the
two scalar fields so that the Lagrangian can be written as
\begin{equation}
\begin{split}
    &\frac{g^2}{4} \mathcal {L}_{eff} = - \frac{1}{2}\left( \partial_\mu \phi^A \right)^2 +
    \frac{\lambda_1}{4}
    \left[ \phi^a \phi^a - \phi^a_0 \phi^a_0
    \right]^2 - \frac{\lambda_1}{4} \left( \phi^a_0 \phi^a_0  \right)^2 + \\
    &\frac{\lambda_2}{4}
    \left[ \phi^m \phi^m - \phi^m_0 \phi^m_0
    \right]^2 - \frac{\lambda_2}{4} \left( \phi^m_0 \phi^m_0  \right)^2 +
    \left( \phi^a \phi^a \right) \left( \phi^m \phi^m \right)
\end{split}
\label{sec4-270}
\end{equation}
where $\phi^A_0 = (\phi ^a _0 , \phi ^m _0)$ are some constants.
The mass terms come from the cross terms of the second and fourth
terms above. Next we drop the constant terms in the Lagrangian
which gives
\begin{equation}
    \frac{g^2}{4} \mathcal {L}_{eff} = - \frac{1}{2}\left( \partial_\mu \phi^A \right)^2 +
    \frac{\lambda_1}{4} \left[ \phi^a \phi^a - \phi^a_0 \phi^a_0
    \right]^2  +
    \frac{\lambda_2}{4}
    \left[ \phi^m \phi^m -  \phi^m_0 \phi^m_0
    \right]^2 +
    \left( \phi^a \phi^a \right) \left( \phi^m \phi^m \right) .
\label{sec4-275}
\end{equation}
In the next subsection we will investigate numerically if this
Lagrangian leads to a solution which is well behaved in some
finite range, and which therefore could be viewed as a glueball.
As a first step we write down the field equations that give an
effective scalar field description of QCD
\begin{eqnarray}
  \partial_\mu \partial^\mu \phi^a &=&
  - \phi^a \left[ 2 \phi^m \phi^m + \lambda_1
  \left(
    \phi^a \phi^a- \phi^a_0 \phi^a_0
  \right) \right],
\label{sec4-280}\\
  \partial_\mu \partial^\mu \phi^m &=&
  - \phi^m \left[ 2 \phi^a \phi^a + \lambda_2
  \left(
    \phi^m \phi^m- \phi^m_0 \phi^m_0
  \right) \right].
\label{sec4-290}
\end{eqnarray}
In ref. \cite{bazeia} a similar coupled scalar field system was
investigated and shown to have soliton solutions for some specific
choice for the potential terms. In the next subsection it will be
shown that the Lagrangian in \eqref{sec4-275} has solutions are
well behaved in some finite range. We will interpret these
solutions as glueballs since the original starting Lagrangian was
the pure SU(3) gauge theory. The importance of the quantization
procedure that led to the Lagrangian in \eqref{sec4-275} is
evident. For classical, pure SU(3) theory one has no finite energy
solutions -- there are no classical glueballs. Our approximate,
nonperturbative quantization method transformed the original pure
Yang-Mills into an effective interacting scalar Lagrangian of
\eqref{sec4-275}. In the next section solutions which have a
soliton-like character in some finite range are investigated, and
it is suggested that this solutions may give a model for
glueballs.

\subsection{Soliton/Glueball solution for the $\phi^A$ field}

We begin our investigation of the field equations \eqref{sec4-280} \eqref{sec4-290}
by taking the following ansatz for the scalar fields
\begin{eqnarray}
    \phi^a (r) &=& \frac{\phi(r)}{\sqrt{6}} , \quad a=1,2,3 ,
\label{sec4-300}\\
  \phi^m (r) &=& \frac{f(r)}{\sqrt{10}} , \quad m=4,5,6,7,8 .
\label{sec4-310}
\end{eqnarray}
This ansatz implies that the components $\phi^a$ and the
components $\phi^m$ have different behavior from one another.
Substituting equations \eqref{sec4-300} \eqref{sec4-310} into
equations \eqref{sec4-280} \eqref{sec4-290} gives the following
coupled, nonlinear equations
\begin{eqnarray}
    \phi'' + \frac{2}{r} \phi' &=& \phi
    \left[
      f^2 + \lambda_1 \left( \phi^2 - m^2 \right)
    \right],
\label{sec4-320}\\
    f'' + \frac{2}{r} f' &=& f
    \left[
      \phi^2 + \lambda_2 \left( f^2 - \mu^2 \right)
    \right]
\label{sec4-330}
\end{eqnarray}
The primes denote differentiation with respect to $r$, we have
redefined $\lambda_{1,2} / 2 \rightarrow \lambda_{1,2}$, and
$2\phi_0^a \phi_0^a = m^2$ and $2 \phi_0^m \phi_0^m = \mu^2$; $m,
\mu$ are constants which will be calculated by solving equations
\eqref{sec4-320} and \eqref{sec4-330}.  We have not found
analytical solutions to these equations and so we investigated
them numerically. The preliminary numerical investigation showed
that these equations do not have regular solutions, $f(x) ,
\phi(x)$,  for an arbitrary choice of the $m, \mu$ parameters.
However, certain discrete values of these parameters appear to
give solutions which are well behaved at least for some finite
range. Thus we will solve equations \eqref{sec4-320}
\eqref{sec4-330} as a nonlinear eigenvalue problem for eigenstates
$\phi(x), f(x)$ and eigenvalues $m, \mu$, {\it i.e.} we will
calculate the parameters $m, \mu$ such that the solutions
$\phi(r)$ and $f(r)$ are regular functions for some finite range.
\par
The solutions of \eqref{sec4-330} will depend on the initial
conditions of the ansatz functions ($\phi(0), f(0), \phi'(0)$,
$f'(0)$) and on $\lambda_{1,2}$. We will rescale the solutions by
dividing equations \eqref{sec4-320} \eqref{sec4-330} by
$\phi^3(0)$, and then introducing the dimensionless radius
$x=r\phi(0)$. Finally by redefining $\phi(x)/\phi(0) \rightarrow
\phi(x)$, $f(x)/\phi(0) \rightarrow f(x)$ and $m/\phi(0)
\rightarrow m$, $\mu/\phi(0) \rightarrow \mu$ we arrive at the
rescaled equations
\begin{eqnarray}
    \phi'' + \frac{2}{x} \phi' &=& \phi
    \left[
      f^2 + \lambda_1 \left( \phi^2 - m^2 \right)
    \right],
\label{sec4-340}\\
    f'' + \frac{2}{x} f' &=& f
    \left[
      \phi^2 + \lambda_2 \left( f^2 - \mu^2 \right)
    \right].
\label{sec4-350}
\end{eqnarray}
where now the primes denote differentiation with respect to $x$.
We will look for solutions to \eqref{sec4-340} \eqref{sec4-350}
which are finite at the origin, and which approach finite values
as $x \rightarrow \infty$. In particular we would like the
asymptotic values to approach $\phi (\infty ) =m$ and $f(\infty)
=0$. We have not been able to numerically integrate the equations
out to arbitrary $x$, but we have found solutions with $\phi (x_F)
=m$ and $f(x_F) =0$, where $x_F$ is some finite $x$ coordinate. To
implement the numerical solution of \eqref{sec4-340}
\eqref{sec4-350} we used the following boundary conditions
\begin{eqnarray}
    \phi(0) &=& 1, \quad \phi'(0) = 0 ,
\label{sec4-360}\\
  f(0) &=& f_0, \quad f'(0) = 0 .
\label{sec4-370}
\end{eqnarray}
These boundary conditions will give solutions which have
soliton-like profiles for the region $0<x<x_F$. We now rewrite
equation \eqref{sec4-350} in the following form
\begin{equation}
    -\left( f'' + \frac{2}{x} f' \right) + f V_{eff} =
    \left( \lambda_2 \mu^2 \right) f
\label{sec4-380}
\end{equation}
where we have introduced an effective potential
\begin{equation}
    V_{eff} = \left( \phi^2 + \lambda_2 f^2 \right).
\label{sec4-390}
\end{equation}
Thus equation \eqref{sec4-380} is now in the form of the
Schr\"odinger equation. It will have regular solutions provided
$V_{eff}$ has a deep enough ``potential well''. Taking into
account the desired boundary conditions of $\phi (x)$ and $f(x)$
for $x \rightarrow x_F$ ({\it i.e.} $\phi (x_F) \rightarrow m$ and
$f (x_F) \rightarrow 0$) we want $V_{eff} (x_F) > V_{eff} (0)$ or
$m^2 > 1+\lambda _2 f_0 ^2$. We would like to set $x_F = \infty$,
but numerically we have only been able to integrate the equations
\eqref{sec4-340} \eqref{sec4-350} out to some finite $x_F$. If
$m^2$ is too close to $1+ \lambda _2 f_0 ^2$ the ``potential well"
may be too shallow and no solution exist. However, assuming $m^2$
is large enough a regular solution will exist and the ``energy
levels", $\lambda_2 \mu^2$, will be quantized.

\subsection{Numerical solution}

We will use the following numerical method for solving equations
\eqref{sec4-340} \eqref{sec4-350}: input some arbitrary zeroth
order approximation for the function $f(x)$ (which is denoted
$f_0(x)$) and solve equation \eqref{sec4-340} for the zeroth order
$\phi$ ($\phi _0$)
\begin{equation}
    \phi_0'' + \frac{2}{x} \phi_0 ' = \phi_0
    \left[
      f^2_0 + \lambda_1 \left( \phi^2_0 - m^2_0 \right)
    \right]
\label{sec4-400}
\end{equation}
where $m_0$ is the zeroth order approximation for the parameter $m$. The
boundary conditions are given in \eqref{sec4-360}.
This regular solution $\phi_0(x)$ is then used to calculate the next
order $f(x)$ ansatz function. Taking $g_0 (x)$ and substituting it into equation
\eqref{sec4-350} we now have an equation for the next order in $f(x)$.
\begin{equation}
    f_1'' + \frac{2}{x} f_1' = f_1
    \left[
      \phi^2_0 + \lambda_2 \left( f^2_1 - \mu^2_1 \right)
    \right]
\label{sec4-410}
\end{equation}
with the boundary conditions of equation \eqref{sec4-370} (for the present
numerical calculations we take $f_0 (0)= \sqrt{0.6}$). Numerically solving
\eqref{sec4-410} then gives the first approximation $f_1(x)$ which is then substituted
into equation \eqref{sec4-340}
\begin{equation}
    \phi_1'' + \frac{2}{x} \phi_1 ' = \phi_1
    \left[
      f^2_1 + \lambda_1 \left( \phi^2_1 - m^2_1 \right)
    \right].
\label{sec4-420}
\end{equation}
This equation is then solved for the first approximation in
$\phi_1 (x)$ which is then put back into the next approximate
equation for $f(x)$ and so on {\it i.e.}
\begin{equation}
\phi_i '' + \frac{2}{x} \phi_i ' = \phi_i
    \left[
      f^2_{i-1} + \lambda_1 \left( \phi^2_i - m^2_i \right)
    \right]
\label{sec4-430}
\end{equation}
and
\begin{equation}
    f_i'' + \frac{2}{x} f_i' = f_i
    \left[
      \phi^2_i + \lambda_2 \left( f^2_i - \mu^2_i \right)
    \right].
\label{sec4-450}
\end{equation}
At each step we find values $m^2_i$ and $\mu^2_i$ which approach
the true eigenvalues values ${m^*}^2$ and ${\mu^*}^2$.

\subsubsection{A more detailed description of the numerical calculations}

\begin{figure}[h]
  \begin{minipage}[t]{.45\linewidth}
  \begin{center}
    \fbox{
    \includegraphics[height=5cm,width=5cm]{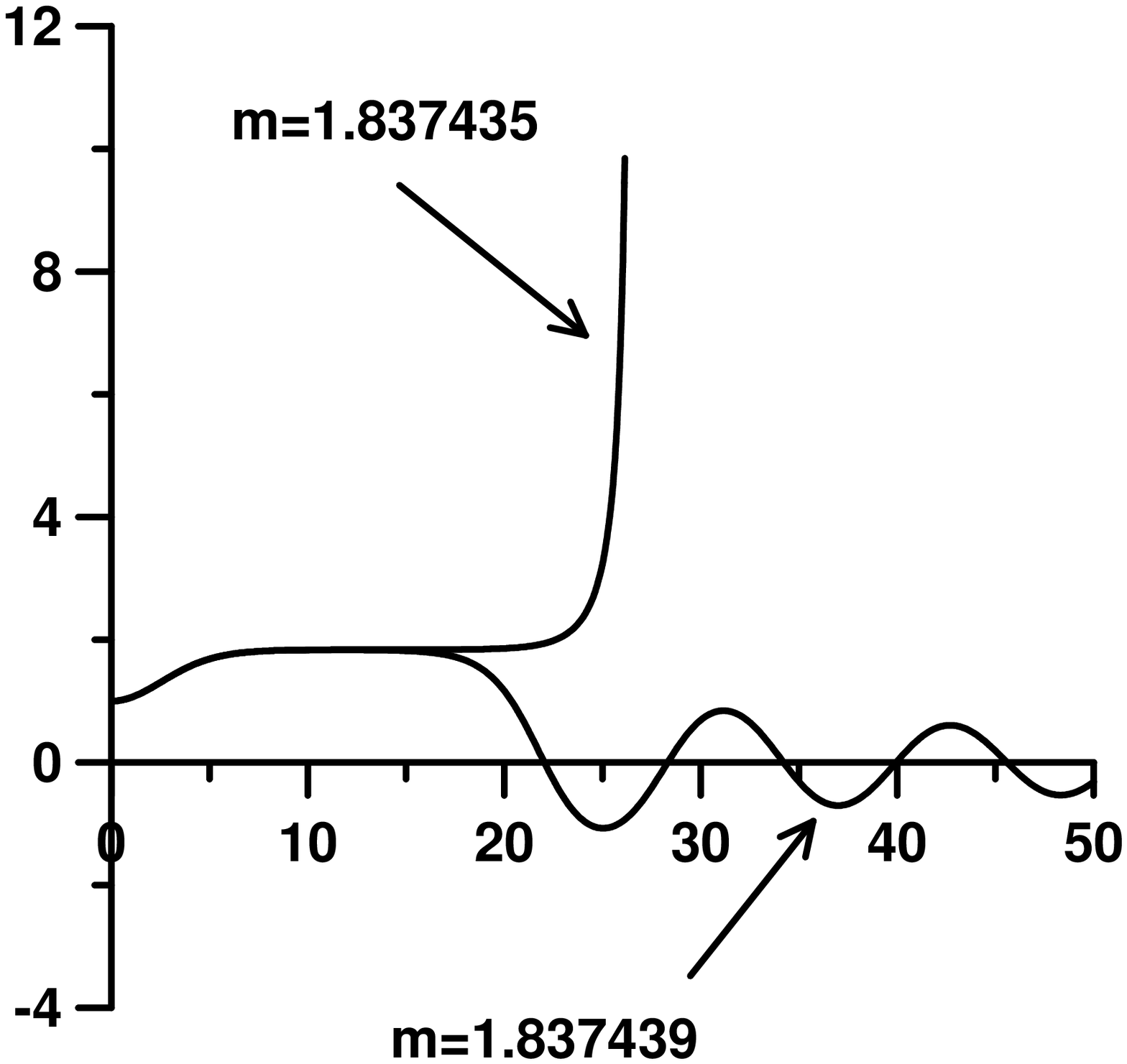}}
    \caption{The typical singular solution for equation \eqref{sec4-400}.
    $\lambda_1=0.1$.}
    \label{fig:phi-sing}
  \end{center}
  \end{minipage}\hfill
  \begin{minipage}[t]{.45\linewidth}
  \begin{center}
    \fbox{
    \includegraphics[height=5cm,width=5cm]{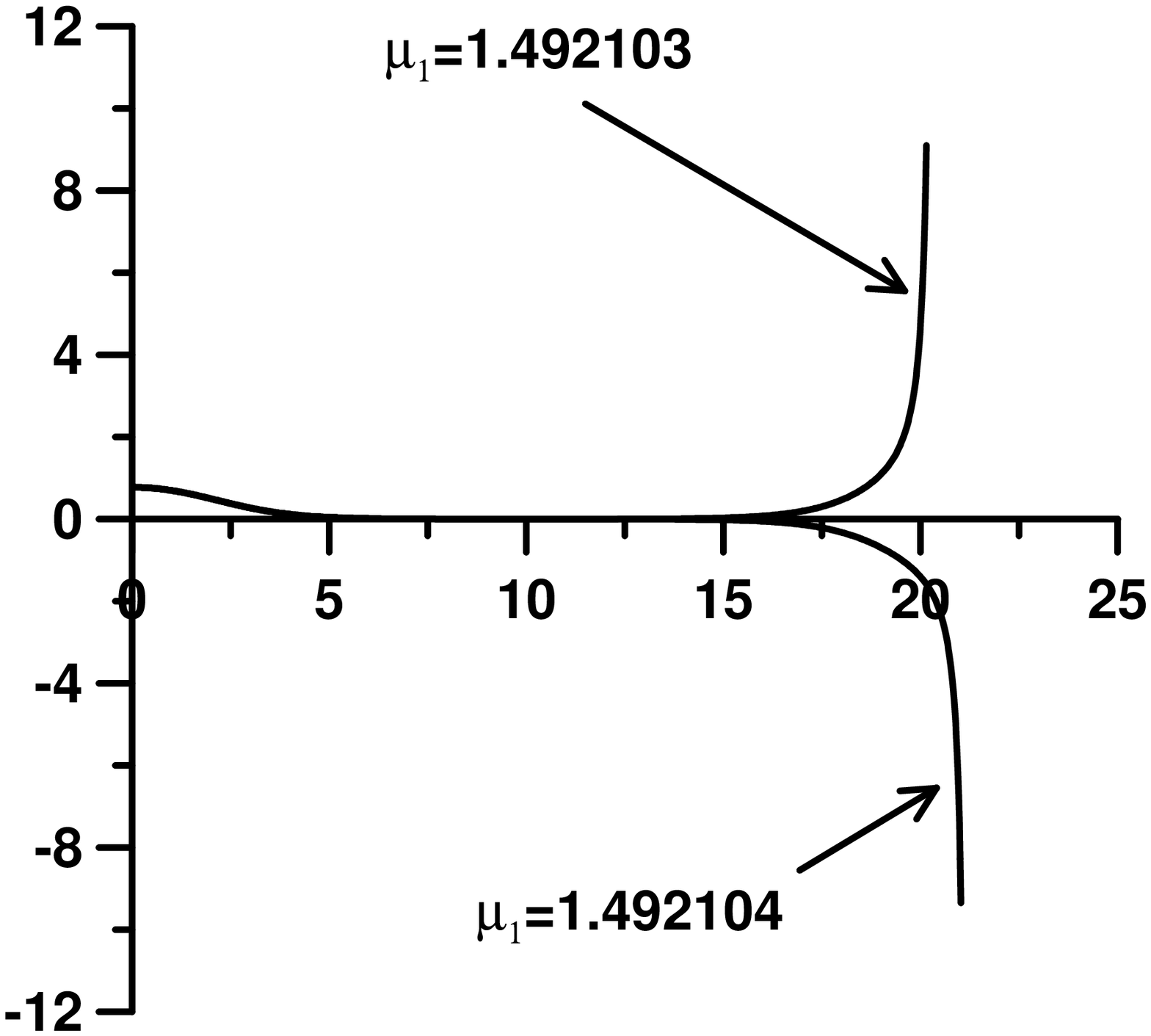}}
    \caption{The typical singular solution for equation \eqref{sec4-410}.
    $\lambda_2=1.0$.}
    \label{fig:f-sing}
  \end{center}
  \end{minipage}
\end{figure}

For the numerical solution of equation \eqref{sec4-400}
we make an initial arbitrary choice of the zeroth order approximation
for $f(x)$ as
\begin{equation}
    f_0(x) = \frac{\sqrt{0.6}}{\cosh^2{\frac{x}{4}}}.
\label{sec4-460}
\end{equation}
The reason for this choice is that the $\cosh$ function in \eqref{sec4-460}
has a soliton-like shape, and we are looking for soliton-like solutions.
Typical solutions for arbitrary values of $m_0$ are presented in
Fig.\ref{fig:phi-sing}.
From the figure one sees that when $m_0 < m_0^*$ ({$m_0^*$ is the
unknown parameter which is conjectured to give a regular solution)
the solution $\phi_0(x)$ is singular. Near to the singularity the
equation \eqref{sec4-400} has the form
\begin{equation}
    \phi_0'' \approx \lambda_1 \phi_0^3
\label{sec4-470}
\end{equation}
and the following approximate solution
\begin{equation}
    \phi_0(x) \approx \sqrt{\frac{2}{\lambda_1}} \frac{1}{x_0 - x}
\label{sec4-480}
\end{equation}
where $x_0$ is some constant depending on $m_0$. On the other hand
for $m_0 > m_0^*$ a typical solution is shown in Fig.\ref{fig:f-sing}.
The corresponding asymptotical equation is
\begin{equation}
    \phi_0''(x) + \frac{2}{x}   \phi_0' \approx -
    \left( \lambda_1 m^2 \right) \phi_0
\label{sec4-490}
\end{equation}
which has the following approximate solution
\begin{equation}
    \phi_0(x) \approx \phi_\infty
    \frac{\sin{\left(x \sqrt{\lambda_1 m^2} + \alpha\right)}}{x}
\label{sec4-500}
\end{equation}
where $\phi_\infty$ and $\alpha$ are constants. We conjecture that
there may exist some value, $m^*_0$, which yields an exceptional,
non-singular solution. Within some numerical accuracy such a
solution is presented in Fig.\ref{fig:phi-reg}. For this value
$m^*_0$ the equation \eqref{sec4-400} has the following
asymptotical behavior
\begin{equation}
    \phi_0''(x) + \frac{2}{x}   \phi_0' \approx
    2 \lambda_1 \left( m^*_0 \right)^2
    \left( \phi_0 - m^*_0 \right)
\label{sec4-510}
\end{equation}
and the corresponding asymptotical solution is
\begin{equation}
    \phi_0(x) \approx m^*_0 + \beta
    \frac{e^{-\left(x \sqrt{2
    \lambda_1 \left( m^*_0 \right)^2}\right)}}{x}
\label{sec4-520}
\end{equation}
where $\phi_\infty$ and $\beta$ are some constants.
\par
The next step is finding the first approximation for the $f_1(x)$
function. The equation is
\begin{equation}
  f_1'' + \frac{2}{x} f' = f_1
  \left[
    \phi_0^2 + \lambda_2 \left( f_1^2 - \mu^2_1 \right)
  \right].
\label{sec4-530}
\end{equation}
From the previous calculations we use the exceptional regular
solution $\phi_0(x)$ with the asymptotical behavior
\eqref{sec4-520}. Then the numerical investigation shows that for
an arbitrary $\mu$ there are two different singular solutions for
$f_1 (x)$ which are presented in Fig.\ref{fig:f-sing}.
\begin{figure}[h]
  \begin{minipage}[t]{.45\linewidth}
  \begin{center}
    \fbox{
    \includegraphics[height=5cm,width=5cm]{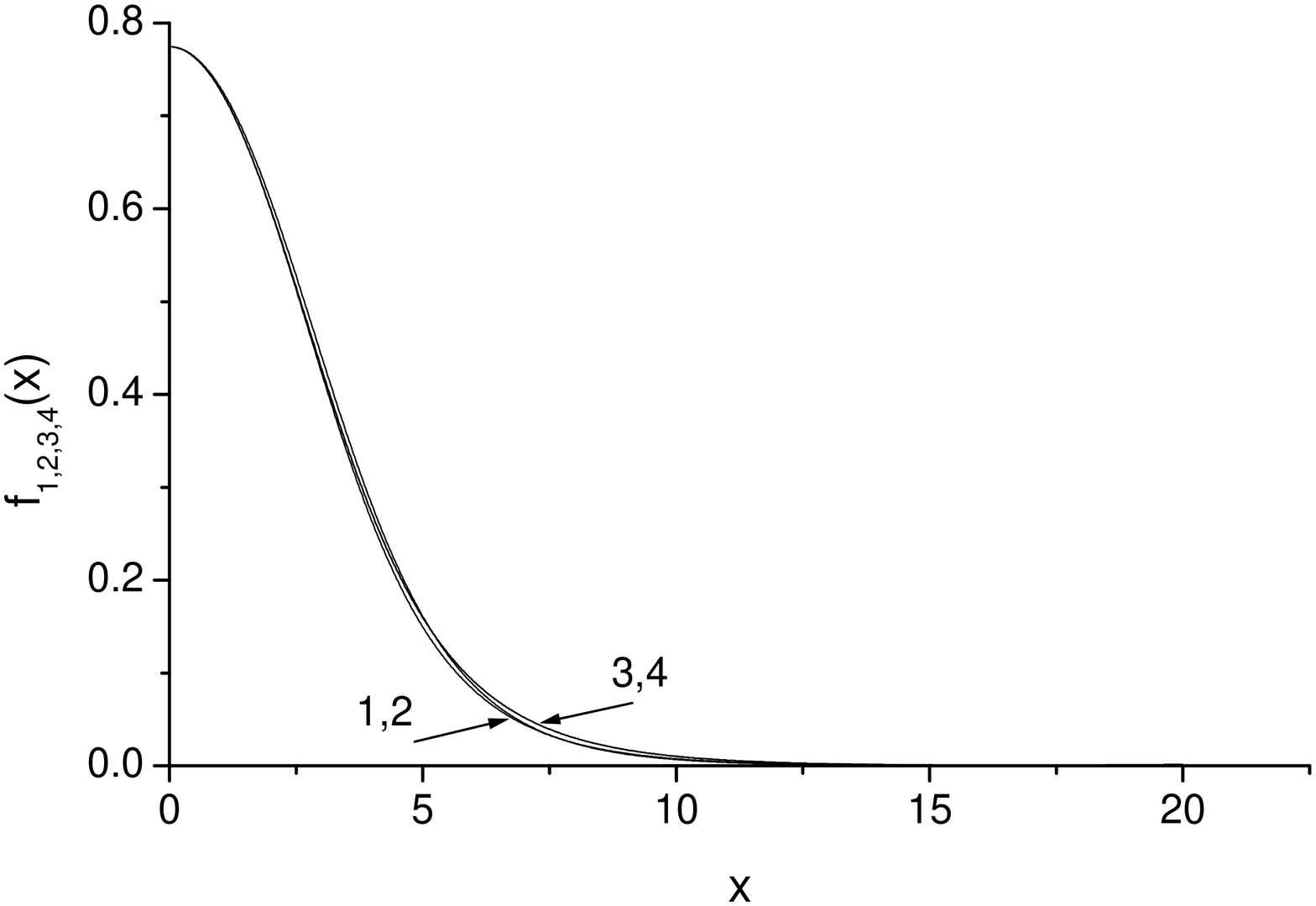}}
    \caption{The iterative functions $f_{1,2,3,4}(x)$.}
    \label{fig:f-reg}
  \end{center}
  \end{minipage}\hfill
  \begin{minipage}[t]{.45\linewidth}
  \begin{center}
    \fbox{
    \includegraphics[height=5cm,width=5cm]{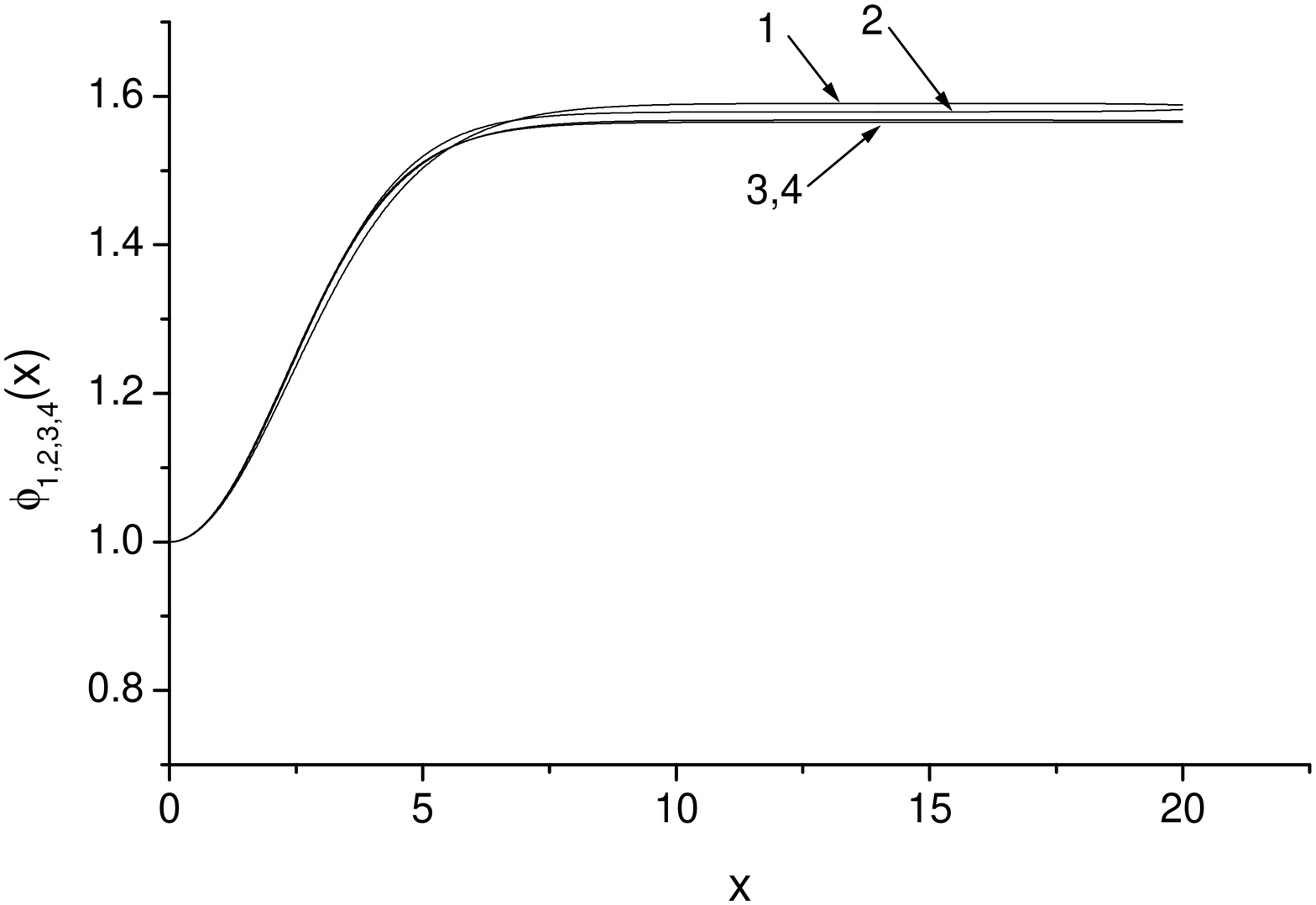}}
    \caption{The iterative functions $\phi_{1,2,3,4}(x)$.}
    \label{fig:phi-reg}
  \end{center}
  \end{minipage}
\end{figure}
As with equation \eqref{sec4-470} the singular behavior of the
function $f_1(x)$ is given by
\begin{eqnarray}
    f_1(x) &\approx & \sqrt{\frac{2}{\lambda_2}} \frac{1}{x - x_0}
    \quad \text {by} \quad \mu_1 < \mu^*_1 ,
\label{sec4-540}\\
  f_1(x) &\approx & - \sqrt{\frac{2}{\lambda_2}} \frac{1}{x - x_0}
    \quad \text {by} \quad \mu_1 > \mu^*_1 .
\label{sec4-550}
\end{eqnarray}
This again may indicate that there is some intermediate special
value for $\mu _1$ ({\it i.e.} $\mu_1 = \mu^*_1$) for which the
solution, $f^*_1(x)$, is regular and has the following
asymptotical behavior
\begin{equation}
    f^*_1(x) \approx f_\infty
    \frac{e^{- x \sqrt{\phi^2_\infty - \lambda_2 \mu^2}}}{x}
\label{sec4-560}
\end{equation}
where $f_\infty$ is some parameter.
The next step is substituting the first approximation $f^*_1(x)$ into
equation \eqref{sec4-340} to find the regular, exceptional
solution $\phi^*_1(x)$ with eigenvalue $m^*_1$; then $\phi^*_1(x)$ is substituted
into equation \eqref{sec4-350} to find the regular exceptional
solution $f^*_2(x)$ with eigenvalue $\mu = \mu^*_2$ and so on.
\par
The result of these calculations is presented in Fig's.
\ref{fig:f-reg}, \ref{fig:phi-reg} and Table \ref{table2}. We see
that there is the convergence $\phi^*_i(x) \rightarrow \phi^*(x)$,
$f^*_i(x) \rightarrow f^*(x)$, $m^*_i \rightarrow m^*$ and
$\mu^*_i \rightarrow \mu^*$; $f^*(x), \phi^*(x)$ are the
eigenstates and $m^*, \mu^*$ are eigenvalues of the nonlinear
eigenvalue problem \eqref{sec4-320} \eqref{sec4-330}.
\par
From the figures it can be seen that the numerical integration of
the coupled, nonlinear equations was only carried out to a finite
$x$ (roughly $x=20$). Beyond this it became increasingly difficult
to adjust the mass values, $m^* , \mu ^*$ in order to obtain the
desired soliton-like behavior. At this stage it is not certain
whether the solution can be made to be well behaved out to
arbitrary $x$. However, the analogy between our system and the
Schr\"odinger Equation given in \eqref{sec4-380} indicates that
there should be some parameters which give regular solutions out
to $x=\infty$. However based on the numerical results it is at
present not certain as to the asymptotic behavior of the ansatz
functions.

This numerical method was tested by applying it to the standard,
analytic soliton solution of $\lambda \phi ^4$ theory in 1 spatial
dimensions. Details of this can be found in Appendix \ref{app2}.
\begin{table}[h]
    \begin{center}
        \begin{tabular}{|c|c|c|c|c|}\hline
          i & 1 & 2& 3 & 4 \\ \hline
            $m^*_i$ & 1.590537877\ldots & 1.579111774\ldots & 1.565224661\ldots
            & 1.5679008386\ldots \\ \hline
            $\mu^*_i$ & 1.222955\ldots & 1.2287\ldots & 1.22902\ldots
            & 1.229019\ldots \\ \hline
        \end{tabular}
    \end{center}
    \caption{The iterative parameters $m^*_i$ and $\mu^*_i$.}
    \label{table2}
\end{table}

Finally, we remark that according to Derrick's Theorem
\cite{derrick} our solution, if it turns out to be regular for
$0<x<\infty$, it will not be stable. For pure scalar field
theories as in \eqref{sec4-275} absolutely stable solutions exist
only in 1 or 2 spatial dimensions. Thus our solution is at best
quasi-stable. This is in fact the reason for choosing different
couplings ({\it i.e.} $\lambda _1, \lambda _2 , 1$) in
\eqref{sec4-130}. This resulted in the potential for the
interacting scalar fields of \eqref{sec4-275} having global and
local minima. Our solution is in one of the local minima. Thus
although it is not absolutely stable it should be quasi-stable so
long as there is a large enough barrier between the local and
global minima. The can be accomplished by requiring that $\lambda
_1$ and $\lambda _2$ be significantly different. This however
raises the problem of why one should develop two different scales
from a theory that originally only had a single scale. We will
leave this open question for future work. 
\par 
The final remark is that we can add the constant 
$\lambda_2 \left( \phi^m_0 \phi^m_0 \right)^2$ (which have not any influence 
on the field equations \eqref{sec4-280} \eqref{sec4-290}) to the Lagrangian 
\eqref{sec4-275} we will have the finite energy of glueball. The profiles 
of the enrgy with respect to $\lambda_{1,2}$ and with different $\phi_0$ 
are presented in fig's \ref{gl_eps1} \ref{gl_eps2}.
\begin{figure}[h]
\begin{minipage}[t]{.45\linewidth}
    \begin{center}
    \fbox{
        \includegraphics[height=5cm,width=5cm]{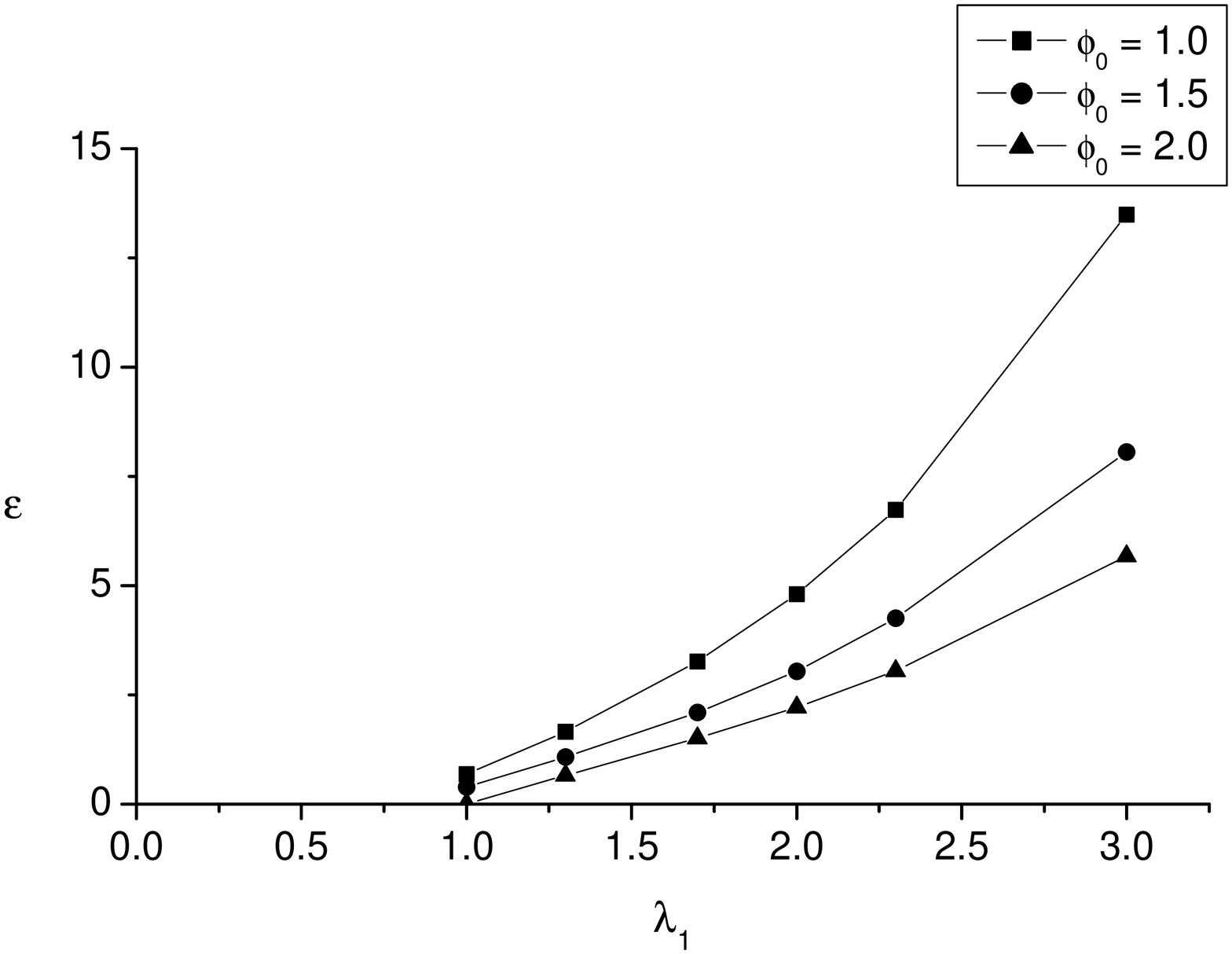}}
        \caption{The profile of the glueball energy with respect to $\lambda_1$.}
        \label{gl_eps1}
    \end{center}
\end{minipage}\hfill
\begin{minipage}[t]{.45\linewidth}
    \begin{center}
    \fbox{
        \includegraphics[height=5cm,width=5cm]{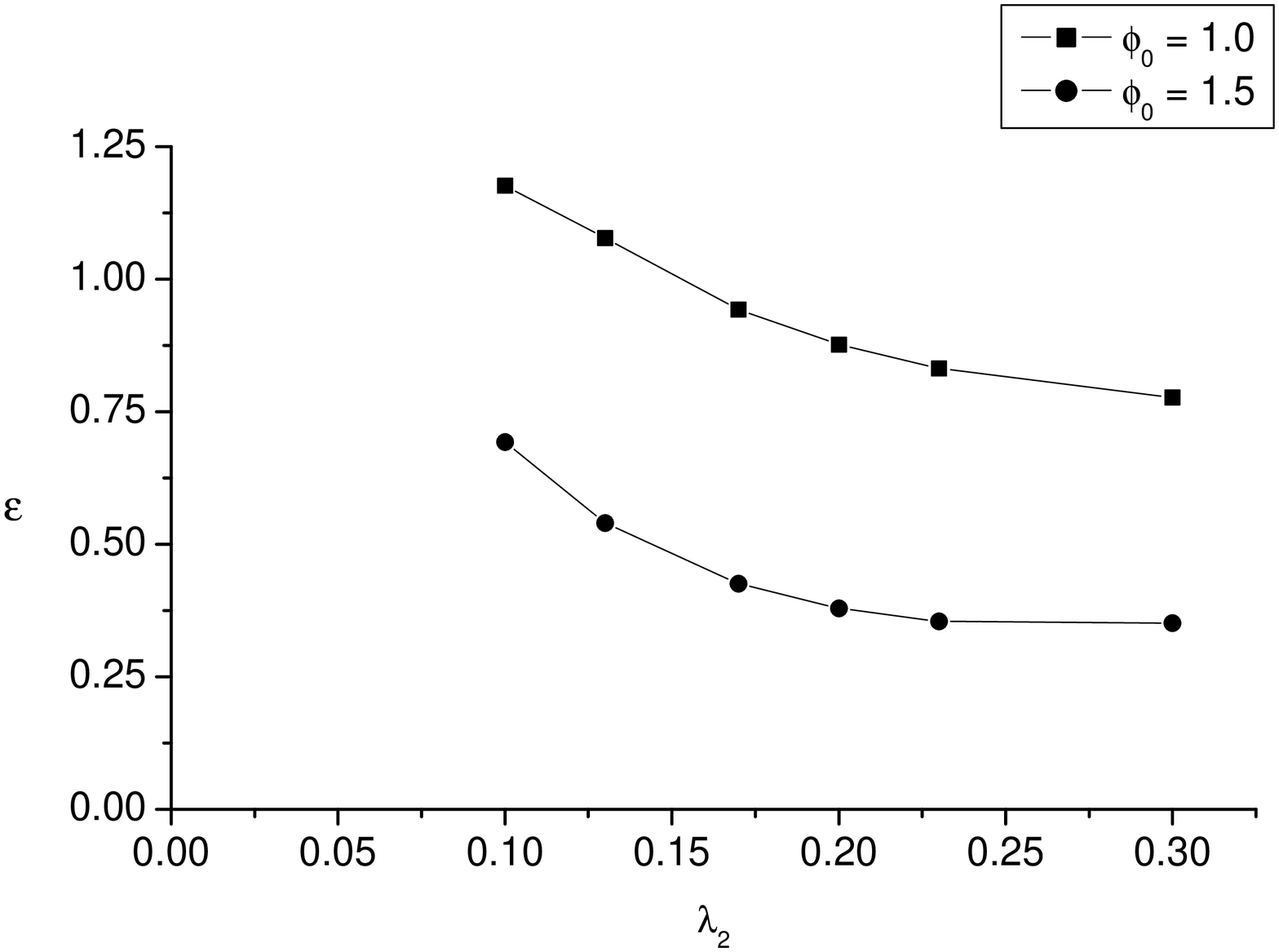}}
        \caption{The profile of the glueball energy with respect to $\lambda_2$.}
        \label{gl_eps2}
    \end{center}
\end{minipage}\hfill
\end{figure}

\section{The second case: ordered and disordered phases -- flux tube.}
\label{fluxtube}

\subsection{Reduction to a small subgroup}

In this section we will describe the second case where one has
both an ordered and disordered phase. We reiterate that we have
not found an exact analytical method for solving the full
equations \eqref{sec2-1-20} \eqref{sec2-1-30} for all the Green's
functions. Our basic approach in this case is to give some
physically reasonable scheme for cutting off the infinite set of
equations for the Green's functions. In addition our approximate
calculations will involve the decomposition of the initial gauge
group into a smaller gauge group: $SU(3) \rightarrow SU(2)$.
Physically we will find that this reduction splits the initial
degrees of freedom (the SU(3) gauge potential components) into
SU(2) and coset components. The SU(2) components will represent
the ordered phase while the coset components, which again will be
represented via an effective scalar field, will represent the
disordered phase. After these approximations one can perform
analytical calculations which suggest the formation of a color
electric flux tube. This is of interest since an important feature
of the confinement phenomenon in the dual superconducting picture
of the QCD vacuum is the formation of electric flux tubes between
the confined quarks.
\par
First we define the reduction of SU(3) to SU(2) following
the conventions of ref. \cite{Kondo1}.
Starting with the SU(3) gauge group with generators $T^B$, we define the SU(3)
gauge fields, $\mathcal{A}_\mu=\mathcal{A}^B_\mu T^B$. SU(2) is a subgroup
of SU(3) and SU(3)/SU(2) is a coset. Then the gauge field
$\mathcal{A}_\mu$ can be decomposed as
\begin{eqnarray}
  \mathcal{A}_\mu & = & \mathcal{A}^B_\mu T^B = a^a_\mu T^a + A^m_\mu T^m ,
\label{sec5-10}\\
  a^a_\mu & \in & SU(2) \quad \text{and} \quad A^m_\mu \in SU(3)/SU(2)
\label{sec5-20}
\end{eqnarray}
where the indices $a,b,c \ldots $ belongs to the subgroup SU(2) and
$m,n, \ldots $to the coset SU(3)/SU(2); $B$ are SU(3) indices.
Based on this the field strength can be decomposed as
\begin{equation}
  \mathcal{F}^B_{\mu\nu} T^B = \mathcal{F}^a_{\mu\nu}T^a +
  \mathcal{F}^m_{\mu\nu}T^m
\label{sec5-30}
\end{equation}
where
\begin{eqnarray}
  \mathcal{F}^a_{\mu\nu} & = & h^a_{\mu\nu} + \Phi^a_{\mu\nu}
  \; \; \; \in SU(2) ,
\label{sec5-35}\\
  h^a_{\mu\nu} & = & \partial_\mu a^a_\nu - \partial_\nu a^a_\mu +
  g \epsilon^{abc}a^b_\mu a^c_\nu \; \; \; \in SU(2) ,
\label{sec5-40}\\
  \Phi^a_{\mu\nu} & = & g f^{amn} A^m_\mu A^n_\nu \; \; \; \in SU(2) ,
\label{sec5-50}\\
  \mathcal{F}^m_{\mu\nu} & = & F^m_{\mu\nu} + G^m_{\mu\nu} \; \; \;
  \in SU(3)/SU(2) ,
\label{sec5-60}\\
  F^m_{\mu\nu} & = & \partial_\mu A^m_\nu - \partial_\nu A^m_\mu +
  g f^{mnp} A^n_\mu A^p_\nu \; \; \; \in SU(3)/SU(2) ,
\label{sec5-70}\\
  G^m_{\mu\nu} & = & g f^{mnb}
  \left(
  A^n_\mu a^b_\nu - A^n_\nu a^b_\mu
  \right) \; \; \; \in SU(3)/SU(2)
\label{sec5-80}
\end{eqnarray}
where $f^{ABC}$ are the structure constants of SU(3),
$\epsilon^{abc} = f^{abc}$ are the structure constants of SU(2).
The SU(3) Yang-Mills field equations can be decomposed as
\begin{eqnarray}
  d_\nu \left( h^{a\mu\nu} +\Phi^{a\mu\nu} \right) & = &
  -g f^{amn} A^m_\nu
  \left(
  F^{n\mu\nu} + G^{n\mu\nu}
  \right) ,
\label{sec5-90}\\
  D_\nu \left( F^{m\mu\nu} + G^{m\mu\nu} \right) & = &
  - g f^{mnb}
  \left[
  A^n_\nu \left( h ^{b\mu\nu} + \Phi^{b\mu\nu} \right) -
  a^b_\nu \left( F^{n\mu\nu} + G^{n\mu\nu} \right)
  \right]
\label{sec5-100}
\end{eqnarray}
where $d_\nu [\cdots]^a = \partial_\nu [\cdots]^a +
f^{abc} a^b_\nu [\cdots]^c$ is the covariant derivative on the
subgroup SU(2) and
$D_\nu [\cdots]^m = \partial_\nu [\cdots]^m +
f^{mnp} A^n_\nu [\cdots]^p$. The Heisenberg quantization procedure gives us
\begin{eqnarray}
  d_\nu \left( \widehat h^{a\mu\nu} + \widehat \Phi^{a\mu\nu} \right) & = &
  -g f^{amn} \widehat A^m_\nu
  \left(
  \widehat F^{n\mu\nu} + \widehat G^{n\mu\nu}
  \right) ,
\label{sec5-110}\\
  D_\nu \left( \widehat F^{m\mu\nu} + \widehat G^{m\mu\nu} \right) & = &
  - g f^{mnb}
  \left[
  \widehat A^n_\nu \left( \widehat h ^{b\mu\nu} + \widehat \Phi^{b\mu\nu} \right) -
  \widehat a^b_\nu \left( \widehat F^{n\mu\nu} + \widehat G^{n\mu\nu} \right)
  \right]
\label{sec5-120}
\end{eqnarray}

\subsection{Basic assumptions}

It is evident that an exact quantization is impossible for eqs.
\eqref{sec5-110} \eqref{sec5-120}. Thus we look for some
simplification in order to obtain equations which can be analyzed.
Our basic aim is to cut off the set of infinite equations using
some simplifying assumptions. For this purpose we will propose
simple but physically reasonable ans\"atzen for the 2 and 4-points
Green's functions -- $\left\langle A^m_\mu(y) A^n_\nu(x)
\right\rangle$, $\left\langle a^a_\alpha (x) a^b_\beta (y)
A^m_\mu(z) A^n_\nu(u) \right\rangle$ and $\left\langle A^m_\alpha
(x) A^n_\beta (y) A^p_\mu(z) A^q_\nu(u) \right\rangle$ -- which
are involved in averaged Lagrangian \eqref{sec4-30}. As was
mentioned earlier we assume that there are two phases
\begin{enumerate}
\item The gauge field components $a^a_\mu$ (a=1,2,3 $a^a_\mu \in SU(2)$)
      belonging to the small subgroup SU(2) are in an ordered phase. Mathematically
      this means that
\begin{equation}
  \left\langle a^a_\mu (x) \right\rangle  =(a^a _{\mu} (x))_{cl}.
\label{sec5-130}
\end{equation}
      The subscript means that this is a classical field. Thus we are
      treating these components as effectively classical gauge fields
      in the first approximation.
\item The gauge field components $A^m_\mu$ (m=4,5, ... , 8 and
      $A^m_\mu \in SU(3)/SU(2)$) belonging to the coset SU(3)/SU(2) are in
      a disordered phase ({\it i.e.} they form a condensate), but have
      non-zero energy. In mathematical terms this means that
\begin{equation}
  \left\langle A^m_\mu (x) \right\rangle = 0,
  \quad \text{but} \quad
  \left\langle A^m_\mu (x) A^n_\nu (x) \right\rangle \neq 0 .
\label{sec5-140}
\end{equation}
      Later we will postulate a specific, and physically reasonable form for the non-zero term.
\item There is no correlation between the ordered (classical) and disordered (quantum) phases
\begin{equation}
  \left\langle f(a^a_\mu) g(A^m_\nu) \right\rangle =
  f(a^a_\mu)  \left\langle g(A^m_\mu) \right\rangle
\label{sec5-150}
\end{equation}
Later we will give a specific form for this for the ``mixed'' 4-point Green's function,\\
$\left\langle a^a_\alpha (x) a^b_\beta (y) A^m_\mu(z) A^n_\nu(u) \right\rangle$.
\end{enumerate}

\subsection{Derivation of an effective Lagrangian}

Our quantization procedure will deviate from the Heisenberg method
in that we will take the expectation of the Lagrangian rather than
for the equations of motions. Thus we will obtain an effective
Lagrangian rather than approximate equations of motion. The
Lagrangian we obtain from the original SU(3) pure gauge theory
will turn out to be an effective SU(2) Yang-Mills-Higgs system
which has monopole solutions. The averaged Lagrangian is
\begin{equation}
  \mathcal{L} = - \frac{1}{4}\left\langle \mathcal{F}^A_{\mu\nu} \mathcal{F}^{A\mu\nu}
  \right\rangle =
- \frac{1}{4} \left( \left\langle \mathcal{F}^a_{\mu\nu}
\mathcal{F}^{a\mu\nu} \right\rangle + \left\langle
\mathcal{F}^m_{\mu\nu} \mathcal{F}^{m\mu\nu} \right\rangle \right)
\label{sec5-160}
\end{equation}
here $\mathcal{F}^{a\mu\nu}$ and $\mathcal{F}^{m\mu\nu}$ are defined by
equations \eqref{sec5-30}-\eqref{sec5-80}.

\subsubsection{Calculation of
$\left\langle \mathcal{F}^a_{\mu\nu} \mathcal{F}^{a\mu\nu} \right\rangle$}

We begin by calculating the first term on the r.h.s. of equation
\eqref{sec5-160}
\begin{equation}
  \left\langle \mathcal{F}^a_{\mu\nu} \mathcal{F}^{a\mu\nu} \right\rangle =
  \left\langle h^a_{\mu\nu} h^{a\mu\nu} \right\rangle +
  2\left\langle h^a_{\mu\nu} \Phi^{a\mu\nu} \right\rangle +
  \left\langle \Phi^a_{\mu\nu} \Phi^{a\mu\nu} \right\rangle .
\label{sec5-170}
\end{equation}
Immediately we see that the first term on the r.h.s. of this equation is
the SU(2) Lagrangian as we assume that $a^a_\mu$ and $h^a_{\mu\nu}$ are
effectively classical quantities and consequently
\begin{equation}
  \left\langle h^a_{\mu\nu} h^{a\mu\nu} \right\rangle \approx
  h^a_{\mu\nu} h^{a\mu\nu} .
\label{sec5-180}
\end{equation}
The second term in equation \eqref{sec5-170} is
\begin{equation}
  \left\langle h^a_{\mu\nu} \Phi^{a\mu\nu} \right\rangle =
  g f^{amn}
  \left\langle
  \left(
  \partial_\mu a^a_\nu - \partial_\nu a^a_\mu
  \right)
  A^{m\mu} A^{n\nu}
  \right\rangle +
  g f^{abc} f^{amn}
  \left\langle
  a^b_\mu a^c_\nu A^{m\mu} A^{n\nu}
  \right\rangle .
\label{sec5-190}
\end{equation}
Using assumptions 1 and 3 from the previous section these terms become
\begin{equation}
  \left\langle a^a_\alpha (x) A^m_\mu (y) A^n_\nu (z) \right\rangle =
  a^a_\alpha (x) \left\langle A^m_\mu (y) A^n_\nu (z) \right\rangle =
  \eta_{\mu\nu} a^a_\alpha (x) \mathcal{G}^{mn} (y,z)
\label{sec5-200a}
\end{equation}
and
\begin{equation}
  \left\langle a^a_\alpha (x) a^b_\beta (y) A^m_\mu (z) A^n_\nu (u) \right\rangle
  = a^a_\alpha (x) a^b_\beta (y) \eta_{\mu\nu} \mathcal{G}^{mn} (z,u)
\label{sec5-210}
\end{equation}
The function $\mathcal{G}^{mn} (x,y)$ is the 2-point correlator (Green's
function) for the disordered phase. Because of the bosonic character of the
coset gauge fields $\mathcal{G}^{mn} (x,y)$ must be symmetric under exchange
of these fields. Also by assumption 2 of the last section this expectation should
be non-zero. We take the form for this 2-point correlator to be
\begin{equation}
  \left\langle A^m_\mu (y) A^n_\nu (x) \right\rangle =
  - \frac{1}{3}\eta_{\mu\nu} f^{mpb} f^{npc}
  \phi^b (y) \phi^c (x) =
  - \eta_{\mu\nu} \mathcal{G}^{mn} (y,x)
\label{sec5-220}
\end{equation}
with
\begin{equation}
  \mathcal{G}^{mn} (y,x) = \frac{1}{3} f^{mpb} f^{npc} \phi^b (y) \phi^c (x)
\label{sec5-230}
\end{equation}
here $\phi^a$ is a real SU(2) triplet scalar fields. Thus we have
replaced the coset gauge fields by an effective scalar field,
which will be the scalar field in our effective SU(2)-scalar
system. Aside from the factor of $-\frac{1}{3}$ this is similar to
the approximation made for the 2-point correlator in the last
section in \eqref{sec4-50}. The factor of $- \frac{1}{3}$ is
introduced so that the effective scalar field, $\phi$, will have
the correct coefficient for the kinetic energy term. If we were to
take the scalar field to be constant ($\phi ^a (x) \approx
const.$) then \eqref{sec5-220} and \eqref{sec5-230} would
represent an effective mass-like, condensation term  of the coset
gauge fields. With this we find that the middle term vanishes
\begin{equation}
\begin{split}
  \left\langle 
    h^a_{\mu\nu} \Phi^{a\mu\nu} \right\rangle = g \eta^{\mu\nu}
    \biggl ( f^{amn} \left(\partial_\mu a^a_\nu (x) - \partial_\nu a^a_\mu (x) \right)
    \mathcal{G}^{mn} (x,x) &+ \\
   g f^{amn} f^{abc} a^b_\mu (x) a^c_\nu (x) \mathcal{G}^{mn} (x,x)
  \biggl )
  &= 0
\end{split}
\label{sec5-240}
\end{equation}
The last term which is quartic in the coset gauge fields will be considered at
the end. Up to this point the SU(2) part of the Lagrangian is
\begin{equation}
  \left\langle \mathcal{F}^a_{\mu\nu} \mathcal{F}^{a\mu\nu} \right\rangle =
  \left\langle h^a_{\mu\nu} h^{a\mu\nu} \right\rangle +
  g^2 f^{anp} f^{an'p'}
  \left\langle
  A^n_\mu A^p_\nu A^{n'\mu} A^{p'\nu}
  \right\rangle .
\label{sec5-250}
\end{equation}

\subsubsection{Calculation of
$\left\langle \mathcal{F}^m_{\mu\nu} \mathcal{F}^{m\mu\nu} \right\rangle$}

Next we work on the coset part of the Lagrangian
\begin{equation}
\begin{split}
  &\left\langle \mathcal{F}^m_{\mu\nu} \mathcal{F}^{m\mu\nu} \right\rangle = \\
  &\left\langle
   \left[
    \left(
    \partial_\mu A^m_\nu - g f^{mnb} A^n_\nu a^b_\mu
    \right) -
    \left(
    \partial_\nu A^m_\mu - g f^{mnb} A^n_\mu a^b_\nu
    \right) +
    g f^{mnp} A^n_\mu A^p_\nu
   \right]^2
  \right\rangle = \\
  &2\left\langle
    \left(
    \partial_\mu A^m_\nu - g f^{mnb} A^n_\nu a^b_\mu
    \right) ^2
    \right\rangle - \\
  &2\left\langle
    \left(
    \partial_\mu A^m_\nu - g f^{mnb} A^n_\nu a^b_\mu
    \right)
    \left(
    \partial^\nu A^{m\mu} - g f^{mn'b'} A^{n'\mu} a^{b'\nu}
    \right)
   \right\rangle + \\
   &4g \left\langle
    \left(
    \partial_\mu A^m_\nu - g f^{bmn'} A^{n'}_\nu a^b_\mu
    \right)
    f^{mnp} A^{n\mu} A^{p\nu}
   \right\rangle + \\
   &g^2 f^{mnp} f^{mn'p'}
   \left\langle
   A^n_\mu A^p_\nu A^{n'\mu} A^{p'\nu}
   \right\rangle .
\end{split}
\label{sec5-260}
\end{equation}
First we calculate
\begin{equation}
\begin{split}
  2 \left\langle \left( \partial_\mu A^m_\nu \right)^2 \right\rangle =
  2 \left\langle
  \partial_{y\mu} A^m_\nu (y) \partial^\mu_x A^{m\nu} (x)
  \right\rangle \Bigr |_{y=x} &= \\
  2 \partial_{y\mu} \partial^\mu_x
  \left\langle A^m_\nu (y) A^{m\nu} (x) \right\rangle \Bigr |_{y=x} = 
  - \frac{2}{3} \eta^\nu_\nu f^{mpb} f^{mpc} \partial_\mu \phi^b \partial^\mu \phi^c &=\\
  - \frac{8}{3} \partial_\mu \phi^a \partial^\mu \phi^a & .
\label{sec5-270}
\end{split}
\end{equation}
Analogously
\begin{equation}
  -2 \left\langle
  \partial_\mu A^m_\nu (y) \partial^\nu A^{m\mu} (x)
  \right\rangle \Bigr |_{y=x} =
  \frac{2}{3} \eta^\mu_\nu f^{mpb} f^{mpc} \partial_\mu \phi^b \partial^\nu \phi^c =
  \frac{2}{3} \partial_\mu \phi^a \partial^\mu \phi^a .
\label{sec5-280}
\end{equation}
The next term is
\begin{equation}
\begin{split}
  - 4 g  \left\langle
   \left(
   \partial_\mu A^m_\nu
   \right) f^{amn} A^{n\nu} a^{a\mu}
  \right\rangle =
  - 4 g f^{amn} a^{a\mu}
  \left\langle
   \left(
   \partial_\mu A^m_\nu
   \right) A^{n\nu}
  \right\rangle &= \\
  - 4 g f^{amn} a^{a\mu} (x)
  \left\langle
   \left(
   \partial_{y\mu} A^m_\nu (y)
   \right) A^{n\nu} (x)
  \right\rangle \Bigr |_{y=x} &= \\
  \frac{4}{3} g \eta^\nu_\nu f^{amn} a^{a\mu} f^{mpb} f^{npc}
  \partial_\mu \phi^b \phi^c =
  - \frac{16}{3} g a^{a\mu} \left( f^{amn} f^{cnp} f^{bpm} \right)
  \partial_\mu \phi^b \phi^c &= \\
  \frac{8}{3} g \epsilon^{abc} a^{a\mu} \partial_\mu \phi^b \phi^c &
\end{split}
\label{sec5-290}
\end{equation}
where the identity $f^{amn}f^{cnp} f^{bpm} = \frac{1}{2} \epsilon^{acb}$ was
used. Analogously
\begin{equation}
\begin{split}
   4 g  \left\langle
   f^{amn} A^n_\nu (y) a^a_\mu (y)
   \left(
   \partial^\nu_x A^{m\mu} (x)
   \right)
  \right\rangle \Bigr |_{y=x} =
  \frac{4}{3} g \left( f^{amn} f^{bnp} f^{cpm} \right) a^{a\mu}
  \phi^b \partial^\mu \phi^c   & =\\
  - \frac{2}{3} g \epsilon^{abc} a^{a\mu} \partial_\mu \phi^b \phi^c & .
\end{split}
\label{sec5-300}
\end{equation}
Using \eqref{sec5-220} the next term is
\begin{equation}
\begin{split}
  2 g^2 \left\langle
  f^{dmn} A^n_\nu a^d_\mu f^{d'mn'} A^{n'\nu} a^{d'\mu}
  \right\rangle =
 2 g^2 f^{dmn} a^d_\mu f^{d'mn'}  a^{d'\mu}
\left\langle  A^n_\nu  A^{n'\nu} \right\rangle &= \\
  - \frac{2}{3} \eta^\nu_\nu
  \left( f^{dmn} f^{d'mn'} f^{npb} f^{n'pc} \right)
  \left( a^d_\mu a^{d'\mu} \phi^b \phi^c \right) =
  - \frac{8}{3} g^2 E^{d'dbc}  a^d_\mu a^{d'\mu} \phi^b \phi^c
   &
\end{split}
\label{sec5-310}
\end{equation}
here $E^{d'dbc} = f^{d'n'm} f^{dmn} f^{bnp} f^{cpn'}$ and its components are
\begin{equation}
\begin{split}
  E^{aaaa} = E^{1111} = E^{2222} = E^{3333}= & \frac{1}{4}\\
  E^{aabb} = E^{1122} = E^{1133} = E^{2211} =
  E^{2233} = E^{3311} = E^{3322} =&  \frac{1}{4}\\
  E^{abab} = -E^{1212} = E^{1221} = -E^{1313} =   E^{1331} =  & \\
  -E^{2121} = E^{2112} =
  -E^{1313} = E^{3113} = -E^{3232} = E^{3223} = & \frac{1}{4} .
\end{split}
\label{sec5-320}
\end{equation}
We now note that $E^{d'dbc} a^d_\mu a^{d'\mu} \phi^b \phi^c
 = \frac{1}{4}( \epsilon ^{abc} \epsilon^{ab'c'} a_{\mu} ^b \phi ^c a^{b' \mu}
\phi ^{c'} + a_{\mu} ^b \phi ^b a^{c \mu} \phi ^c)$. Thus \eqref{sec5-310} becomes
\begin{equation}
  2 g^2 \left\langle
  f^{dmn} A^n_\nu a^d_\mu f^{d'mn'} A^{n'\nu} a^{d'\mu}
  \right\rangle =
- \frac{2 g^2}{3} (\epsilon ^{abc} \epsilon^{ab'c'} a_{\mu} ^b \phi ^c a^{b' \mu}
\phi ^{c'} + a_{\mu} ^b \phi ^b a^{c \mu} \phi ^c )
\label{sec5-330}
\end{equation}
Analogously
\begin{equation}
-2   g^2 \left\langle
  f^{dmn} A^n_\nu a^d_\mu f^{d'mn'} A^{n'\mu} a^{d'\nu}
  \right\rangle =
  \frac{g^2}{6} (\epsilon^{abc} \epsilon^{ab'c'} a^b_\mu \phi^c a^{b'\mu} \phi^{c'}
 + a_{\mu} ^b \phi ^b a^{c \mu} \phi ^c )
\label{sec5-340}
\end{equation}
Next we consider terms with three coset fields. The term involving
the derivative is
\begin{equation}
f^{mnp} \langle  (\partial _{y \mu} A^m _\nu (y)) A^{n\mu} (x) A^{p\nu} (z) \rangle
\label{sec5-350}
\end{equation}
Since the gauge fields must be symmetric under exchange, and because of the
antisymmetry of the of $f^{mnp}$ this term vanishes. Next the terms involving
three coset fields and one $SU(2)$ field we will approximate as
\begin{equation}
\begin{split}
f^{bmn'} f^{mnp} a^b_\mu \langle A^{n'}_\nu  A^{n\mu} A^{p\nu}  \rangle \approx
\frac{1}{3} f^{bmn'} f^{mnp} a^b_\mu \biggl ( \langle A^{n'}_\nu
\rangle \langle A^{n\mu} A^{p\nu}  \rangle &+ \\
\langle A^{n'}_\nu  A^{n\mu} \rangle \langle A^{p\nu}  \rangle +
\langle A^{n'}_\nu  A^{p\nu} \rangle \langle A^{n\mu} \rangle \biggl ) &
\label{sec5-360}
\end{split}
\end{equation}
By the second assumption in the previous section, $\langle A ^m _\mu (x) \rangle =0$, this
term also vanishes. Thus
\begin{equation}
\begin{split}
  \left\langle \mathcal{F}^m_{\mu\nu} \mathcal{F}^{m\mu\nu} \right\rangle
  = - 2 \partial_\mu \phi^{a} \partial^\mu \phi^a +
  2 g \epsilon^{abc} a^{a\mu} \partial_\mu \phi^{b} \phi^c  -
  \frac{g^2}{2} \epsilon^{abc} \epsilon^{ab'c'} a^b_\mu \phi^c a^{b'\mu} \phi^{c'}
& \; \\
 - \frac{g^2}{2} a_{\mu} ^b \phi ^b a^{c \mu} \phi ^c +
  g^2 f^{mnp} f^{mn'p'}
  \left\langle A^n_\mu A^p_\nu A^{n'\mu} A^{p'\nu} \right\rangle &= \\
 - 2 \left(
  \partial_\mu \phi^a + \frac{g}{2} \epsilon^{abc} a^b_\mu \phi^c
  \right)^2 -
  \frac{g^2}{2} a_{\mu} ^b \phi ^b a^{c \mu} \phi ^c +
  g^2 f^{mnp} f^{mn'p'}
  \left\langle A^n_\mu A^p_\nu A^{n'\mu} A^{p'\nu} \right\rangle .
\end{split}
\label{sec5-370}
\end{equation}
The full averaged Lagrangian is
\begin{equation}
\begin{split}
    - \frac{1}{4}  \left\langle \mathcal{F}^A_{\mu\nu} \mathcal{F}^{A\mu\nu} \right\rangle = &
  - \frac{1}{4}  h^a_{\mu\nu} h^{a\mu\nu} +
  \frac{1}{2}   \left(
  \partial_\mu \phi^a + \frac{g}{2} \epsilon^{abc} a^b_\mu \phi^c
  \right)^2 +
  \frac{g^2}{2} a_{\mu} ^b \phi ^b a^{c \mu} \phi ^c - \\
  &\frac{1}{4}  g^2 f^{Anp} f^{An'p'}
  \left\langle A^n_\mu A^p_\nu A^{n'\mu} A^{p'\nu} \right\rangle .
\end{split}
\label{sec5-380}
\end{equation}
where we have collected the quartic terms from eq. \eqref{sec5-250} and \eqref{sec5-370}
together into \\
$f^{Anp} f^{An'p'} \left\langle A^n_\mu A^p_\nu A^{n'\mu} A^{p'\nu} \right\rangle$.

\subsubsection{The quartic term}

In this section we show that the quartic term -- $  f^{Anp} f^{An'p'}
  \left\langle A^n_\mu A^p_\nu A^{n'\mu} A^{p'\nu} \right\rangle $ --
from eq. \eqref{sec5-380} becomes an effective $\lambda \phi ^4$
interaction term for the effective scalar field introduced in  eq. \eqref{sec5-230}.
Just as in eq. \eqref{sec5-230} where a quadratic gauge field expression was
replaced by a quadratic effective scalar field expression, here we replace
the quartic gauge field term by a quartic scalar field term
\begin{equation}
\begin{split}
  \left\langle
  A^m_\alpha (x) A^n_\beta (y) A^p_\mu (z) A^q_\nu (u)
  \right\rangle =
  \left(
  E^{mnpq}_{1,abcd} \eta_{\alpha\beta} \eta_{\mu\nu} +
  E^{mpnq}_{2,abcd} \eta_{\alpha\mu} \eta_{\beta\nu} +
  E^{mqnp}_{3,abcd} \eta_{\alpha\nu} \eta_{\beta\mu}
  \right)& \\
  \phi^a (x) \phi^b(y) \phi^c (z) \phi^d(u) &
\label{sec5-390}
\end{split}
\end{equation}
here $E^{mnpq}_{1,abcd}, E^{mpnq}_{2,abcd}, E^{mqnp}_{3,abcd}$ are constants.
Because of the bosonic character of the gauge fields in \eqref{sec5-390} the
indices of these constants in conjunction with the indices of the
$\eta _{\alpha \beta}$'s must reflect symmetry under exchange of
the fields.  The simplest choice that satisfies this requirement is
\begin{equation}
\begin{split}
  \left\langle
  A^m_\alpha (x) A^n_\beta (y) A^p_\mu (z) A^q_\nu (u)
  \right\rangle =
  \left(
  \delta^{mn}\delta^{pq} \eta_{\alpha\beta} \eta_{\mu\nu} +
  \delta^{mp}\delta^{nq} \eta_{\alpha\mu} \eta_{\beta\nu} +
  \delta^{mq}\delta^{np} \eta_{\alpha\nu} \eta_{\beta\mu}
  \right) & \\ 
  e_{abcd} \phi^a (x) \phi^b(y) \phi^c (z) \phi^d(u) &
\label{sec5-400}
\end{split}
\end{equation}
This choice of taking the constants from eq. \eqref{sec5-390} to be products
of Kronecker deltas and fixing $a=b=c=d$  for the lower indices, satisfies
the bosonic character requirement for the gauge fields, and is equivalent
to the reduction used for the quartic term in \cite{dzhsin6, dzhsin7}.
Evaluating eq. \eqref{sec5-400} at one spacetime point ({\it i.e.} $x=y=z=u$)
and contracting the indices to conform to quartic term in eq. \eqref{sec5-380}
gives
\begin{equation}
\begin{split}
  \left\langle
  A^m_\alpha (x) A^n_\beta (x) A^p_\mu (x) A^q_\nu (x)
  \right\rangle =
  \left(
  \delta^{mn}\delta^{pq}  \eta_{\mu\nu} \eta^{\mu\nu} +
  \delta^{mp}\delta^{nq} \eta_{\mu} ^{\mu} \eta_{\nu} ^{\nu} +
  \delta^{mq}\delta^{np} \eta_{\mu} ^{\nu} \eta_{\nu} ^{\mu}
  \right) & \\
  \left( \phi^a (x) \phi^a(x) \right)^2 &
\label{sec5-410}
\end{split}
\end{equation}
where the constants $e_{abcd}$ are chosen that at the point  $x=y=z=u$
\begin{equation}
\label{sec5-420}
    e_{abcd} = \frac{1}{3}
    \left(
    \delta_{ab} \delta_{cd} + \delta_{ac} \delta_{bd} +
    \delta_{ad} \delta_{bc}
    \right) .
\end{equation}
This expression can be further simplified to
\begin{equation}
  \left\langle
  A^m_\mu A^n_\nu A^{p\mu} A^{q\nu}
  \right\rangle =
  \left(
  4 \delta^{mn}\delta^{pq}   +
  16 \delta^{mp}\delta^{nq}  +
  4  \delta^{mq}\delta^{np}
  \right)
  \left( \phi^a (x) \phi^a(x) \right)^2 .
\label{sec5-430}
\end{equation}
Substituting this into the original quartic term of eq. \eqref{sec5-380} yields
\begin{equation}
\begin{split}
  f^{Anp} f^{An'p'}
  \left\langle A^n_\mu A^p_\nu A^{n'\mu} A^{p'\nu} \right\rangle &= \\
  (4 f^{Ann} f^{An'n'} + 16 f^{Anp} f^{Anp} + 4 f^{Anp} f^{Apn})
  \left( \phi^a (x) \phi^a(x) \right)^2  & = \\
  12  f^{Anp} f^{Anp}  \left( \phi^a (x) \phi^a(x) \right)^2
\label{sec5-440}
\end{split}
\end{equation}
where the antisymmetry property of the structure constants has been used.
Using the explicit expression for the structure constants
($f^{123} =1  ; \; \; f^{147}=f^{246}=f^{257} =
f^{345}=f^{516}=f^{637} = \frac{1}{2} ; \; \; f^{458} =f^{678} =\frac{\sqrt{3}}{2}$
plus those related to these by permutations), and recalling that the index
$A$ runs from $1-8$ while the indices $n, p$ run from $4-8$ one can show that
$f^{Anp} f^{Anp}= 12$ ($f^{123}$
and related constants do not contribute to this expression). Combining
these results transforms the quartic term in eq. \eqref{sec5-380} as
\begin{equation}
 \frac{1}{4}  g^2 f^{Anp} f^{An'p'}
  \left\langle A^n_\mu A^p_\nu A^{n'\mu} A^{p'\nu} \right\rangle =
36 g^2 \left( \phi^a (x) \phi^a(x) \right)^2
\equiv \frac{\lambda}{4} \left( \phi^a (x) \phi^a(x) \right)^2
\label{sec5-450}
\end{equation}
One can show that the simplification \eqref{sec4-110} is also applicable here and will 
lead to the same result. 
This has transformed the quartic gauge field term of the coset fields into a quartic
interaction term for the effective scalar field. Substituting this result back into the
averaged Lagrangian of eq. \eqref{sec5-380} we find
\begin{equation}
\begin{split}
- \frac{1}{4}  \left\langle \mathcal{F}^A_{\mu\nu} \mathcal{F}^{A\mu\nu} \right\rangle =
  - \frac{1}{4}  h^a_{\mu\nu} h^{a\mu\nu} +
  \frac{1}{2}   \left(
  \partial_\mu \phi^a + \frac{g}{2} \epsilon^{abc} a^b_\mu \phi^c
  \right)^2 +
  \frac{g^2}{2} a_{\mu} ^b \phi ^b a^{c \mu} \phi ^c & - \\
  \frac{\lambda}{4} \left( \phi^a (x) \phi^a(x) \right)^2 & 
\label{sec5-460}
\end{split}
\end{equation}
The original pure SU(3) gauge theory has been transformed into an SU(2) gauge
theory coupled to an effective triplet scalar field. This is similar to the
Georgi-Glashow \cite{gg} Lagrangian except for the presence of the term
$\frac{g^2}{2} a_{\mu} ^b \phi ^b a^{c \mu} \phi ^c$ and
the absence of a negative mass term for $\phi ^a$ of the form
$m^2 \phi^a (x) \phi^a(x)$.

The Georgi-Glashow Lagrangian is known to have topological monopole solutions
\cite{thp} which have the form
\begin{equation}
\phi ^a =  \frac{x^a f(r)}{g r^2} \; \;  ; \; \; a^a_0 = 0 \; \;  ; \; \;
a^a _i = \frac{\epsilon _{aib} x^b [1- h(r)]}{g r^2}
\label{sec5-465}
\end{equation}
where $f(r)$ and $h(r)$ are functions determined by the field equations.
For this form of the scalar and SU(2) gauge fields the term,
$\frac{g^2}{2} a_{\mu} ^b \phi ^b a^{c \mu} \phi ^c$,
vanishes from the Lagrangian in eq. \eqref{sec5-460} by the antisymmetry
of $\epsilon _{aib}$ and the symmetry of $x^a x^b$. Thus for the monopole
ansatz of eq. \eqref{sec5-465} the Lagrangian in \eqref{sec5-460} becomes
equivalent to the Georgi-Glashow Lagrangian minus only the mass term for
the scalar field.

In the present work we simply postulate that the effective scalar field develops
a negative mass term of the form $-\frac{m^2}{2} (\phi ^a \phi ^a )$ which is
added by hand to the Lagrangian of
\eqref{sec5-460} to yield
\begin{equation}
\begin{split}
  - \frac{1}{4}  h^a_{\mu\nu} h^{a\mu\nu} +
  \frac{1}{2}   \left(
  \partial_\mu \phi^a + \frac{g}{2} \epsilon^{abc} a^b_\mu \phi^c
  \right)^2 + \frac{m^2}{2} (\phi ^a \phi ^a )
  - \frac{\lambda}{4} \left( \phi^a (x) \phi^a(x) \right)^2 & + \\
  \frac{g^2}{2} a_{\mu} ^b \phi ^b a^{c \mu} \phi ^c &
\label{sec5-470}
\end{split}
\end{equation}
The scalar field now has the standard symmetry breaking form and this effective
Lagrangian has finite energy 't Hooft-Polyakov solutions (the last term should
not alter the monopole construction since it vanishes under the ansatz in
\eqref{sec5-465}).
\par 
At the end of this section we have to note the following problem. 
The covariant deravitives 
$D_{\mu} a^a_\nu = \partial_\mu a^a_\nu + g \epsilon^{abc} a^b_\mu a^c_\nu$ and 
$D_{\mu} \phi^a = \partial_\mu \phi^a + \frac{g}{2} \epsilon^{abc} a^b_\mu \phi^c $ 
are different. One can avoid this problem if we suppose that there is a correlation 
between ordered phase $a^a_\mu$ and disordered phase $A^m_\mu$ 
\begin{equation}
\begin{split}
  \left\langle 
  	a^{a_1}_{\mu_1}(x_1) \cdots a^{a_n}_{\mu_n}(x_n) 
  	A^m_\alpha (y) A^n_\beta (z) 
  \right\rangle = 
 	k_n a^{a_1}_{\mu_1}(x_1) \cdots a^{a_n}_{\mu_n}(x_n)
  \left\langle 
  	A^m_\alpha (y) A^n_\beta (z) 
  \right\rangle .
\label{sec5-480}
\end{split}
\end{equation}
If we suppose that $k_1 = 2$ and $k_2 = 4$ then the rhs of equations 
\eqref{sec5-290} \eqref{sec5-300} will be multiplied on 2 and equations 
\eqref{sec5-310} \eqref{sec5-330} \eqref{sec5-340} will be multiplied on 
4. Consequently the full Lagrangian will be 
\begin{equation}
\begin{split}
    - \frac{1}{4}  \left\langle \mathcal{F}^A_{\mu\nu} \mathcal{F}^{A\mu\nu} \right\rangle = &
  - \frac{1}{4}  h^a_{\mu\nu} h^{a\mu\nu} +
  \frac{1}{2}   \left(
  \partial_\mu \phi^a + g \epsilon^{abc} a^b_\mu \phi^c
  \right)^2 + \frac{m^2}{2} (\phi ^a \phi ^a ) \\
  &- \frac{\lambda}{4} \left( \phi^a (x) \phi^a(x) \right)^2 + 
  \frac{g^2}{2} a_{\mu} ^b \phi ^b a^{c \mu} \phi ^c 
\end{split}
\label{sec5-380}
\end{equation}
This Lagrangian we will investigate numerically on the next sections.

Our final result given in \eqref{sec5-470} depends on several crucial assumptions
({\it e.g.} the existence of the negative mass term $-\frac{m^2}{2} (\phi ^a \phi ^a )$).
In the next section we will investigate a non-topological flux tube solution to this
effective Lagrangian.

\subsection{Flux tube equations}

We will start from the SU(2) Yang - Mills - Higgs field equations \eqref{sec5-470}
with broken gauge symmetry
\begin{eqnarray}
  \mathcal{D}_\nu h^{a\mu\nu} &=& g \epsilon^{abc} \phi^b
  \mathcal{D}^\mu \phi^c - \left( m^2 \right)^{ab} a^{b\mu} ,
\label{sec5-475}\\
  \mathcal{D}_\mu \mathcal{D}^\mu \phi^a &=& -\lambda \phi^a
  \left(
  \phi^b \phi^b - \phi^2_\infty
  \right)
\label{sec5-480}
\end{eqnarray}
here $h_{a\mu\nu} = \partial_\mu a^a_\nu - \partial_\nu a^a_\mu
+ g \epsilon^{abc} a^b_\mu a^c_\nu$ is the field tensor for the SU(2) gauge
potential $a^a_\mu$; $a,b,c = 1,2,3$ are the color indices;
$D_\nu [\cdots ]^a = \partial_\nu [\cdots ]^a +
g \epsilon^{abc} a^b_\mu  [\cdots ]^c$ is the gauge derivative; $\phi^a$
is the Higgs field; $\lambda , g$ and $\phi_\infty$ are some constants;
$\left( m^2 \right)^{ab}$ is a masses matrix which
destroys the gauge invariance of the Yang - Mills - Higgs theory,
here we choose $\left( m^2 \right)^{ab} = diag \left\{ m^2_1, m^2_2, 0 \right\}$.
With a standard SU(2)-Higgs symmetry breaking Lagrangian the masses
developed would be identical, and would be related to the gauge coupling and
vacuum expectation of the field -- $m_a \simeq \phi _{\infty} g$. In the present case
we are postulating that the masses of the two gauge bosons are different. This was done
in order to find electric flux tube solutions for the system \eqref{sec5-475} \eqref{sec5-480}. 
For equal
masses electric flux tubes did not exist for the ansatz used. In numerically solving
\eqref{sec5-475} \eqref{sec5-480} the two mass come out close to one another, but are not equal. 
This splitting of the gauge bosons mass over and above the Higgs mechanism explicitly
breaks the gauge invariance, and at present we have no good justification for this
step other than the fact that it leads to the existence of electric flux tubes of the kind that
are very important in confinement. 
\par
We will use the following ansatz
\begin{equation}
    a^1_t(\rho) = \frac{f(\rho)}{g} ; \quad a^2_z(\rho) = \frac{v(\rho)}{g} ;
    \quad \phi^3(\rho) = \frac{\phi(\rho)}{g}
\label{sec5-490}
\end{equation}
here $z, \rho , \varphi$ are cylindrical coordinates. Substituting this 
into the Yang - Mills - Higgs equations \eqref{sec1-10} \eqref{sec5-480} gives us
\begin{eqnarray}
    f'' + \frac{f'}{x} &=& f \left( \phi^2 + v^2 - m^2_1 \right),
\label{sec5-500}\\
    v'' + \frac{v'}{x} &=& v \left( \phi^2 - f^2 - m^2_2 \right),
\label{sec5-510}\\
    \phi'' + \frac{\phi'}{x} &=& \phi \left[ - f^2 + v^2
    + \lambda \left( \phi^2 - \phi^2_\infty \right)\right]
\label{sec5-520}
\end{eqnarray}
here we redefined $\phi /\phi_0 \rightarrow \phi$, $f /\phi_0  \rightarrow f$,
$v /\phi_0  \rightarrow v$, $\phi_\infty /\phi_0 \rightarrow \phi_\infty$,
$m_{1,2} /\phi_0  \rightarrow m_{1,2}$, $\rho \phi_0  \rightarrow x$;
$\phi_0 = \phi(0)$.
Similar equations with the presence of $A^a_\varphi$ and
without masses $m_{1,2}$ were investigated in ref's \cite{Obukhov}
\cite{dzhsin8} where it was shown that the corresponding solutions
becomes singular either on a finite distance from the axis $\rho = 0$ or
on the infinity. The solutions with the masses $m_{1,2}$ and a magnetic
field $H_z$ were obtained in ref. \cite{dzhun4} and the result is that
the following versions of the flux tube
exist: (1) the flux tube filled with electric/magnetic fields
on the background of an external constant magnetic/electric field; (2)
the Nielsen-Olesen flux tube dressed with transversal color electric
and magnetic fields.

\subsection{Numerical algorithm}

As in section \ref{glueball} we will solve the equations set
\eqref{sec5-500}-\eqref{sec5-520} with the iterative procedure
which is described in more detail in ref. \cite{dzhun1}. Eq. \eqref{sec5-510} for
the $i ^{th}$ step has the following form
\begin{equation}
    v_i'' + \frac{v_i'}{x} = v_i \left( \phi^2_{i-1} - f^2_{i-1} - m^2_{2,i} \right),
\label{sec5-530}\\
\end{equation}
here the functions $f_{i-1}$ and $\phi_{i-1}$ were defined at the
$(i-1) ^{th}$ step and the null approximation for the functions $f(x)$
and $\phi(x)$ is $\phi_0(x) = 1.3 - 0.3/\cosh^2 (x/4)$ and $f_0(x)
= 0.2/\cosh^2(x)$. Again these forms are chosen because they are already
similar to the form of solution which is being sought. The numerical investigation 
for eq. \eqref{sec5-530} shows that there is a number $m^*_{2,i}$ for
which: for $m_{2,i} > m^*_{2,i}$ the solution is singular $v_i(x)
\rightarrow - \infty$ by $x \rightarrow \infty$ and for $m_{2,i} <
m^*_{2,i}$ the solution is also singular $v_i(x) \rightarrow +
\infty$ by $x \rightarrow \infty$. In Fig. \ref{fig:v-sing} the
singular solutions are presented.
\begin{figure}[h]
  \begin{center}
    \fbox{
    \includegraphics[height=5cm,width=5cm]{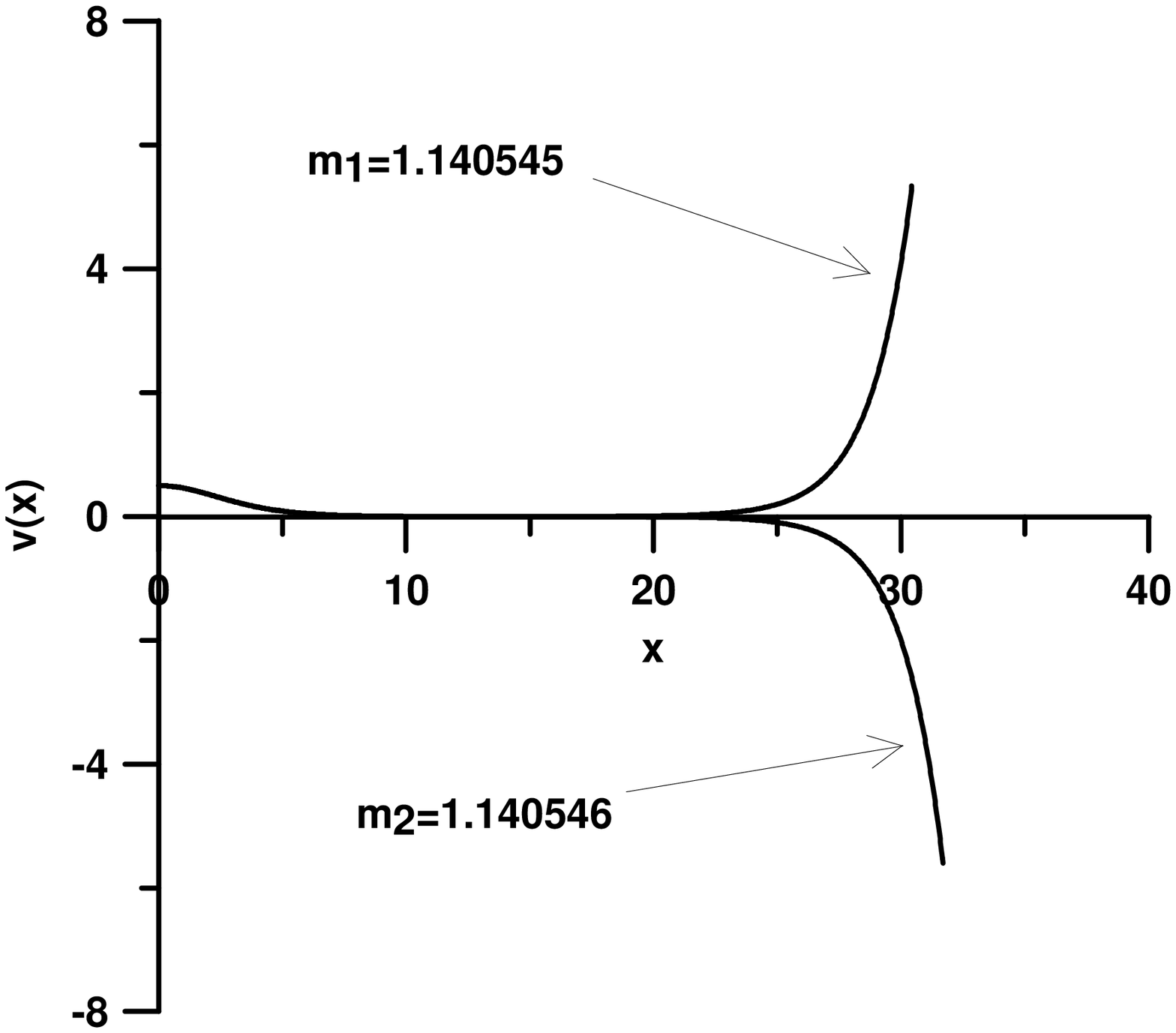}
         }
    \caption{The typical singular solution $v_i(x)$ for equation \eqref{sec5-530}}.
    \label{fig:v-sing}
  \end{center}
\end{figure}
The asymptotical form of the equation for the singular solution is
\begin{equation}
    v_{i,s}'' + \frac{v_{i,s}'}{x} \approx v_{i,s} \left( \phi^2_{i-1} - m^2_{2,i} \right)
\label{sec5-533}
\end{equation}
and corresponding asymptotical singular solutions are
\begin{equation}
    v_i \approx \pm a_{\pm}\frac{e^{x \sqrt{\phi^2_{i-1} - m^2_{2,i}}}}{\sqrt{x}}
\label{sec5-536}
\end{equation}
where $a_{\pm}$ are some constants. This indicates that there may exist 
some value, $m_{2,i} = m^*_{2,i}$, for which the solution
$v_i(x)$ is regular. The number $m^*_{2,i}$ is found with the method of
the iterative approximation as for  $m^* _i$ and $\mu ^*_i$ in the
glueball case. It is necessary that the functions $f_{i-1}(x) \rightarrow 0$ and
$\phi_{i-1}(x) \rightarrow \phi^*_{\infty, i-1}$ as
$x \rightarrow \infty$ in order to have acceptable solutions.
\par
Eq. \eqref{sec5-500} at this step has the following form
\begin{equation}
    f_i'' + \frac{f_i'}{x} = f_i \left( \phi^2_{i-1} + v^2_i - m^2_{1,i} \right),
\label{sec5-540}\\
\end{equation}
here the function $\phi_{i-1}$ was defined on $(i-1)^{th}$ step and the
function $v_i$ is the solution of eq. \eqref{sec5-530}. The
numerical investigation for eq. \eqref{sec5-540} shows that there
is a number $m^*_{1,i}$ for which: for $m_{1,i} > m^*_{1,i}$ the
solution is singular $f_i(x) \rightarrow - \infty$ by $x
\rightarrow \infty$ and for $m_{1,i} < m^*_{1,i}$ the solution is
also singular $f_i(x) \rightarrow + \infty$ by $x \rightarrow
\infty$. The asymptotic singular behavior for $f_i(x)$ is the
same as in the previous case. This implies that there may be some
particular value $m_{1,i} = m^*_{1,i}$ for which the solution $f_i(x)$ is 
regular. The number $m^*_{1,i}$ is found with the method of
iterative approximation. It is necessary that the functions
$v_i(x) \rightarrow 0$ and $\phi_{i-1}(x) \rightarrow
\phi^*_{\infty,i}$ as $x \rightarrow \infty$ in order to have acceptable solutions.
\par
The next step in the $i^{th}$ iteration is solving of eq.
\eqref{sec5-520}
\begin{equation}
    \phi_i'' + \frac{\phi_i'}{x} = \phi_i \left[ -f^2_i + v^2_i +
    \lambda \left( \phi^2_i - \phi^2_{\infty ,i} \right)\right]
\label{sec5-550}
\end{equation}
here the functions $f_i$ and $v_i$ are the solutions of eqns.
\eqref{sec5-530} \eqref{sec5-540}. The numerical investigation for
this equation shows that there is a number $\phi^*_{\infty ,i}$
for which: for $\phi_{\infty ,i} > \phi^*_{\infty ,i}$ the
function $\phi_i (x)$ oscillates with decreasing amplitude, for
$\phi_{\infty ,i} < \phi^*_{\infty ,i}$ the function $\phi_i (x)
\rightarrow + \infty$ by $x \rightarrow x_0$. The singular
solution in this case is similar to the solution presented on Fig.
\ref{fig:phi-sing}. The number $\phi^*_{\infty ,i}$ is intermediate 
between the values that give the singular and oscillating solutions.
For this value it is possible that a regular solution exists. To have acceptable
behavior for the solution $\phi _i (x)$  it is necessary that the functions
$v_i(x) \rightarrow 0$ and $f_i(x) \rightarrow 0$ as $x \rightarrow \infty$.
\par
For the initial values we chose the following values:
$f(0) = 0.2$,  $v(0) = 0.5$ and $\phi(0) = 1.0$, and
$f'(0)=v'(0)=\phi ' (0) =0$. 
The set of equations \eqref{sec5-500}-\eqref{sec5-520}
has one independent parameter $\lambda$, which we set to
$\lambda = 0.2$ to perform the numerical solution of the equations. 
The other parameters of the equations ($m^*_{(1,2), i}$ and
$\phi^*_{\infty ,i}$ ) are determined in the process of solving the
equations.
\par
The iterative process described above gives $m^*_{(1,2), i}$ and
$\phi^*_{\infty ,i}$ presented on Table \ref{table3}.
\begin{table}[h]
    \begin{center}
        \begin{tabular}{|c|c|c|c|c|}\hline
          i & 1 & 2& 3& 4 \\ \hline
            $m^*_{1,i}$ & 1.194482\ldots & 1.20214\ldots & 1.20106\ldots
            & 1.2011\ldots \\ \hline
            $m^*_{2,i}$ & 1.1405457\ldots & 1.1517068\ldots & 1.151087\ldots
            & 1.150935\ldots\\ \hline
            $\phi^*_{\infty ,i}$ & 1.2360886\ldots & 1.23189178\ldots & 1.23253772\ldots
            & 1.23260797\ldots\\ \hline
        \end{tabular}
    \end{center}
    \caption{The iterative parameters $m^*_{(1,2), i}$ and $\phi^*_{\infty ,i}$.}
    \label{table3}
\end{table}
The functions $f_i(x), v_i(x)$ and $\phi_i(x)$ for $i=1,2,3,4$ are presented
in figs. \ref{fig1}, \ref{fig2} and \ref{fig3a}. As in the glueball analysis we have
only numerically integrated the solution out to finite $x$ so it is not clear if this solution
is truely well behaved as $x \rightarrow \infty$. For the range of $x$ where we have found
the solution it behaves as desired. 
\begin{figure}[h]
\begin{minipage}[t]{.45\linewidth}
    \begin{center}
    \fbox{
        \includegraphics[height=5cm,width=5cm]{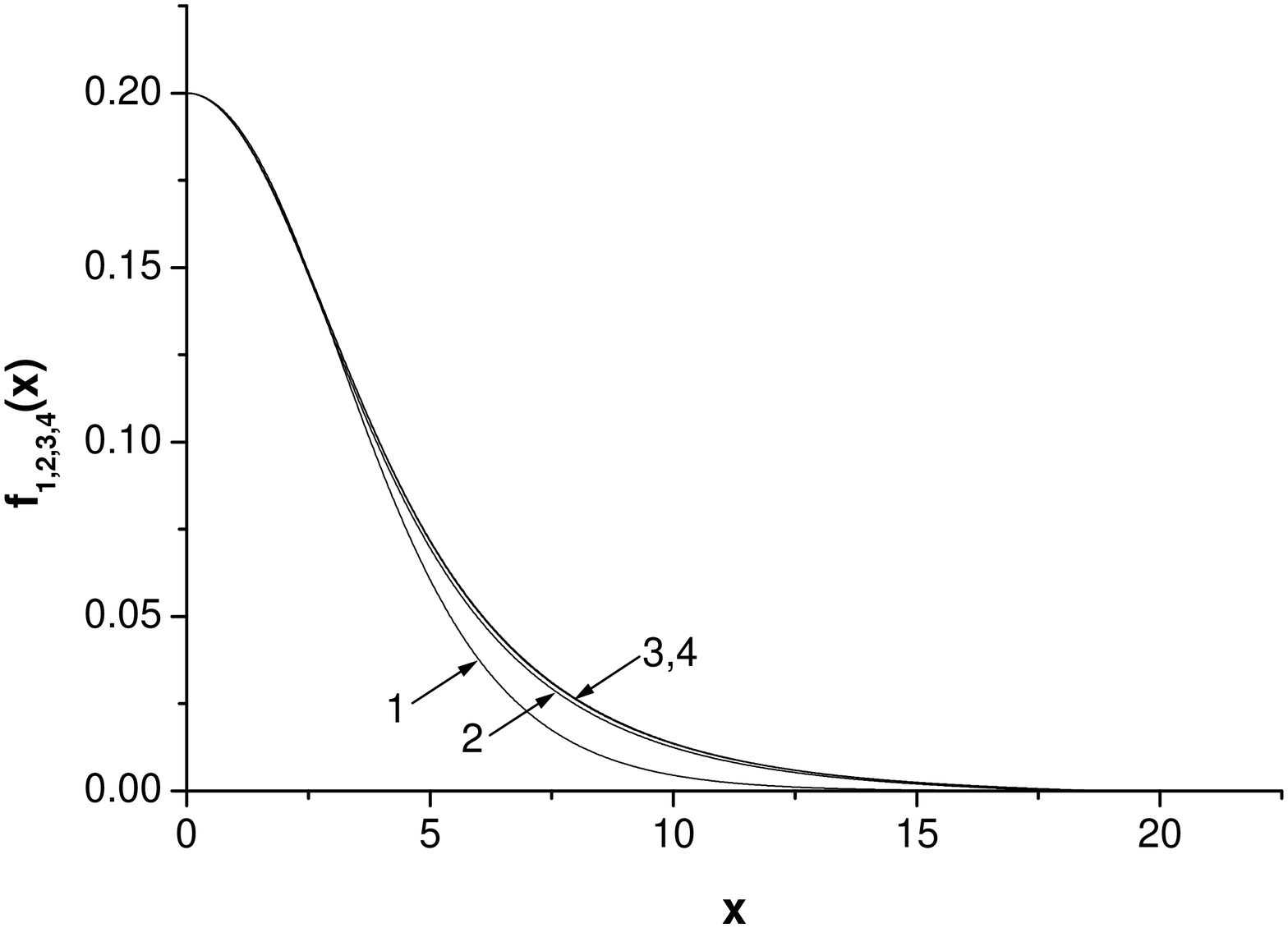}}
        \caption{The first four iterations for the $A^1_t$ gauge potential component.}
        \label{fig1}
    \end{center}
\end{minipage}\hfill
\begin{minipage}[t]{.45\linewidth}
    \begin{center}
    \fbox{
        \includegraphics[height=5cm,width=5cm]{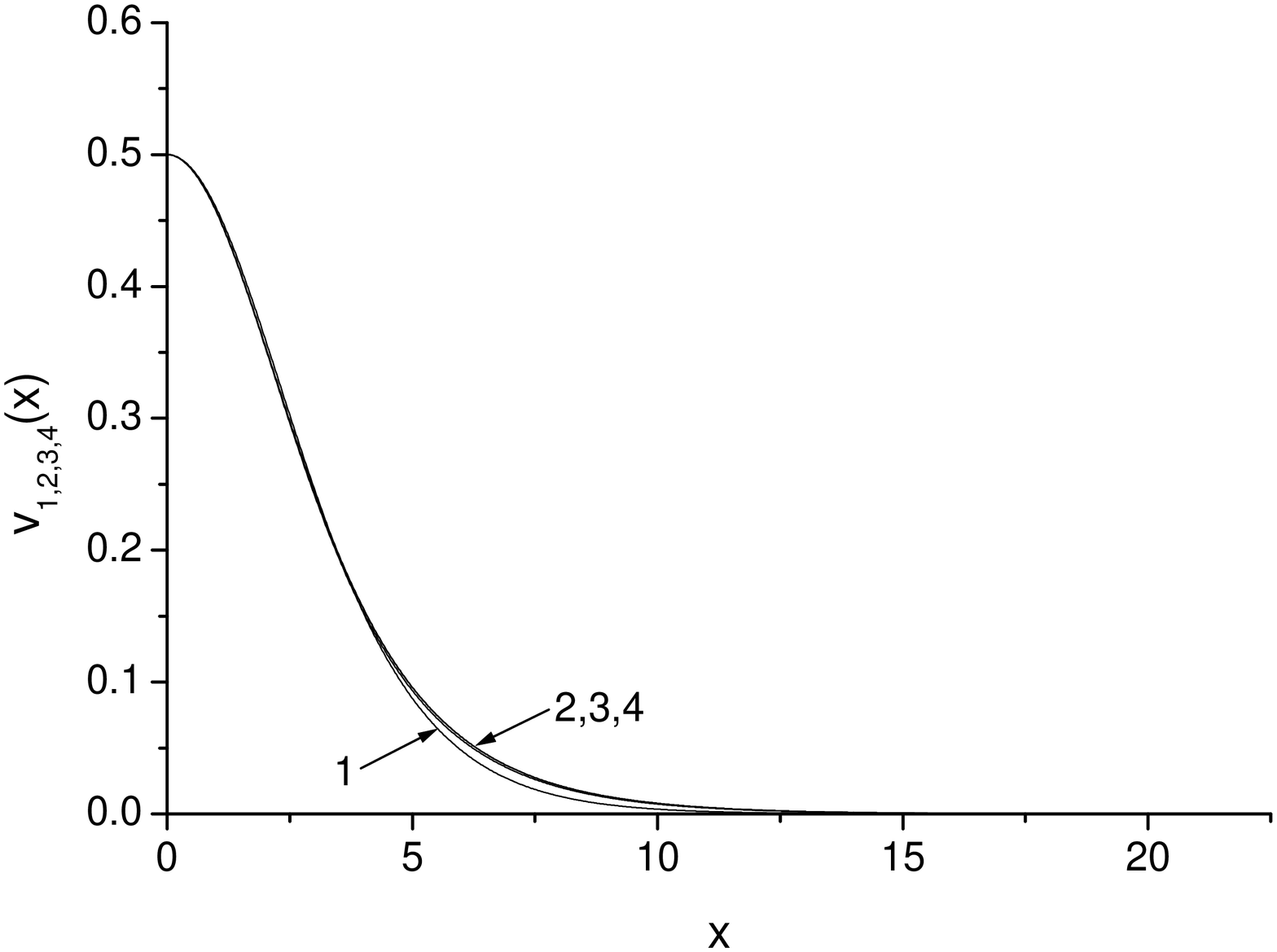}}
        \caption{The first four iterations for the $A^2_z$ gauge potential component.}
        \label{fig2}
    \end{center}
\end{minipage}\hfill
\end{figure}

\begin{figure}[h]
\begin{minipage}[t]{.45\linewidth}
    \begin{center}
    \fbox{
        \includegraphics[height=5cm,width=5cm]{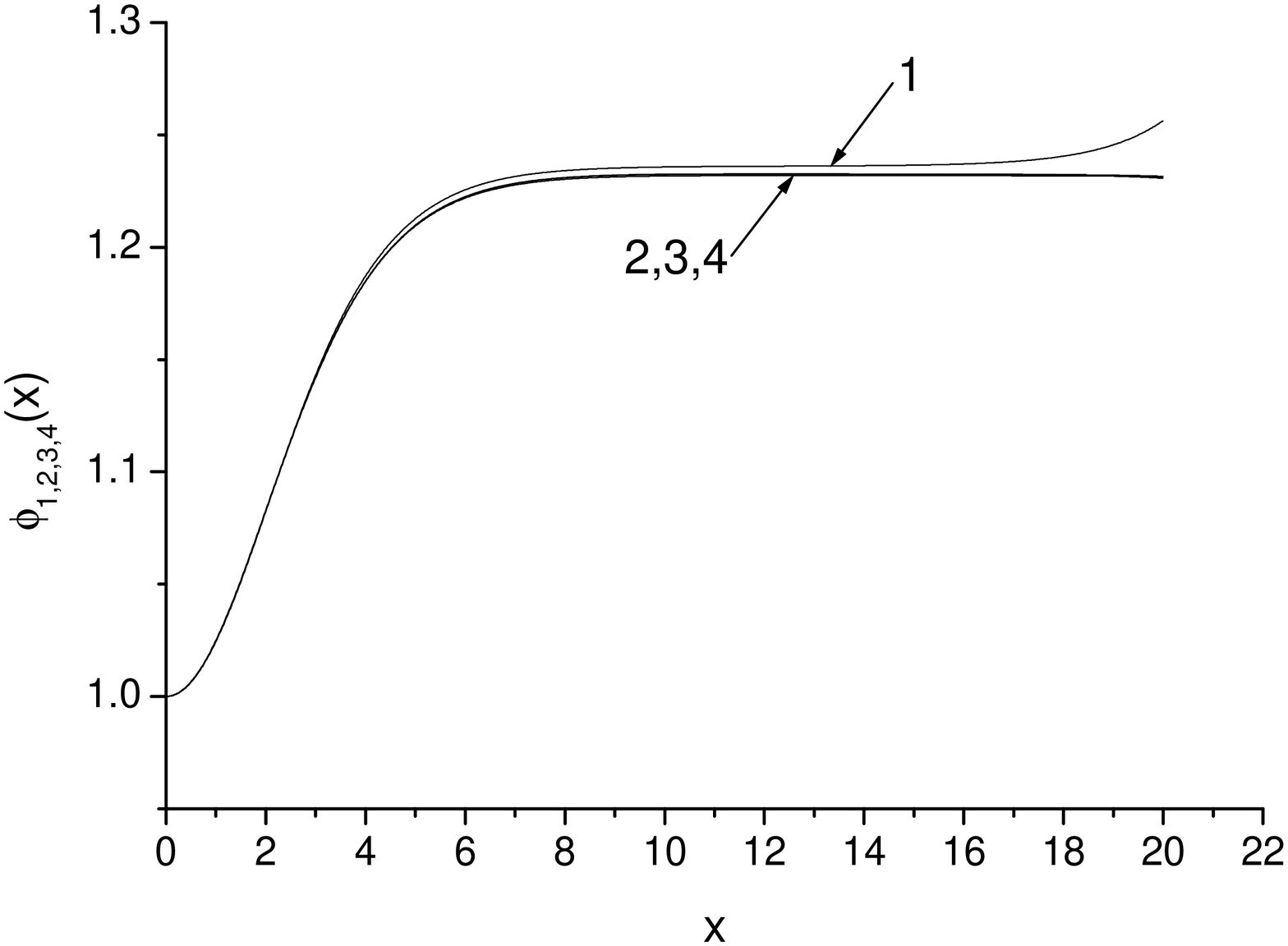}}
    \caption{The first three iterations for the $\phi^3$ scalar field.}
    \label{fig3a}
    \end{center}
\end{minipage}\hfill
\begin{minipage}[t]{.45\linewidth}
    \begin{center}
    \fbox{
        \includegraphics[height=5cm,width=5cm]{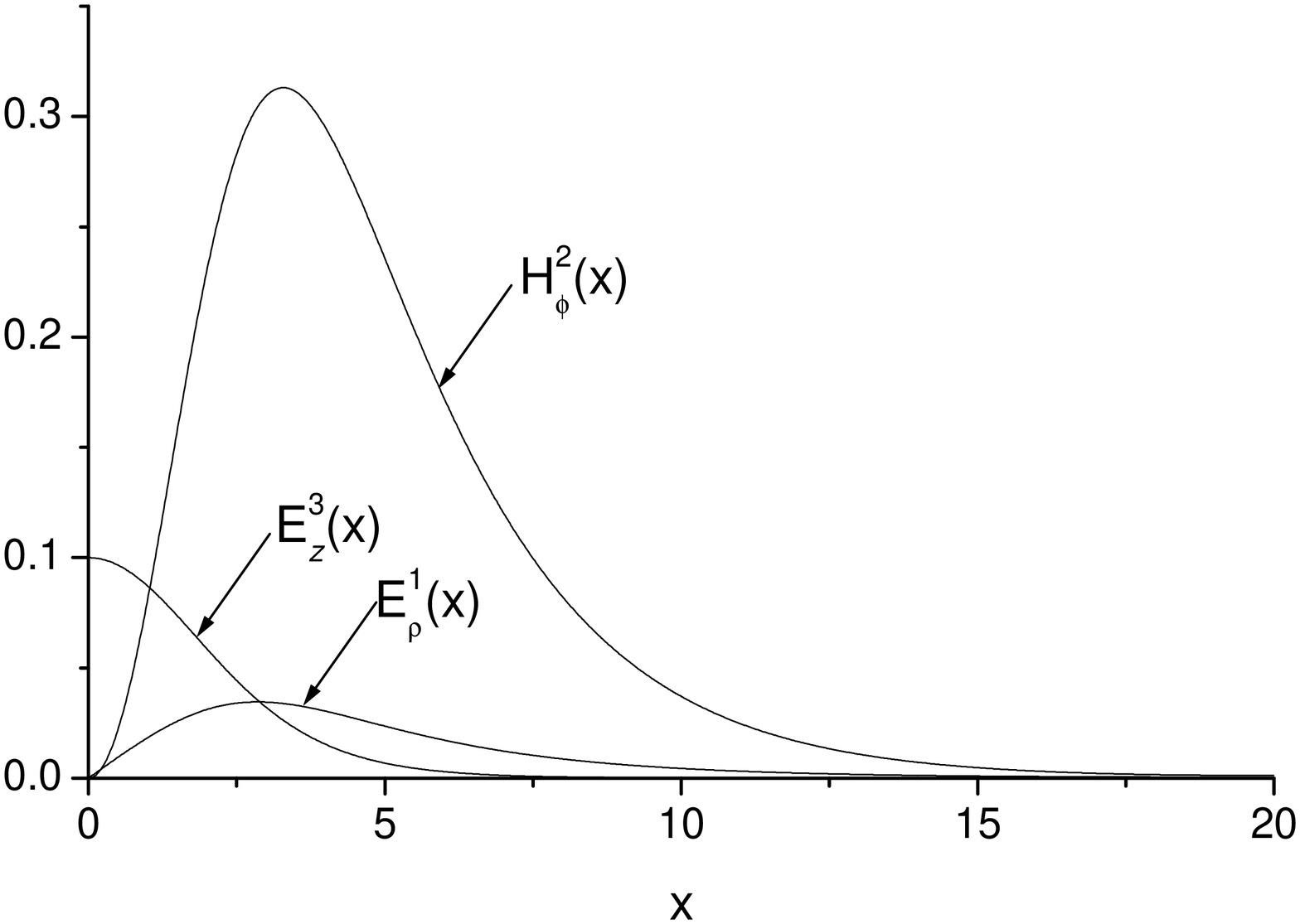}}
    \caption{The color electric $E^3_z(x)$, $E^1_\rho(x)$ and magnetic
    $H^2_\varphi (x)$ fields}
    \label{fig4}
    \end{center}
\end{minipage}\hfill
\end{figure}
\par
In terms of the ansatz functions the color electric and magnetic fields are
\begin{eqnarray}
    E^3_z(x) &=& F^3_{tz} = \frac{f(x) v(x)}{g}  ,
\label{sec5-560}\\
  E^1_\rho(x) &=& F^1_{t \rho} = -\frac{f'(x)}{g} ,
\label{sec5-570}\\
  H^2_\varphi (x) &=& x \epsilon_{\varphi \rho z} F^{2\rho z} =
  -x \frac{v'(x)}{g}.
\label{sec5-580}
\end{eqnarray}
These shape of these fields is presented in fig. \ref{fig4}.
\par
From eqs. \eqref{sec5-500}-\eqref{sec5-520} one can see that
the asymptotic behavior of the regular solutions $f^*(x)$,
$v^*(x)$ and $\phi^*(x)$ is
\begin{eqnarray}
    f^*(x) &=& f_0 \frac{e^{-x\sqrt{\phi^{*2}_\infty - m_1^{*2}}}}{\sqrt{x}} +
    \cdots ,
\label{sec5-590}\\
    v^*(x) &=& v_0 \frac{e^{-x\sqrt{\phi^{*2}_\infty - m_2^{*2}}}}{\sqrt{x}} +
    \cdots ,
\label{sec5-600}\\
  \phi^*(x) &=& \phi^*_\infty -
  \phi_0 \frac{e^{-x\sqrt{2\lambda \phi^{*2}_\infty}}}{\sqrt{x}} + \cdots
\label{sec5-610}
\end{eqnarray}
where $f_0, v_0$ and $\phi_0$ are constants.
\par
From the Lagrangian in \eqref{sec5-470} the energy density $\epsilon(x)$ in
terms of the ansatz functions is given by 
\begin{eqnarray}
    2 g ^2 \epsilon(x) = {f'}^2(x) + {v'}^2(x) + {\phi '}^2(x) +
    v^2(x) f^2(x) + v^2(x) \phi^2(x) + 
\nonumber \\
    f^2(x) \phi^2(x) + m^*_1 f^2(x) - m^*_2 v^2(x) + \frac{\lambda}{2}
    \left( \phi^2 - \phi^2_\infty \right)^2
\label{sec5-620}
\end{eqnarray}
Its shape is shown in Fig. \ref{fig:en-dens}. The exponentially decreasing  
asymptotic forms indicate that the energy density should go to zero 
at the infinity, and thus the total energy of the system should also be finite.
Since we have only been able to integrate the solution to finite $x$ we can not
say absolutely that the total energy is finite.
In the figs. \ref{en_lam} and \ref{en_phi} the profiles of the linear energy
density (the integral of the energy density over the $\rho$ coordinate) is
shown for different $\lambda$ and $\phi_0$. The linear energy density is 
given by
\begin{equation}
    w = \int 2 \pi \rho \epsilon(\rho) d\rho
\label{sec5-625}
\end{equation}
\begin{figure}[h]
    \begin{center}
    \fbox{
        \includegraphics[height=5cm,width=5cm]{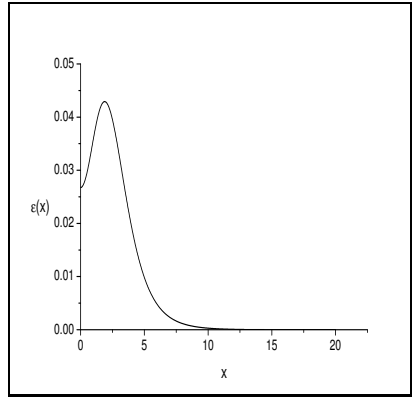}}
    \caption{The energy density $\epsilon(x)$.}
    \label{fig:en-dens}
    \end{center}
\end{figure}
\begin{figure}[h]
  \begin{minipage}[t]{.45\linewidth}
  \begin{center}
    \fbox{
    \includegraphics[height=5cm,width=5cm]{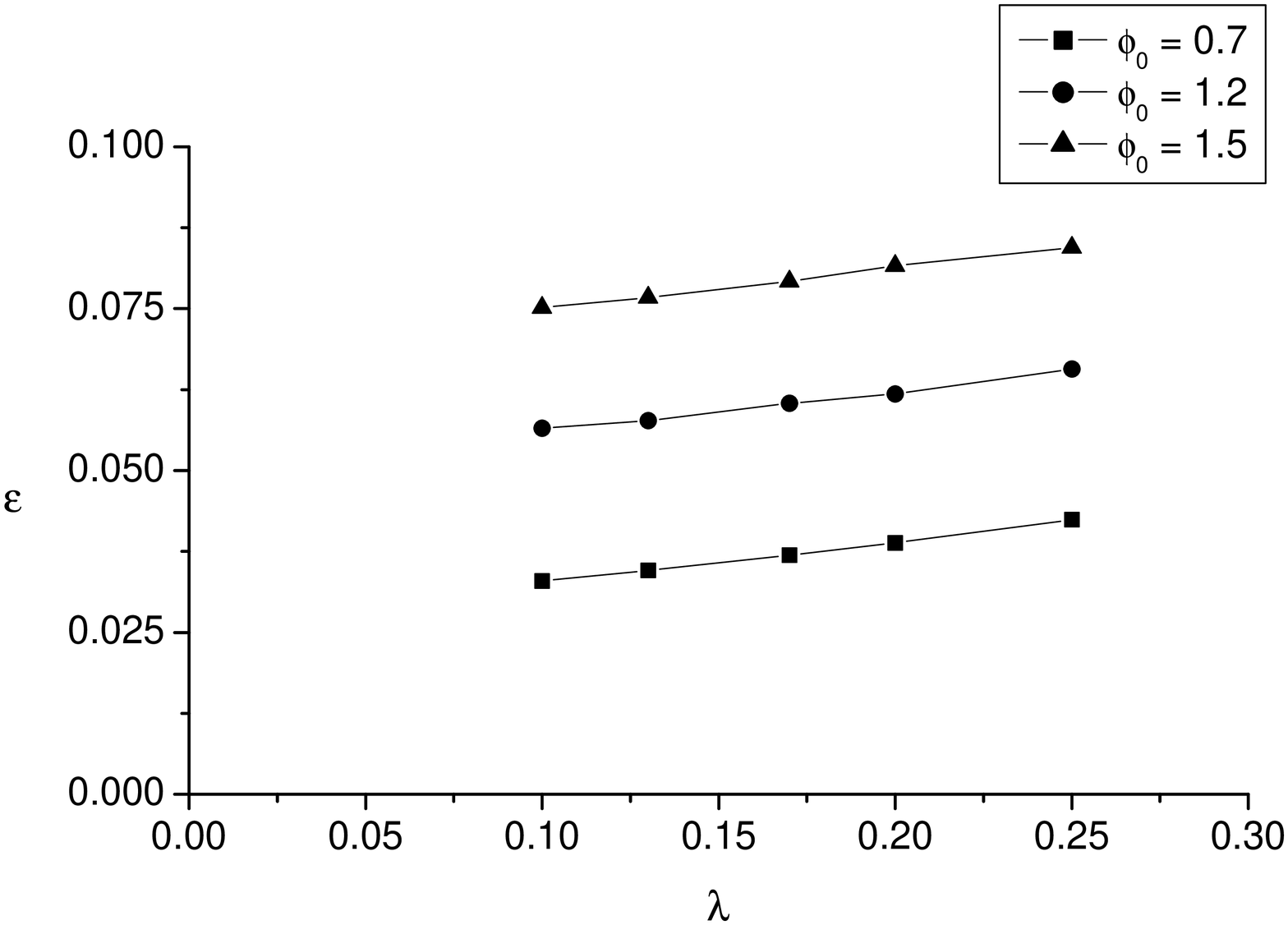}}
    \caption{The profile $\epsilon(\lambda)$ with the different $\phi_0$.}
    \label{en_lam}
  \end{center}
  \end{minipage}\hfill
  \begin{minipage}[t]{.45\linewidth}
  \begin{center}
    \fbox{
    \includegraphics[height=5cm,width=5cm]{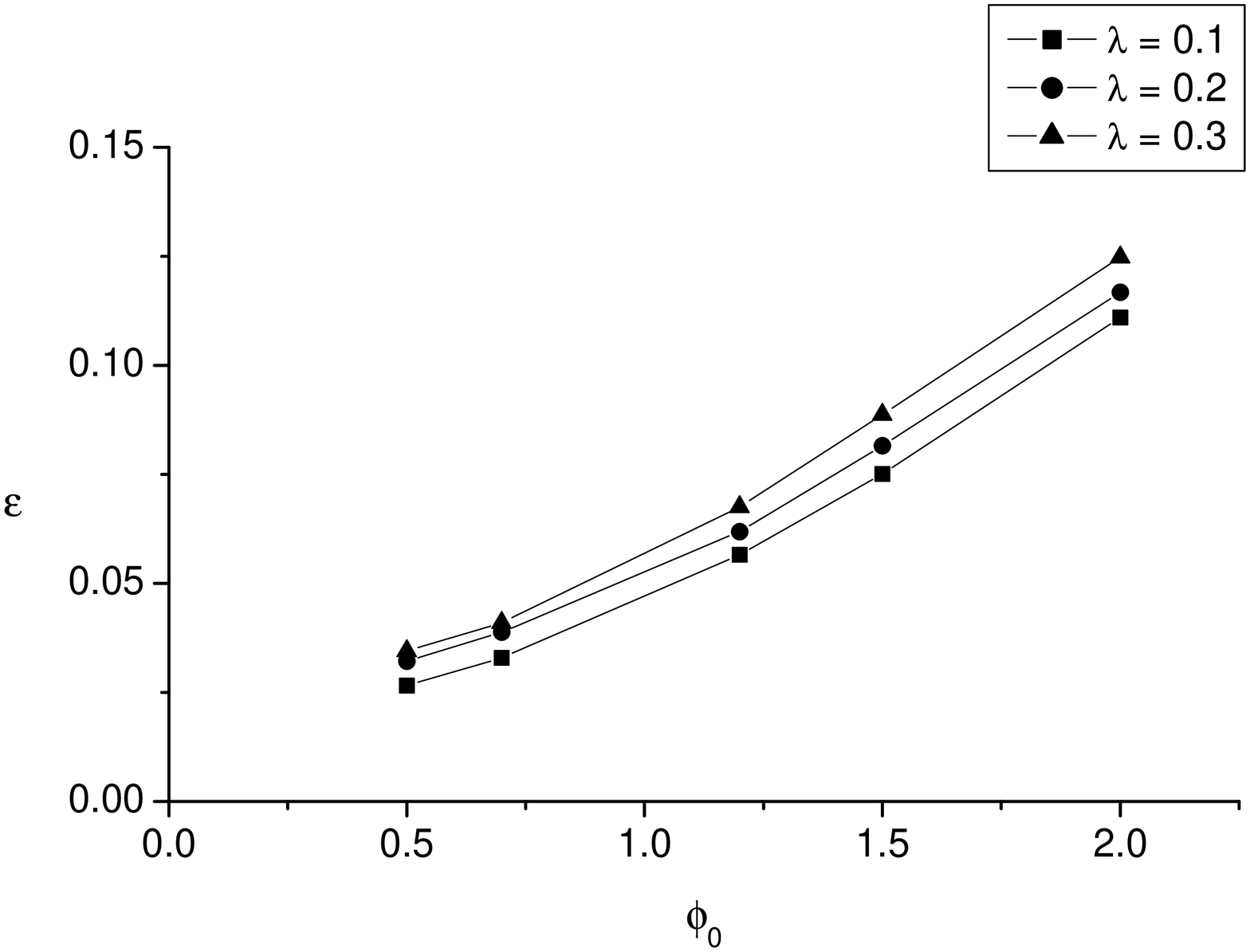}}
    \caption{The profile $\epsilon(\phi_0)$ with the different $\lambda$.}
    \label{en_phi}
  \end{center}
  \end{minipage}
\end{figure}
\par
The flux of the longitudinal electric field is
\begin{equation}
    \Phi = \int E^3_z ds = 2 \pi \int^{\infty}_{0} \rho \frac{f(\rho)v(\rho)}{g}
    d\rho = \frac{2 \pi}{g} \int^{\infty}_{0} x f(x) v(x) dx < \infty.
\label{sec5-630}
\end{equation}
In the figs. \ref{flux_lam1}-\ref{flux_phi2} the profiles of the electric
flux for different $\lambda$ and $\phi_0$ is given. Again the electric
flux was calculated only out to some finite $x$. Eq. \eqref{sec5-630}
assumes that the ansatz functions remain well behaved ({\it i.e.} approach 
zero exponentially as $x \rightarrow \infty$) for larger $x$.
\begin{figure}[h]
  \begin{minipage}[t]{.45\linewidth}
  \begin{center}
    \fbox{
    \includegraphics[height=5cm,width=5cm]{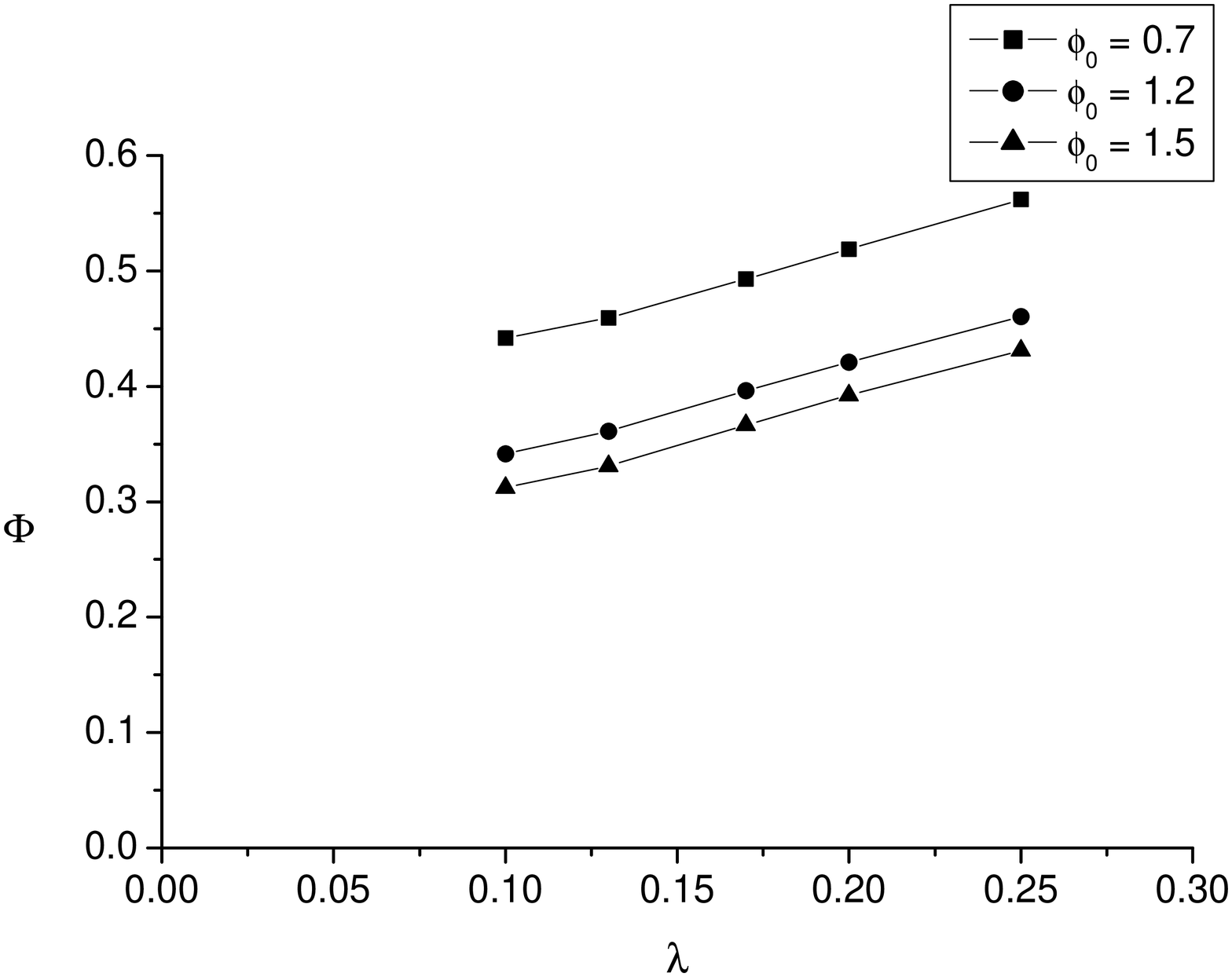}}
    \caption{The profile $\Phi(\lambda)$ with the different $\phi_0$.}
    \label{flux_lam1}
  \end{center}
  \end{minipage}\hfill
  \begin{minipage}[t]{.45\linewidth}
  \begin{center}
    \fbox{
    \includegraphics[height=5cm,width=5cm]{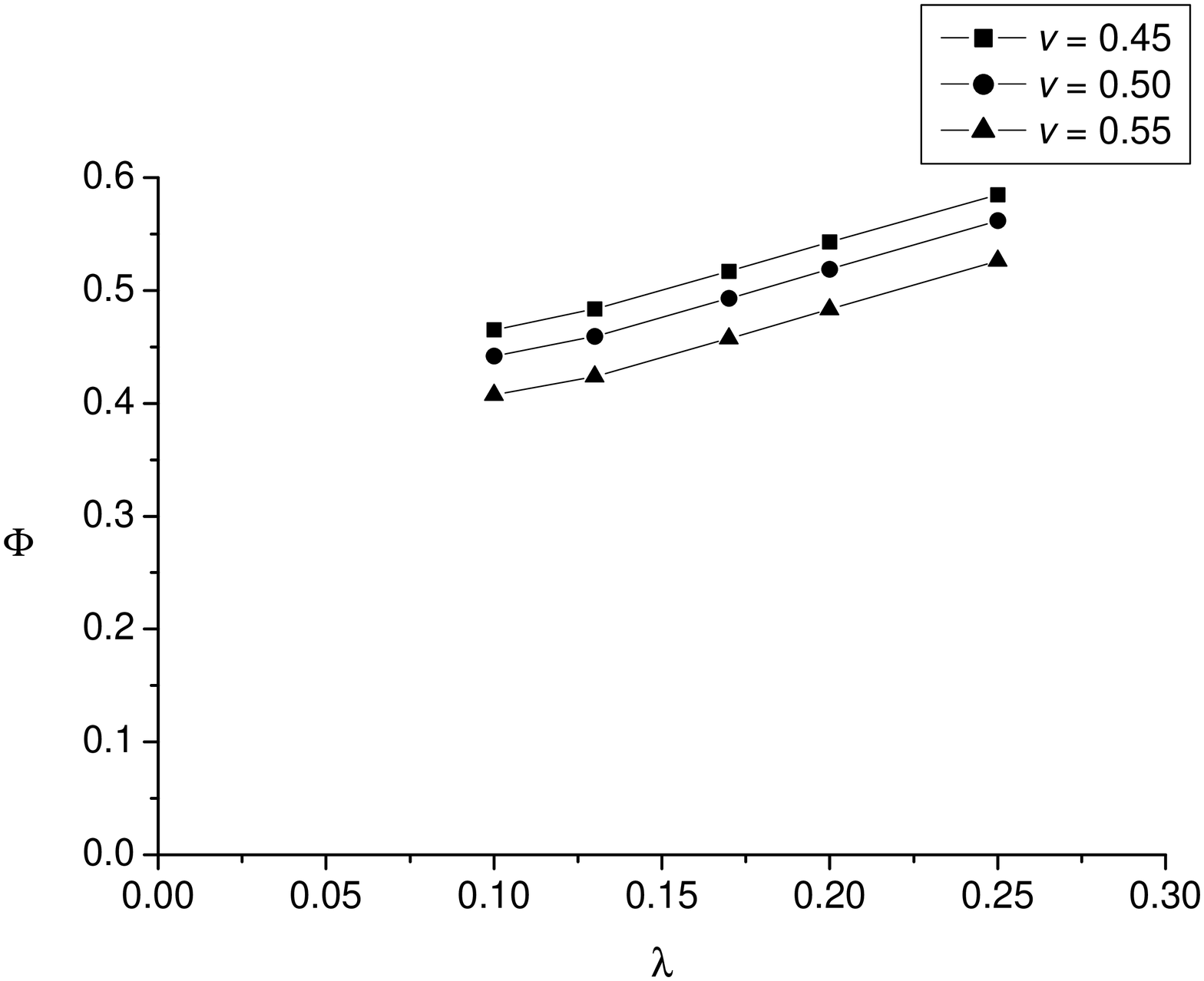}}
    \caption{The profile $\Phi(\lambda)$ with the different $v_0$.}
    \label{flux_lam2}
  \end{center}
  \end{minipage}
\end{figure}

\begin{figure}[h]
  \begin{minipage}[t]{.45\linewidth}
  \begin{center}
    \fbox{
    \includegraphics[height=5cm,width=5cm]{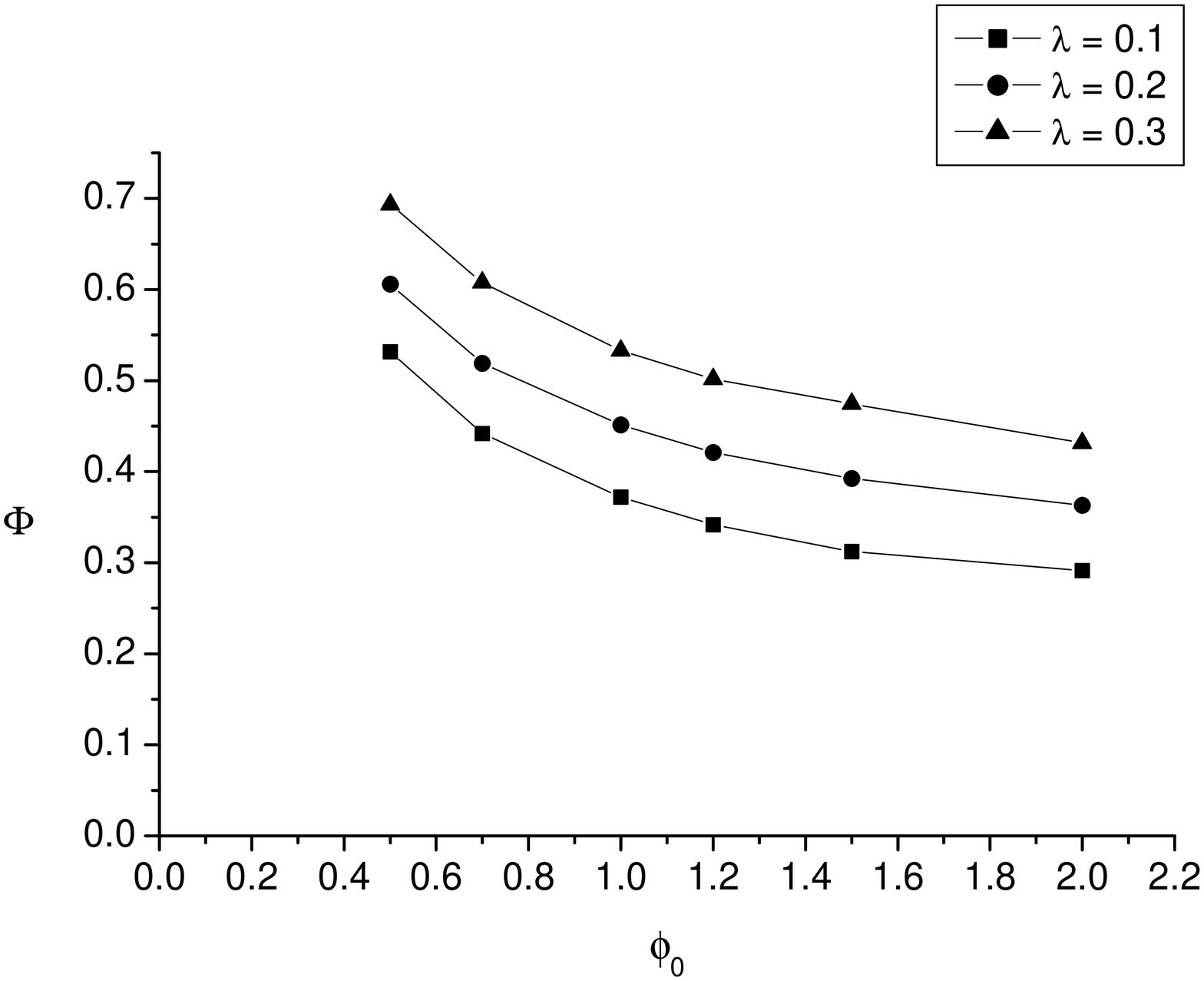}}
    \caption{The profile $\Phi(\phi_0)$ with the different $\phi_0$.}
    \label{flux_phi1}
  \end{center}
  \end{minipage}\hfill
  \begin{minipage}[t]{.45\linewidth}
  \begin{center}
    \fbox{
    \includegraphics[height=5cm,width=5cm]{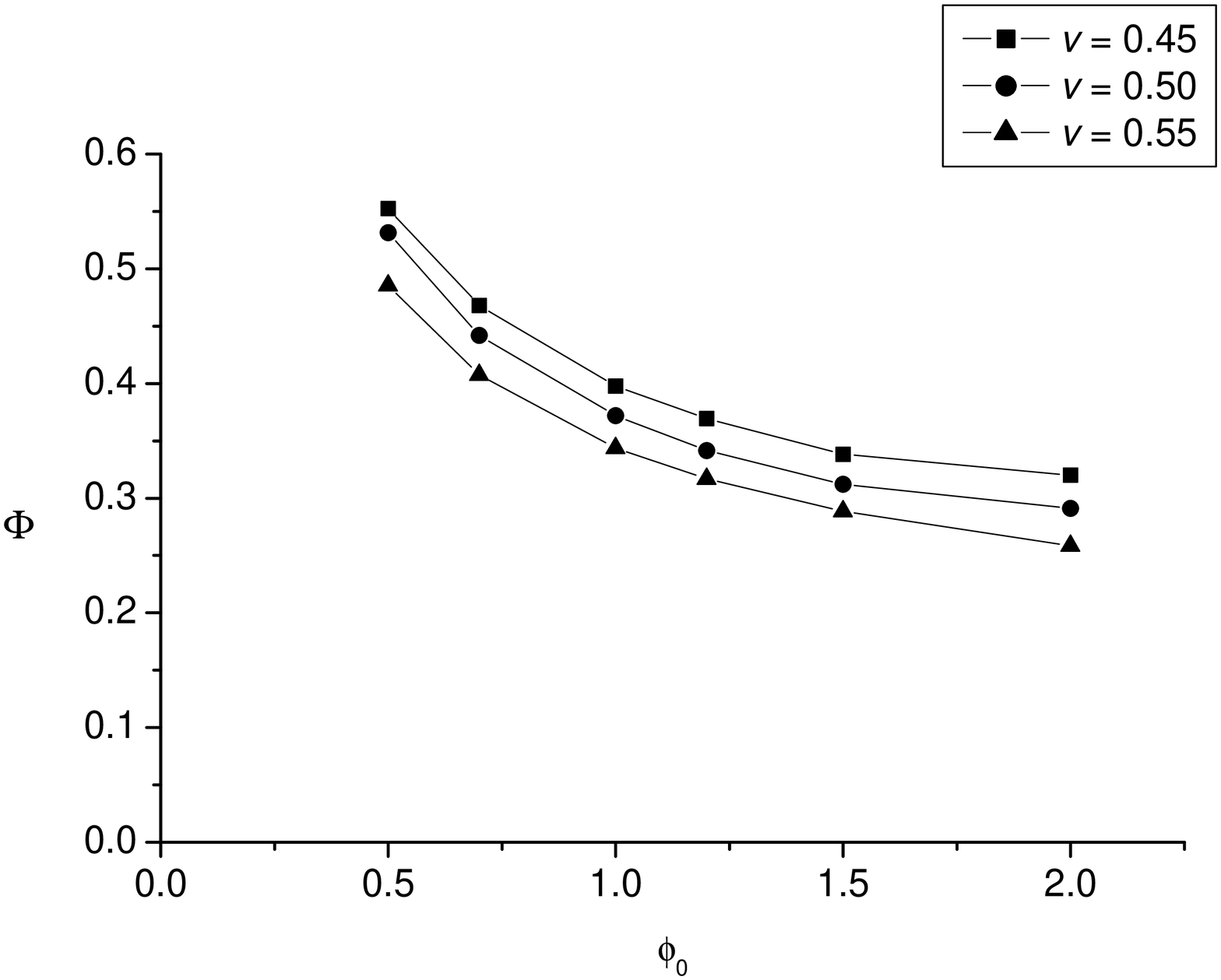}}
    \caption{The profile $\Phi(\phi_0)$ with the different $v_0$.}
    \label{flux_phi2}
  \end{center}
  \end{minipage}
\end{figure}
We see that the flux $\Phi$ depends on the parameters $\lambda,
v(0)$ and $\phi (0)$ since the solutions $f(x)$ and $v(x)$ depend on
these parameters. This is in contrast with such topological
solutions as 't Hooft-Polyakov monopole and Nielsen-Olesen flux
tube which have fixed fluxes. This is the indication of the fact that the colored flux tube
solution is not the consequence of a first order differential equation (the
Bogomolnyi Equation \cite{bogo}) as is the case 
for the topological solutions. The colored 
flux tube is a dynamical object not topological one. This raises questions
about its stability, since this solution is then no longer guranteed to be
stable via topological arguments. 

\section{Discussion and conclusions}

Now we summarize the nonperturbative quantization method of this paper and enumerate 
some features of nonperturbative quantum field theory:
\begin{itemize}
    \item
    The existence and character of stationary, extended objects, such as flux tubes
    or glueballs must be explained using nonperturbative QFT . Such extended objects 
    can not be understood within perturbative QFT as they are not simply a cloud of quanta.
    These objects share some similarity  to a stationary turbulent liquid: there are
    fluctuations in any point within the liquid and simultaneously there exist averaged quantites 
	(velocity, for example) which are correlated at any point as a consequence of the equations of
	motion for the turbulent liquid.
    \item
    It is well known that the ordinary commutations relations like
    \begin{equation}
        \left[
            \widehat \pi(x), \widehat \phi(y)
        \right] = i \delta(x - y)
    \label{sec6-10}
    \end{equation}
are correct only for linear field theory. The algebra of strongly
interacting fields are more complicated and unknown. The algebra
of nonperturbative field operators may be less singular then the
algebra of linear fields in the sense that the Green's functions
for the nonpertubative field operators may be more smooth 
and well behaved. As a consequence the fields
at different spatial points but at one time will be correlated in
contrast with the second (linear) case. The physical explanation
for this is that in the linear case the quantized field at
different points can be correlated only by exchanging
quanta which move with some speed, but in the nonperturbative
case these fields are correlated through nonlinear equations (similar to the 
correlation in a turbulent liquid).
    \item
    Nonperturbative QFT may be important for string theory.
    String theory says nothing about any possible inner structure of the string. By
    studying flux tube solutions such as those presented here one may obtain some
    insight at making a guess as to possible inner structure for strings. Here we would 
    like to cite Polyakov \cite{polyakov}: ``\ldots It might be possible to visualize 
    the superstrings as flux lines of some unknown gauge theory \ldots''
\end{itemize}
\par
As a final remark one can think to apply the present nonperturbative method to other strongly
coupled, nonlinear systems such as gravity. In this way one could see if some insight
into quantum gravity might not be gleaned.

\section{Acknowledgments} DS is supported by a Fulbright Senior Scholars grant, and would
like to thank Prof. Vitaly Melnikov for the invitation to work at PFUR.
VD is very grateful to the ISTC grant KR-677 for the financial support.

\appendix

\section{The flux tube in Euclidean spacetime}
\label{app1}

In the Euclidean spacetime the equations \eqref{sec5-500}-\eqref{sec5-520} can be
simplified. Making the transition to Euclidean spacetime via $f(x) \rightarrow if(x)$ 
we have the following equations
\begin{eqnarray}
    f'' + \frac{f'}{x} &=& f \left( \phi^2 + v^2 - m^2_1 \right),
\label{app1-10}\\
    v'' + \frac{v'}{x} &=& v \left( \phi^2 + f^2 - m^2_2 \right),
\label{app1-20}\\
    \phi'' + \frac{\phi'}{x} &=& \phi \left[ f^2 + v^2
    + \lambda \left( \phi^2 - \phi^2_\infty \right)\right].
\label{app1-30}
\end{eqnarray}
For simplicity one can consider the case $f(x) = v(x)$ and $m_1 = m_2 = m$ which
then leads to only two equations
\begin{eqnarray}
    f'' + \frac{f'}{x} &=& f \left( \phi^2 + f^2 - m^2 \right),
\label{app1-40}\\
    \phi'' + \frac{\phi'}{x} &=& \phi
    \left[ 2 f^2
    + \lambda \left( \phi^2 - \phi^2_\infty \right)\right].
\label{app1-50}
\end{eqnarray}
The numerical method for obtaining a solution is similar to the methods applied in sections
\ref{glueball}, \ref{fluxtube} and the results are presented in figs. \ref{fig:iter-phi} , 
\ref{fig:iter-f} and Table \ref{table1}.
\begin{figure}[h]
  \begin{minipage}[t]{.45\linewidth}
  \begin{center}
    \fbox{
    \includegraphics[height=5cm,width=5cm]{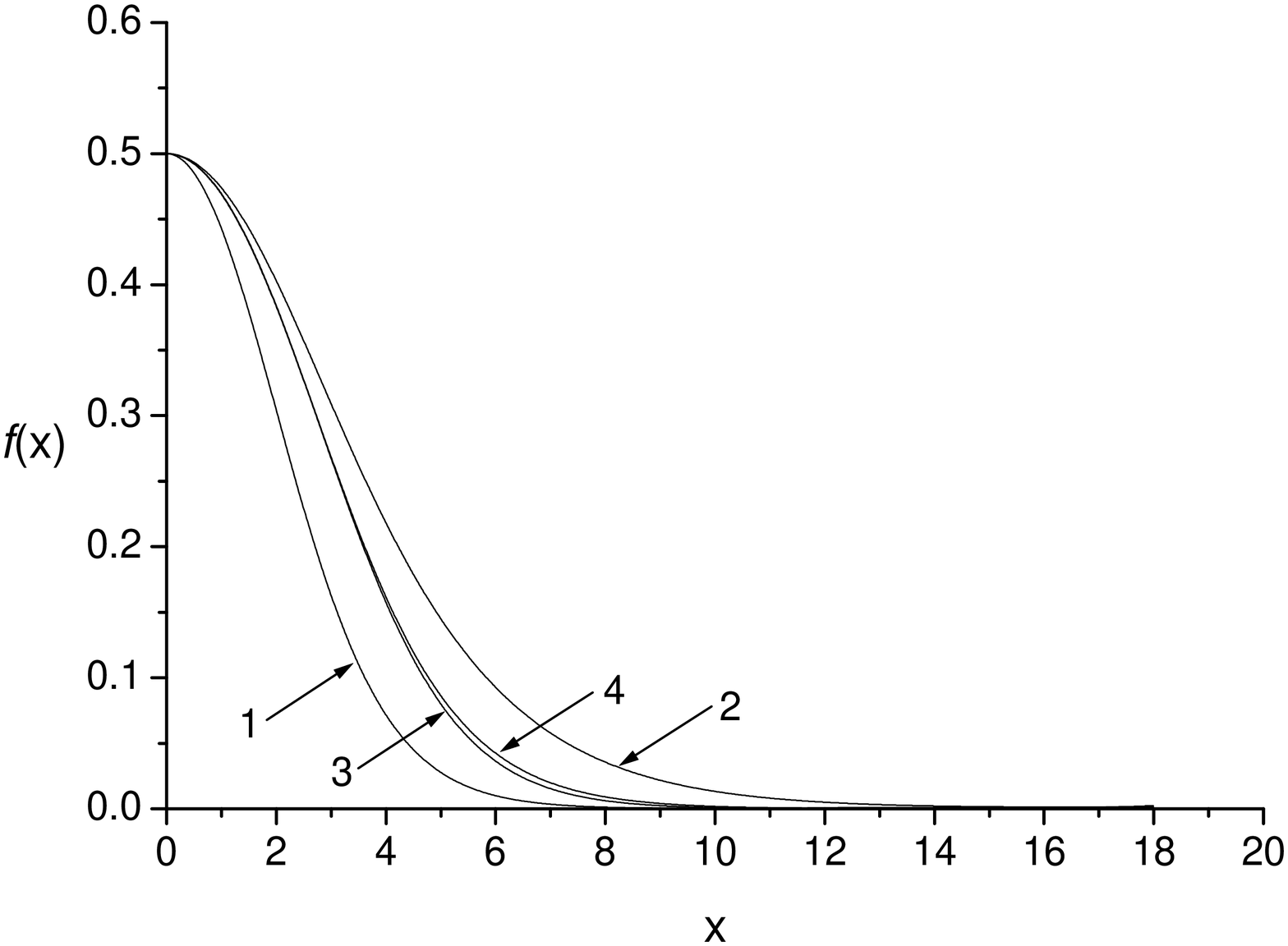}}
    \caption{The iterative solutions $f_i(x)$ for the flux tube in Euclidean spacetime.}
    \label{fig:iter-phi}
  \end{center}
  \end{minipage}\hfill
  \begin{minipage}[t]{.45\linewidth}
  \begin{center}
    \fbox{
    \includegraphics[height=5cm,width=5cm]{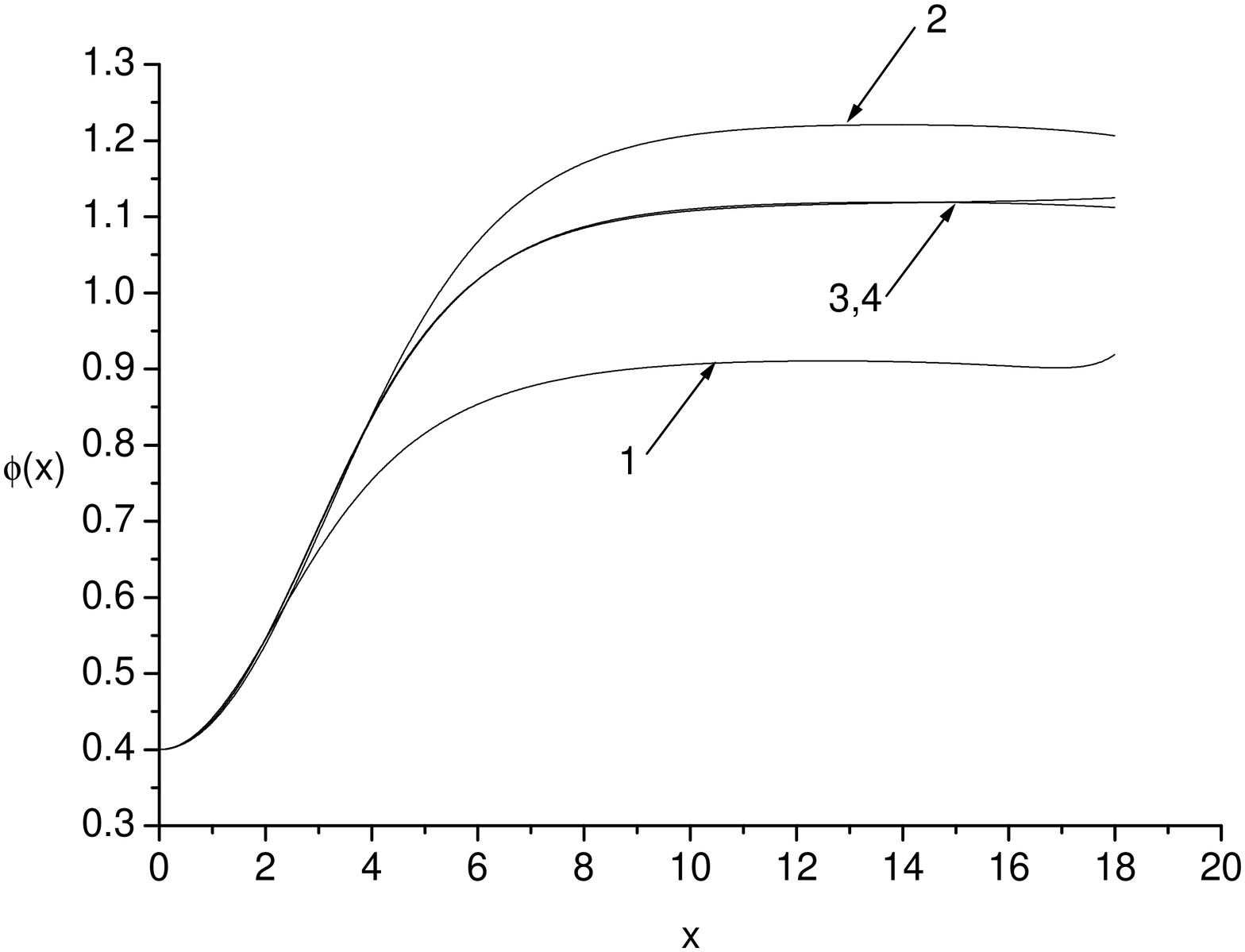}}
    \caption{The iterative solutions $\phi_i(x)$ for the flux tube in Euclidean spacetime.}
    \label{fig:iter-f}
  \end{center}
  \end{minipage}
\end{figure}
\begin{table}[h]
    \begin{center}
        \begin{tabular}{|c|c|c|c|c|}\hline
          i & 1 & 2& 3 & 4 \\ \hline
            $m^*_i$ & 0.9436768025\ldots & 0.79033\ldots & 0.81214284\ldots
            & 0.8139867\ldots \\ \hline
            $\phi^*_{\infty,i}$ & 0.9175\ldots & 1.2248\ldots & 1.1186\ldots
            & 1.12185\ldots \\ \hline
        \end{tabular}
    \end{center}
    \caption{The iterative parameters $m^*_i$ and $\phi^*_{\infty,i}$.}
    \label{table1}
\end{table}
In fig.\ref{fields2} the profile of the color electric and magnetic fields are presented.
\begin{figure}[h]
  \begin{center}
    \fbox{
    \includegraphics[height=5cm,width=7cm]{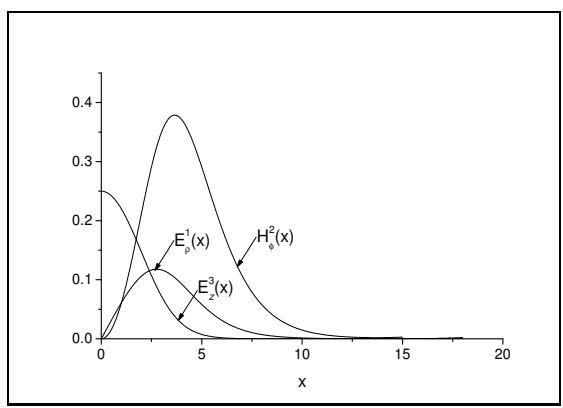}}
    \caption{The profiles of the color electric and magnetic fields.}
    \label{fields2}
  \end{center}
\end{figure}
In figs. \ref{flux_lam_eucl} \ref{flux_phi_eucl} we show the flux $\Phi$
of color electric field
\begin{equation}
    \Phi = \int E^3_z ds = 2 \pi \int^{\infty}_{0} \rho \frac{f(\rho)v(\rho)}{g}
    d\rho = \frac{2 \pi}{g} \int^{\infty}_{0} x f(x) v(x) dx .
\label{app1-60}
\end{equation}
through the tube with respect to parameters $\lambda$ and $\phi_0$.
\begin{figure}[h]
  \begin{minipage}[t]{.45\linewidth}
  \begin{center}
    \fbox{
    \includegraphics[height=5cm,width=5cm]{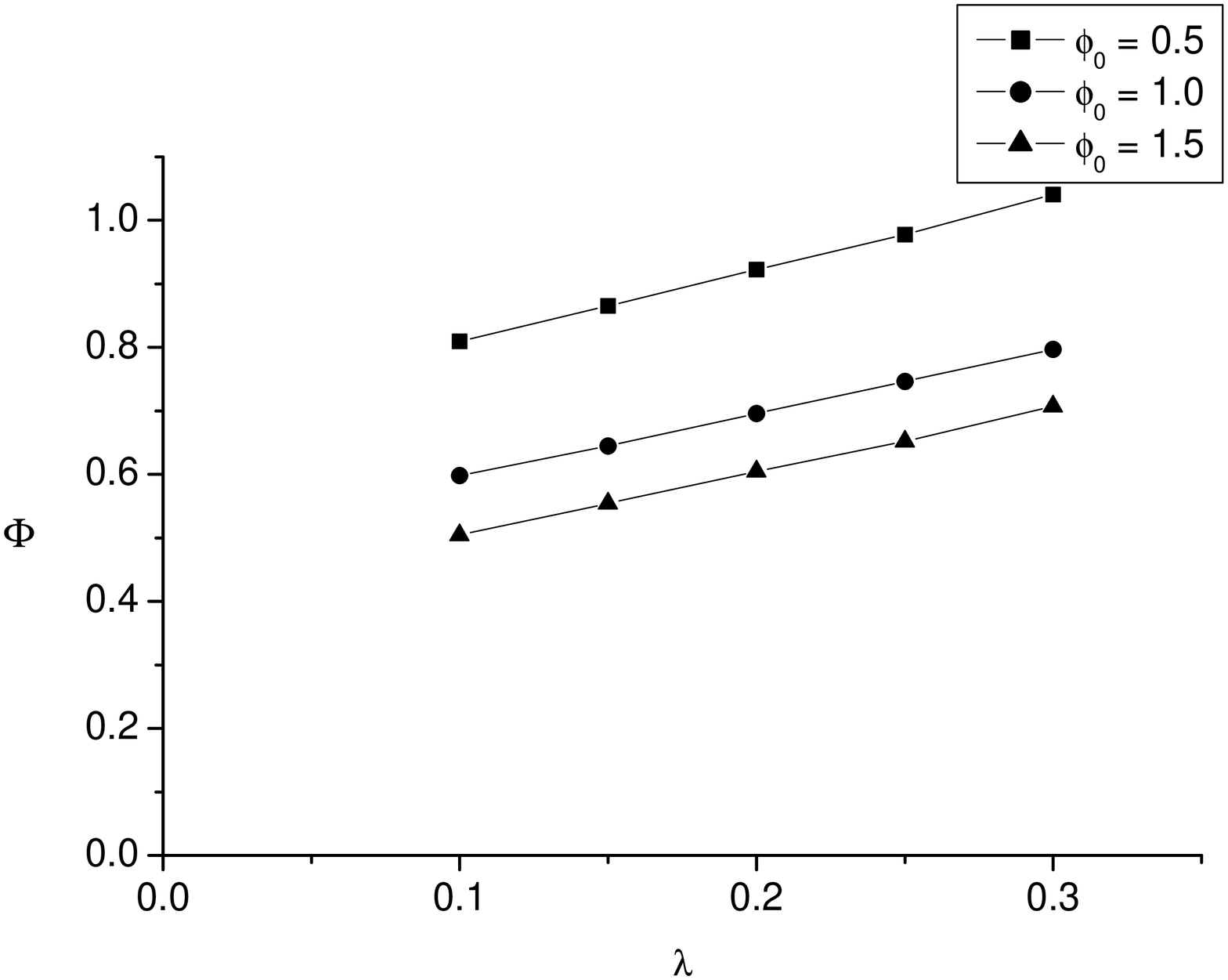}}
    \caption{The flux of the color electric field with respect to $\lambda$ with
    different $\phi_0$.}
    \label{flux_lam_eucl}
  \end{center}
  \end{minipage}\hfill
  \begin{minipage}[t]{.45\linewidth}
  \begin{center}
    \fbox{
    \includegraphics[height=5cm,width=5cm]{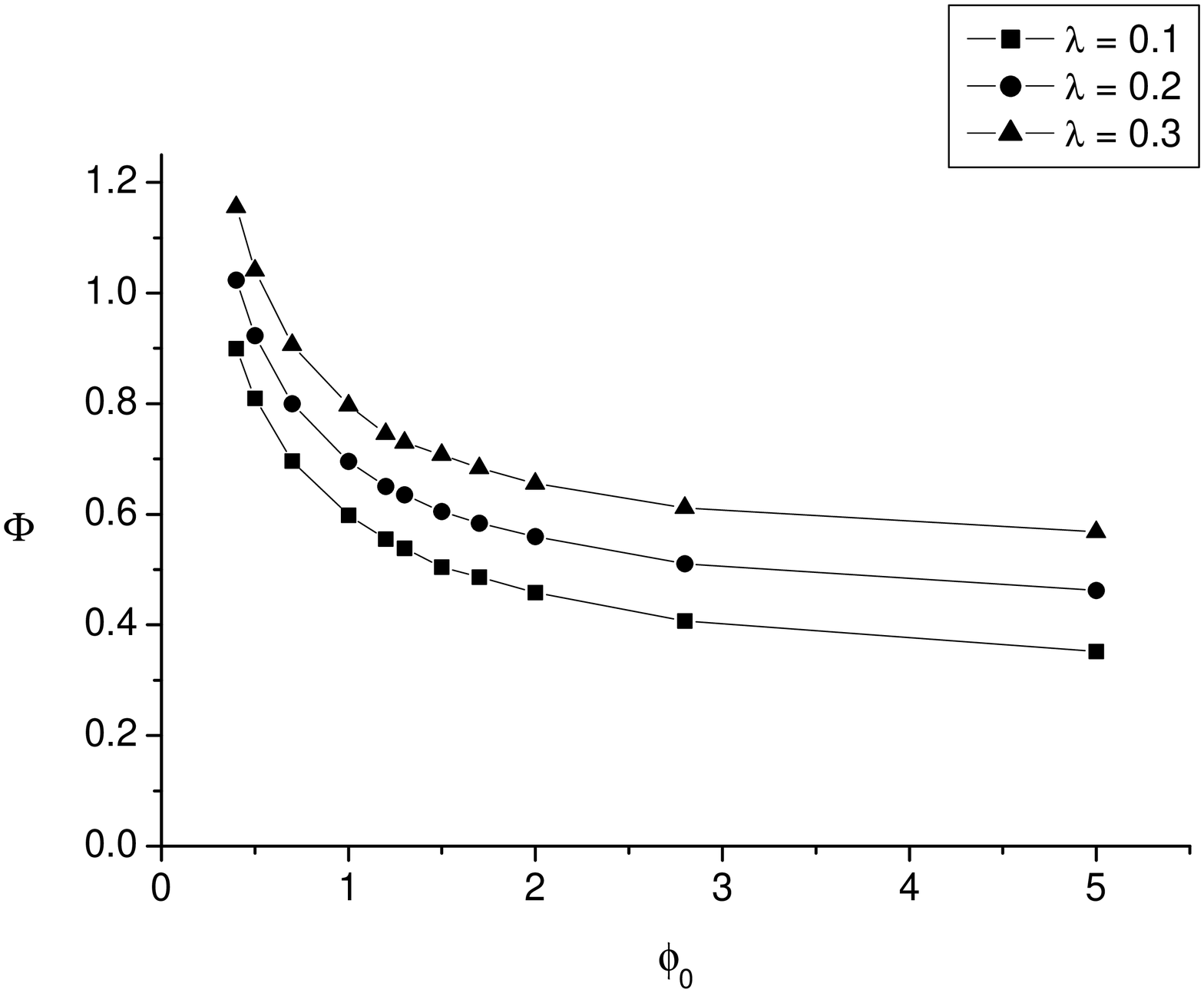}}
    \caption{The flux of the color electric field with respect to $\phi_0$ and with
    different $\lambda$.}
    \label{flux_phi_eucl}
  \end{center}
  \end{minipage}
\end{figure}

\section{The numerical calculations of the 1D $\lambda \phi ^4$ soliton}
\label{app2}

As a test of the numerical method of solving the nonlinear equations
\eqref{sec4-320} \eqref{sec4-330} we will apply this method to a known 
solution namely the $\lambda \phi^4$ soliton in 1D. The equation for
this system is
\begin{equation}
    \frac{d^2 y}{d x^2} = y'' = y \left( 1 - y^2 \right) ~,
\label{app2:10}
\end{equation}
and the solution is
\begin{equation}
    y(x) = \frac{\sqrt{2}}{\cosh x}.
\label{app2:20}
\end{equation}
We rewrite \eqref{app2:10} in the form of the Schr\"odinger equation
\begin{equation}
    -y'' + y V_{eff} = - \lambda y
\label{app2:30}
\end{equation}
where $V_{eff} = -y^2$ and $\lambda = 1$. We will solve this equation 
by an iterative procedure. We start with the equation
\begin{equation}
     -y_1'' + y_1 \left( -y_0^2 \right) =
     - \lambda_1 y_1
\label{app2:50}
\end{equation}
where $y_1(x)$ and $\lambda_1$ are the first approximations of the solutions. 
For the numerical solution we choose the null approximation as
\begin{equation}
    y_0 =  \frac{\sqrt{2}}{\cosh \left( \frac{x}{2} \right)}.
\label{app2:40}
\end{equation}
A typical solution for arbitrary values of the parameter
$\lambda_1$ is presented in fig.\ref{fig:soliton-sing}.
\begin{figure}[h]
  \begin{minipage}[t]{.45\linewidth}
  \begin{center}
    \fbox{
    \includegraphics[height=5cm,width=5cm]{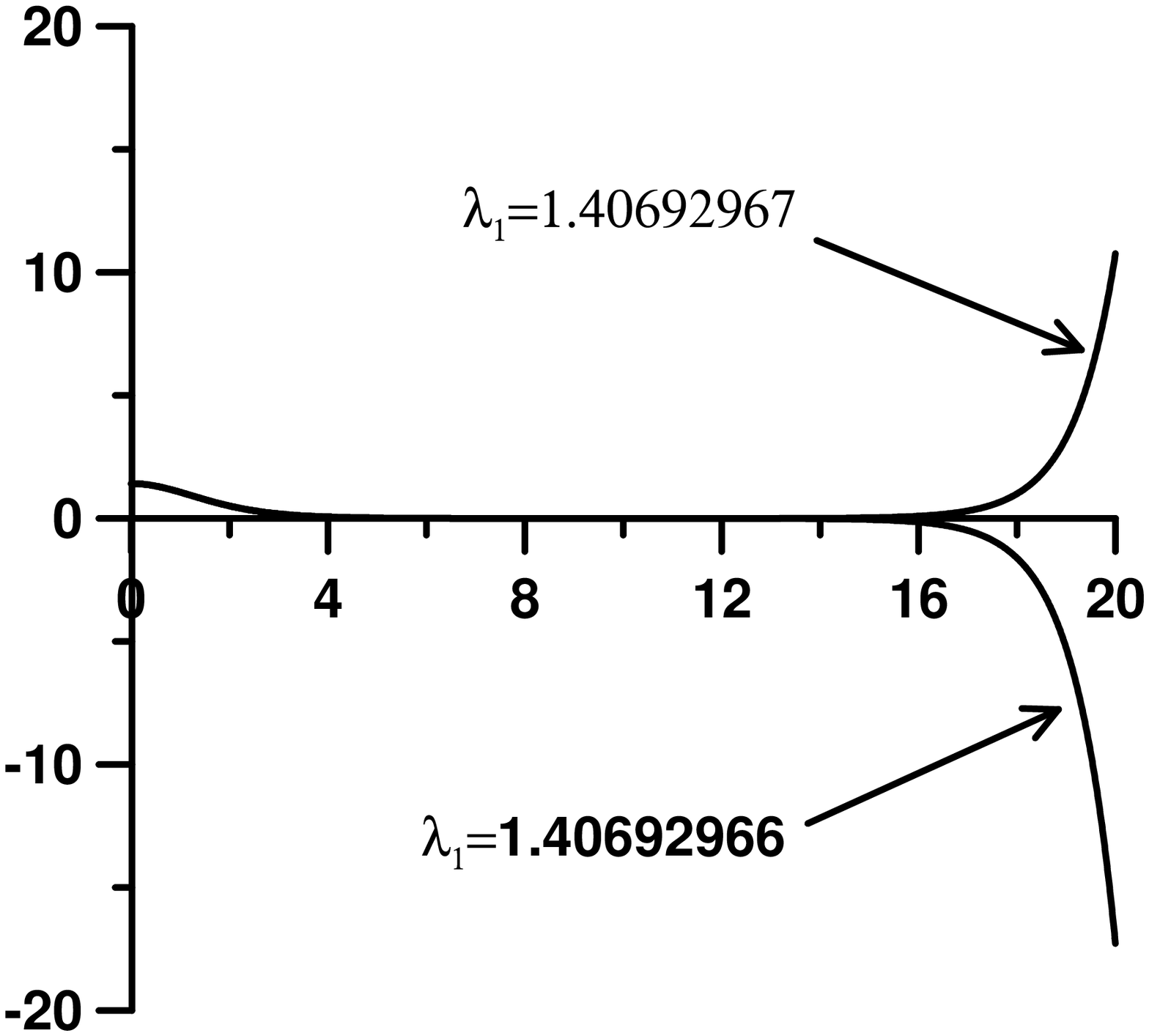}}
    \caption{The singular solutions for the soliton equation.}
    \label{fig:soliton-sing}
  \end{center}
  \end{minipage}\hfill
  \begin{minipage}[t]{.45\linewidth}
  \begin{center}
    \fbox{
    \includegraphics[height=5cm,width=5cm]{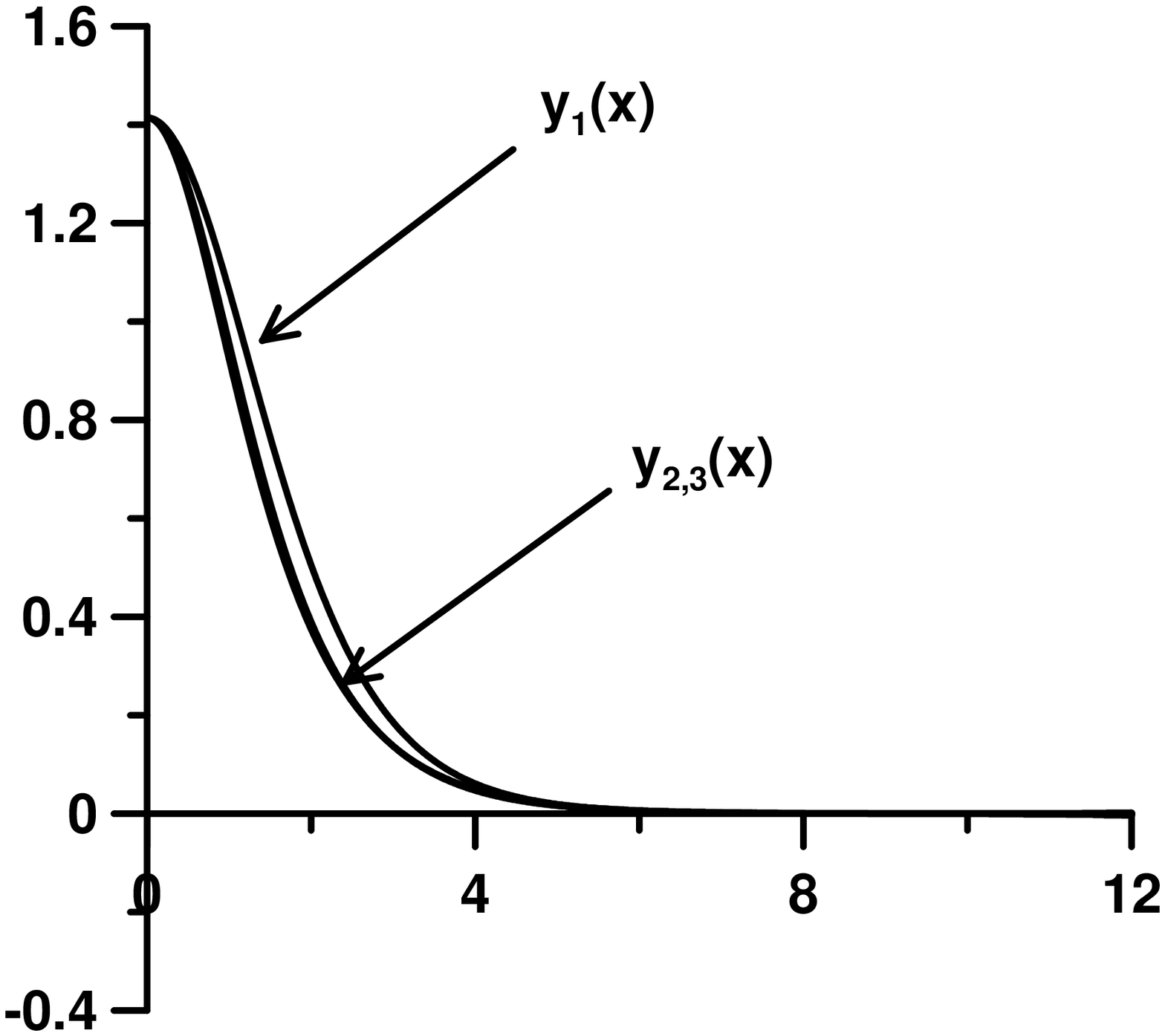}}
    \caption{The iterative functions $y_{1,2,3}$.}
    \label{fig:iter-y}
  \end{center}
  \end{minipage}
\end{figure}
This picture shows that there is a value $\lambda^*_1$ 
for which the solution is exceptional one.
One can find this exceptional solution choosing the appropriate value
of the ``energy level'' $\lambda_1^*$. After which an exceptional solution
$y^*_1(x)$ is substituted into equation for the second approximation
$y_2(x)$
\begin{equation}
     -y_2'' - y_2 \left( y^*_1 \right)^2 = - \lambda_2 y_2
\label{app2:60}
\end{equation}
and so on. The result is presented in Table \ref{table2} and
fig.\ref{fig:soliton-sing}. One can see that $\lambda^*_i \rightarrow 1$
and $y^*_i(x)$ is convergent to $y^*(x)$.
\begin{table}[h]
    \begin{center}
        \begin{tabular}{|c|c|c|c|}\hline
          i & 1 & 2& 3 \\ \hline
            $\lambda^*_i$ & 1.406929666\ldots & 1.148564915\ldots & 1.03968278\ldots
            \\ \hline
        \end{tabular}
    \end{center}
    \caption{The iterative values of the parameter $\lambda$.}
    \label{table4}
\end{table}

\end{document}